\documentclass[sigconf,edbt]{acmart-edbt2018}
\usepackage{booktabs} % For formal tables
\usepackage{times}
\usepackage{epsfig}
\usepackage{subcaption}

\usepackage{amsthm}
\usepackage{amsmath}
\usepackage{amssymb}
\usepackage{comment}
\usepackage{balance}
\usepackage{tikz}
\usepackage{multirow}
\usepackage{url}
\usepackage{romannum}
\usepackage{mathtools}
\usetikzlibrary{arrows}
\usetikzlibrary{shapes,backgrounds}
\usepackage{pgfplots}
\usepgfplotslibrary{fillbetween}

\makeatletter
\newif\if@restonecol
\makeatother

\usepackage[utf8]{inputenc}
\usepackage[ruled, vlined]{algorithm2e}

\usepackage{microtype}
\newcommand{\eat}[1]{}
\newcommand{\attrname}[1]{\textsf{#1}}
\newcommand{\smallattrname}[1]{\textsf{\small #1}}

\newcommand\xqed[1]{%
  \leavevmode\unskip\penalty9999 \hbox{}\nobreak\hfill
  \quad\hbox{#1}}
\newcommand\closedef{\xqed{$\triangle$}}
\renewcommand\qed{\xqed{\mbox{\raggedright \rule[0pt]{1.2ex}{1.2ex}}}}

\newcommand{\algname}[1]{{\small \textsf{#1}}}
\newcommand{\solname}[1]{{\small \textsf{#1}}}

\newcommand{\obj}[1]{\emph{#1}}

\newcommand*{\Perm}[2]{P^{#1}_{#2}}
\newcommand*{\ParetoApprox}[1]{\widehat{\mathcal{P}}_{#1}}

\makeatletter
\newcommand\semiHuge{\@setfontsize\semiHuge{23.72}{27.38}}
\makeatother

\setcopyright{rightsretained}

\acmDOI{}

\acmISBN{978-3-89318-078-3}

\acmConference[EDBT 2018]{21st International Conference on Extending Database Technology (EDBT)}{March 26-29, 2018}{Vienna, Austria} 
\acmYear{2018}

\begin{document}

% Title portion
\title{Continuous Monitoring of Pareto Frontiers on Partially Ordered Attributes for Many Users}

\author{Afroza Sultana}
\affiliation{%
  \institution{University of Texas at Arlington}
  %\city{Arlington} 
  %\state{Texas}
}
%\email{afroza.sultana@mavs.uta.edu}

\author{Chengkai Li}
\affiliation{%
  \institution{University of Texas at Arlington}
  %\city{Arlington} 
  %\state{Texas}
}
%\email{cli@uta.edu}

\begin{abstract}
We study the problem of \emph{continuous object dissemination}---given a large number of users and continuously arriving new objects, deliver an object to all users who prefer the object. Many real world applications analyze users' preferences for effective object dissemination. %%%For instance, Facebook considers user's feedback in selecting news feeds interesting to the user. 
For continuously arriving objects, timely finding users who prefer a new object is challenging. In this paper, we consider an append-only table of objects with multiple attributes and  users' preferences on  individual attributes are modeled as \emph{strict partial orders}. An object is preferred by a user if it belongs to the \emph{Pareto frontier} with respect to the user's partial orders. Users' preferences can be similar. Exploiting shared computation across similar preferences of different users, we design algorithms to find \emph{target users} of a new object. In order to find users of similar preferences, we study the novel problem of clustering users' preferences that are represented as partial orders. We also present an approximate solution of the problem of finding target users which is more efficient than the exact one while ensuring sufficient accuracy. Furthermore, we extend the algorithms to operate under the semantics of sliding window. We present the results from comprehensive experiments for evaluating the efficiency and effectiveness of the proposed techniques.\vspace{-2mm}
\end{abstract}

\maketitle
\section{Introduction}\label{sec:intro}
Many applications serve users better by disseminating objects to the
users according to their preferences.  User preferences can be modeled
via a variety of means including \emph{collaborative
filtering}~\cite{collaborativefiltering}, \emph{top-k
ranking}~\cite{Fagin96,FaginLN01}, \emph{skyline}~\cite{656550}, and general
\emph{preference queries}~\cite{kiessling2002foundations,chomicki2003preference}.
In various scenarios, users' preferences stand or only change occasionally, while the objects
keep coming continuously.  Such scenarios warrant the need for
a capability of continuous monitoring of preferred objects.
While previous studies have made notable contributions on continuous evaluation
of skyline~\cite{wu2007deltasky, lee2007approaching} and top-\emph{k} queries~\cite{yu2012processing}, we note that two important
considerations are missing from prior works:
\begin{list}{$\bullet$}
{ \setlength{\leftmargin}{1em} \setlength{\partopsep}{1pt} \setlength{\topsep}{-\parskip} \setlength{\itemsep}{1pt} \setlength{\parsep}{1pt}}
\item \emph{Many users}: There may be a large number of users and the
users may have similar preferences. Prior studies focus on the query
needs of one user and thus their algorithmic solutions can only be
applied separately on individual users.  A solution can potentially
attain significant query performance gain by leveraging users' \emph{common
preferences}.

\item \emph{Partially ordered attributes}:
Prior works focus on top-\emph{k} and skyline queries.  In
multi-objective optimization, a more general concept than skyline is
\emph{Pareto frontier}. Consider a table of objects with a set of
attributes.  An object is \emph{Pareto-optimal} (i.e., it belongs to
the Pareto frontier) if and only if it is not dominated by any other
object~\cite{barndorff1966distribution,paretoopt}.  Object \obj{y} dominates \obj{x} if and only if \obj{y} is
better than or equal to \obj{x} on every attribute and is
better on at least one attribute.  In defining the \emph{better-than} relations,
most studies on skyline queries assume a total order on the ordinal or numeric values of an
attribute, except for~\cite{sacharidis2009topologically,zhang2010efficient}
which consider strict partial orders.  The psychological nature
of human's preferences determines that it is not always natural to enforce
a total order.  Oftentimes real-world preferences can only be modeled as
strict partial orders~\cite{kiessling2002foundations,chomicki2003preference,sacharidis2009topologically}.
\end{list}

\vspace{1mm} Consider the following motivating applications which monitor
Pareto frontiers on partially ordered attributes for many users.
\begin{list}{$\bullet$}
{ \setlength{\leftmargin}{1em} \setlength{\partopsep}{1pt} \setlength{\topsep}{-\parskip} \setlength{\itemsep}{1pt} \setlength{\parsep}{1pt}}
\item Social network content and news delivery: It is often impossible and unnecessary for a user to keep up with the plethora of updates (e.g., news feeds in Facebook) from their social circles.  When a new item is posted, if the item is Pareto-optimal with respect to a user, it can be displayed above other updates in the user's view.  Similar ideas can be adopted by mass media to ensure their news reaches the right audience.  User preferences can be modeled on content creator, topic, location, and so on.  Enforcing total orders on such attributes is both cumbersome and unnatural.
%%%\vspace{-2mm}{\flushleft \emph{News delivery}}:  A mass media needs to ensure their news reaches the right audience. Each news item is associated with attributes such as source, category and location.  A reader's preference on each attribute is modeled as a partial order. Pareto-optimal items get higher priority to be brought to the reader's attention.
\item Publication alerts: Bibliography servers such as PubMed and Google Scholar can notify users about newly published articles matching their preferences on venues and keywords. Such attributes do not welcome total orders either. 
\item Product recommendation: When a new product becomes available, a retailer can notify customers who may be interested. It can distill customers' preferences on product specifications (e.g., brand, display and memory for laptops) from profiles, past transactions and website browsing logs.  Example~\ref{exp:exp_intro} discusses this application more concretely.\vspace{-2mm}
\end{list}

\begin{example}\label{exp:exp_intro}
Consider an inventory of laptops in Table~\ref{tab:product} and customers' preferences on individual product attributes (\smallattrname{display}, \smallattrname{brand} and \smallattrname{CPU}) modeled as strict partial orders in Table~\ref{tab:customer}.  For an attribute, the corresponding strict partial order is depicted as a directed acyclic graph (DAG), more specifically a Hasse diagram. Given two values \emph{x} and \emph{y} in the attribute's domain, the existence of a path from \emph{x} to \emph{y} in the DAG implies that \emph{x} is preferred to \emph{y}. With respect to customer $c_1$ and attribute \smallattrname{brand}, the path from \emph{Lenovo} to \emph{Toshiba} implies that $c_1$ prefers \emph{Lenovo} to \emph{Toshiba}. There is no path between \emph{Toshiba} and \emph{Samsung}, which indicates $c_1$ is indifferent between the two brands.

The strict partial orders on various attributes together represent a customer's preferences on objects. For instance, $c_1$ prefers $o_2$$=$$\langle 14$, \emph{Apple}, \emph{dual}$\rangle$ to $o_1$$=$$\langle 12$, \emph{Apple}, \emph{single}$\rangle$, since they prefer $13$$-$$15.9$ to $10$$-$$12.9$ on \smallattrname{display} and \emph{dual} to \emph{single} on \smallattrname{CPU}. With regard to $o_1$ and $o_3$$=$$\langle 15$, \emph{Samsung}, \emph{dual}$\rangle$, $c_1$ does not prefer one over the other because, though they prefer $13$$-$$15.9$ to $10$$-$$12.9$ and \emph{dual} to \emph{single}, they prefer \emph{Apple} to \emph{Samsung} on \smallattrname{brand}.

According to the data in Tables~\ref{tab:product} and~\ref{tab:customer}, if the existing products are $o_1$ to $o_{14}$ (ignore $o_{15}$ and $o_{16}$ for now), the Pareto frontiers of $c_1$ and $c_2$ are \{$o_2$\} and \{$o_2$, $o_3$\}, respectively. Suppose $o_{15}$$=$$\langle 16.5$, \emph{Lenovo}, \emph{quad}$\rangle$ just becomes available.  For $c_1$, $o_{15}$ does not belong to the Pareto frontier. It is dominated by $o_2$, because $c_1$ prefers $14$-inch display over $16.5$-inch, \emph{Apple} over \emph{Lenovo}, and \emph{dual}-core CPU over \emph{quad}-core CPU. However, $o_{15}$ is a Pareto-optimal object for $c_2$ since it is not dominated by any other object according to $c_2$'s preferences. It is thus recommended to $c_2$, and the Pareto frontier of $c_2$ is updated to \{$o_2$, $o_3$, $o_{15}$\}.\closedef\vspace{-1.5mm}
\end{example}

\begin{table}[t]
\centering
\scriptsize
\begin{tabular}{|l|*{4}{c|}}\hline
$ $ & \smallattrname{{display}} & \smallattrname{{brand}} & \smallattrname{{CPU}}\\
\hline
\hline
{\scriptsize $o_1$} & 12 & \emph{Apple} & \emph{single}\\
\hline
{\scriptsize $o_2$} & 14 & \emph{Apple} & \emph{dual}\\
\hline
{\scriptsize $o_3$} & 15 & \emph{Samsung} & \emph{dual}\\
\hline
{\scriptsize $o_4$} & 19 & \emph{Toshiba} & \emph{dual}\\
\hline
{\scriptsize $o_5$} & 9 & \emph{Samsung} & \emph{quad}\\
\hline
{\scriptsize $o_6$} & 11.5 & \emph{Sony} & \emph{single}\\
\hline
{\scriptsize $o_7$} & 9.5 & \emph{Lenovo} & \emph{quad}\\
\hline
{\scriptsize $o_8$} & 12.5 & \emph{Apple} & \emph{dual}\\
\hline
{\scriptsize $o_9$} & 19.5 & \emph{Sony} & \emph{single}\\
\hline
{\scriptsize $o_{10}$} & 9.5 & \emph{Lenovo} & \emph{triple}\\
\hline
{\scriptsize $o_{11}$} & 9 & \emph{Toshiba} & \emph{triple}\\
\hline
{\scriptsize $o_{12}$} & 8.5 & \emph{Samsung} & \emph{triple}\\
\hline
{\scriptsize $o_{13}$} & 14.5 & \emph{Sony} & \emph{dual}\\
\hline
{\scriptsize $o_{14}$} & 17 & \emph{Sony} & \emph{single}\\
\hline
\hline
{\scriptsize $o_{15}$} & 16.5 & \emph{Lenovo} & \emph{quad}\\
\hline
\hline
\hline
{\scriptsize $o_{16}$} & 16 & \emph{Toshiba} & \emph{single}\\
\hline
\end{tabular}
\caption{\small Product table.}\vspace{-6mm}
\label{tab:product}
\end{table}

\begin{table}[t]
\centering
\scriptsize
\begin{tabular}{|l|*{4}{@{\hskip-1pt}c@{\hskip-1pt}|}}\hline
$ $ & \smallattrname{\small{display}} & \smallattrname{\small{brand}} & \smallattrname{\small{CPU}}\\
\hline
\hline
{\small{$c_1$}} & \begin{tikzpicture}
  \node (one) at (0,0) {$13$$-$$15.9$};
  \node (two) at (0,-0.6) {$10$$-$$12.9$};
  \node (three) at (-0.55,-1.2) {$16$$-$$18.9$};
  \node (four) at (0.55,-1.2) {$19$$-$\emph{up}};
  \node (five) at (0,-1.8) {$9.9$$-$\emph{under}};
  \draw[->] (one) -- (two);
  \draw[->] (two) -- (three);
  \draw[->] (two) -- (four);
  \draw[->] (three) -- (five);
  \draw[->] (four) -- (five);
\end{tikzpicture} & \begin{tikzpicture}
  \node (one) at (0,0) {\emph{Apple}};
  \node (two) at (0,-0.6) {\emph{Lenovo}};
  \node (three) at (0,-1.3) {\emph{Sony}};
  \node (four) at (-0.75,-1.8) {\emph{Toshiba}};
  \node (five) at (0.75,-1.8) {\emph{Samsung}};
  \draw[->] (one) -- (two);
  \draw[->] (two) -- (three);
  \draw[->] (three) -- (four);
  \draw[->] (three) -- (five);
\end{tikzpicture} & \begin{tikzpicture}
  \node (one) at (0,0) {\emph{dual}};
  \node (two) at (-0.55,-0.6) {\emph{triple}};
  \node (three) at (0.55,-0.6) {\emph{quad}};
  \node (four) at (0,-1.2) {\emph{single}};
  \draw[->] (one) -- (two);
  \draw[->] (one) -- (three);
  \draw[->] (one) -- (three);
  \draw[->] (two) -- (four);
  \draw[->] (three) -- (four);
\end{tikzpicture}\\
\hline
{\small{$c_2$}} & \begin{tikzpicture}
  \node (one) at (0,0) {$13$$-$$15.9$};
  \node (two) at (-0.55,-0.6) {$10$$-$$12.9$};
  \node (three) at (0.55,-0.6) {$16$$-$$18.9$};
  \node (four) at (0,-1.2) {$19$$-$\emph{up}};
  \node (five) at (0,-1.8) {$9.9$$-$\emph{under}};
  \draw[->] (one) -- (two);
  \draw[->] (one) -- (three);
  \draw[->] (two) -- (four);
  \draw[->] (three) -- (four);
  \draw[->] (four) -- (five);
\end{tikzpicture} & \begin{tikzpicture}
  \node (one) at (-0.55,-0.6) {\emph{Apple}};
  \node (two) at (0,0) {\emph{Lenovo}};
  \node (three) at (0,-1.8) {\emph{Sony}};
  \node (four) at (0,-1.2) {\emph{Toshiba}};
  \node (five) at (0.75,-0.6) {\emph{Samsung}};
  \draw[->] (two) -- (one);
  \draw[->] (two) -- (five);
  \draw[->] (one) -- (four);
  \draw[->] (five) -- (four);
  \draw[->] (four) -- (three);
\end{tikzpicture} & \begin{tikzpicture}
  \node (one) at (0,-1.2) {\emph{dual}};
  \node (two) at (0,-0.6) {\emph{triple}};
  \node (three) at (0,0) {\emph{quad}};
  \node (four) at (0,-1.8) {\emph{single}};
  \draw[->] (three) -- (two);
  \draw[->] (two) -- (one);
  \draw[->] (one) -- (four);
\end{tikzpicture}\\
\hline
{\small{$U$}} & \begin{tikzpicture}
  \node (one) at (0,0) {$13$$-$$15.9$};
  \node (two) at (-0.55,-0.6) {$10$$-$$12.9$};
  \node (three) at (0.55,-0.6) {$16$$-$$18.9$};
  \node (four) at (-0.55,-1.2) {$19$$-$\emph{up}};
  \node (five) at (0,-1.8) {$9.9$$-$\emph{under}};
  \draw[->] (one) -- (two);
  \draw[->] (one) -- (three);
  \draw[->] (three) -- (five);
  \draw[->] (two) -- (four);
  \draw[->] (four) -- (five);
\end{tikzpicture} & \begin{tikzpicture}
  \node (one) at (-0.5,0) {\emph{Apple}};
  \node (two) at (0.5,0) {\emph{Lenovo}};
  \node (three) at (0,-0.6) {\emph{Sony}};
  \node (four) at (-1,-0.6) {\emph{Toshiba}};
  \node (five) at (1,-0.6) {\emph{Samsung}};
  \draw[->] (one) -- (four);
  \draw[->] (one) -- (three);
  \draw[->] (two) -- (four);
  \draw[->] (two) -- (three);
  \draw[->] (two) -- (five);
\end{tikzpicture} & \begin{tikzpicture}
  \node (one) at (-0.9,0) {\emph{dual}};
  \node (two) at (0,0) {\emph{triple}};
  \node (three) at (0.9,0) {\emph{quad}};
  \node (four) at (0,-0.6) {\emph{single}};
  \draw[->] (one) -- (four);
  \draw[->] (two) -- (four);
  \draw[->] (three) -- (four);
\end{tikzpicture}\\
\hline
{\small{$\widehat{U}$}} & \begin{tikzpicture}
  \node (one) at (0,0) {$13$$-$$15.9$};
  \node (two) at (0,-0.6) {$10$$-$$12.9$};
  \node (three) at (0,-1.2) {$16$$-$$18.9$};
  \node (four) at (0,-1.8) {$19$$-$\emph{up}};
  \node (five) at (0,-2.4) {$9.9$$-$\emph{under}};
  \draw[->] (one) -- (two);
  \draw[->] (two) -- (three);
  \draw[->] (three) -- (four);
  \draw[->] (four) -- (five);
\end{tikzpicture} & \begin{tikzpicture}
  \node (one) at (-0.75,0) {\emph{Apple}};
  \node (two) at (0.75,0) {\emph{Lenovo}};
  \node (three) at (-0.75,-1.2) {\emph{Sony}};
  \node (four) at (0.75,-1.2) {\emph{Toshiba}};
  \node (five) at (0.75,-0.6) {\emph{Samsung}};
  \draw[->] (one) -- (five);
  \draw[->] (one) -- (three);
  \draw[->] (two) -- (five);
  \draw[->] (two) -- (three);
  \draw[->] (five) -- (four);
\end{tikzpicture} & \begin{tikzpicture}
  \node (one) at (-0.55,0) {\emph{dual}};
  \node (two) at (0.55,-0.6) {\emph{triple}};
  \node (three) at (0.55,0) {\emph{quad}};
  \node (four) at (0.55,-1.2) {\emph{single}};
  \draw[->] (one) -- (four);
  \draw[->] (two) -- (four);
  \draw[->] (three) -- (two);
\end{tikzpicture}\\
\hline
\end{tabular}
\caption{\small User preferences. $U$$=$\{$c_1$,$c_2$\}.}\vspace{-9mm}
\label{tab:customer}
\end{table}

This paper \textbf{formulates the problem of \emph{continuous monitoring of Pareto frontiers}}: given a large number of users and continuously arriving new objects, for each newly arrived object, discover all users for whom the object is Pareto-optimal.  Users' preferences are modeled as strict partial orders, one for each attribute domain of the objects.

It is key to devise an efficient approach to this problem.  The value of a Pareto-optimal object diminishes quickly; the earlier it is found to be worth recommendation, the better.  For instance, a status update in a social network keeps getting less relevant since the moment it is posted; a customer's need for a product may be fulfilled by a less preferred choice, if an even better option was not shown to the customer in time.

A simple, brute-force approach is to, given a newly arrived object, compute for every user if the object belongs to the Pareto frontier with respect to the user's preferences.  This entails continuous maintenance of Pareto frontier for each and every user. The brute-force approach is subject to a clear drawback---repeated and wasteful maintenance of Pareto frontier for every user. 

\textbf{Sharing computation across users}\hspace{2mm} To tackle the aforementioned drawback, we partly resort to sharing computation across users. The challenge lies in the diversity of corresponding partial orders---a Pareto-optimal object with respect to one user may or may not be in the Pareto frontier for another user. Nonetheless, users have common preferences. In Table~\ref{tab:customer}, both $c_1$ and $c_2$ prefer $13-15.9$ inch display the most. Both prefer \emph{Apple} and \emph{Lenovo} to \emph{Toshiba} and \emph{Sony}, and they both prefer \emph{single}-core CPU the least. In Table~\ref{tab:customer}, $U$ is a \emph{virtual user} whose partial orders depict the common preferences of $c_1$ and $c_2$. Intuitively, users having similar preferences can be clustered together.

We thus design algorithms to mitigate repetitive computation via sharing computation across similar preferences of users. To intuitively understand the idea, consider two example scenarios. i) If $o$ is dominated by $o'$ with respect to the common preferences of a set of users, then $o$ is disqualified in Pareto-optimality for all users in the set. In Example~\ref{exp:exp_intro}, consider $o_{16}$$=$$\langle 16, \emph{Toshiba}, \emph{single} \rangle$ as the new object. With respect to $U$, $o_{16}$ is dominated by both $o_2$$=$$\langle 14, \emph{Apple}, \emph{dual} \rangle$ and $o_{15}$$=$$\langle 16.5$, $\emph{Lenovo}$, $\emph{quad} \rangle$. Therefore, $o_{16}$ belongs to the Pareto frontier of neither $c_1$ nor $c_2$. ii) Before the arrival of $o_2$, obviously $o_1$=$\langle 12$, $\emph{Apple}$, $\emph{single} \rangle$ is the only Pareto-optimal object for $U$, $c_1$ and $c_2$. Now consider the entrance of $o_2$. As $o_1$ is dominated by $o_2$ with respect to $U$, $o_1$ is replaced by $o_2$ in the Pareto frontier. This comparison is sufficient to decide that $o_1$ is dominated by $o_2$ for both $c_1$ and $c_2$. %%%Based on these two observations, our algorithms exploit shared computation across preferences of users.

\textbf{Clustering users}\hspace{2mm} To find users sharing similar preferences, we study the novel problem of clustering strict partial orders, which are used to model the preferences of both users and clusters. We measure the similarity between clusters and users by their common preferences. Such similarity measures factor in the different significance of preferences at various levels of the partial orders. Table~\ref{tab:customer4} depicts six customers' preferences on \smallattrname{brand}, in which $c_4$, $c_5$, and $c_6$ prefer \emph{Lenovo} to all other brands except that $c_4$ prefers \emph{Samsung} over \emph{Lenovo}. Consider the objects in Table~\ref{tab:product}. For both $c_5$ and $c_6$, the Pareto frontiers contain $\{o_7, o_{10}, o_{15}\}$, while $c_4$ has $\{o_3, o_5, o_{12}\}$ as its Pareto frontier. We can say that $c_5$ and $c_6$ are more similar than $c_4$ and $c_5$ or $c_4$ and $c_6$. 

\begin{table}[t]
\centering
\scriptsize
\begin{tabular}{|c|*{3}{@{\hskip-1pt}c@{\hskip-1pt}|}}\hline
\begin{tikzpicture}
  \node (one) at (-0.75,0) {\emph{Apple}};
  \node (two) at (-0.75,-0.6) {\emph{Lenovo}};
  \node (three) at (0,-1.2) {\emph{Samsung}};
  \node (four) at (0.5,0) {\emph{Toshiba}};
  \draw[->] (one) -- (two);
  \draw[->] (two) -- (three);
  \draw[->] (four) -- (three);
\end{tikzpicture} & \begin{tikzpicture}
  \node (one) at (-0.75,0) {\emph{Apple}};
  \node (two) at (0,-0.6) {\emph{Lenovo}};
  \node (three) at (0,-1.2) {\emph{Samsung}};
  \node (four) at (0.75,0) {\emph{Toshiba}};
  \draw[->] (one) -- (two);
  \draw[->] (two) -- (three);
  \draw[->] (four) -- (two);
\end{tikzpicture} & \begin{tikzpicture}
  \node (one) at (-0.75,0) {\emph{Apple}};
  \node (two) at (-0.75,-0.6) {\emph{Lenovo}};
  \node (three) at (0,-1.2) {\emph{Samsung}};
  \node (four) at (0.75,0) {\emph{Toshiba}};
  \draw[->] (one) -- (two);
  \draw[->] (two) -- (three);
  \draw[->] (four) -- (three);
\end{tikzpicture}\\
\small{$c_1$} & \small{$c_2$} & \small{$U_1$}\\
\hline
\begin{tikzpicture}
  \node (one) at (0,-4) {\emph{Samsung}};
  \node (two) at (-0.75,-5.2) {\emph{Apple}};
  \node (three) at (0,-4.6) {\emph{Lenovo}};
  \node (four) at (0.75,-5.2) {\emph{Toshiba}};
  \draw[->] (one) -- (three);
  \draw[->] (three) -- (two);
  \draw[->] (three) -- (four);
\end{tikzpicture} & \begin{tikzpicture}
  \node (one) at (0,-4) {\emph{Samsung}};
  \node (two) at (0,-5.2) {\emph{Apple}};
  \node (three) at (0,-4.6) {\emph{Lenovo}};
  \node (four) at (0,-5.8) {\emph{Toshiba}};
  \draw[->] (one) -- (three);
  \draw[->] (three) -- (two);
  \draw[->] (two) -- (four);
\end{tikzpicture} & \begin{tikzpicture}
  \node (one) at (0,-4) {\emph{Samsung}};
  \node (two) at (-0.75,-5.2) {\emph{Apple}};
  \node (three) at (0,-4.6) {\emph{Lenovo}};
  \node (four) at (0.75,-5.2) {\emph{Toshiba}};
  \draw[->] (one) -- (three);
  \draw[->] (three) -- (two);
  \draw[->] (three) -- (four);
\end{tikzpicture}\\
\small{$c_3$} & \small{$c_4$} & \small{$U_2$}\\
\hline
\begin{tikzpicture}
  \node (one) at (0,0) {\emph{Lenovo}};
  \node (two) at (-0.75,-0.6) {\emph{Apple}};
  \node (three) at (0.75,-0.6) {\emph{Toshiba}};
  \node (four) at (0,-1.2) {\emph{Samsung}};
  \draw[->] (one) -- (two);
  \draw[->] (one) -- (three);
  \draw[->] (two) -- (four);
  \draw[->] (three) -- (four);
\end{tikzpicture} & \begin{tikzpicture}
  \node (one) at (0,0) {\emph{Lenovo}};
  \node (two) at (0,-0.6) {\emph{Apple}};
  \node (three) at (-0.75,-1.2) {\emph{Toshiba}};
  \node (four) at (0.75,-1.2) {\emph{Samsung}};
  \draw[->] (one) -- (two);
  \draw[->] (two) -- (three);
  \draw[->] (two) -- (four);
\end{tikzpicture} & \begin{tikzpicture}
  \node (one) at (0,-7) {\emph{Lenovo}};
  \node (two) at (-0.75,-7.6) {\emph{Apple}};
  \node (three) at (0.75,-7.6) {\emph{Toshiba}};
  \node (four) at (-0.75,-8.2) {\emph{Samsung}};
  \draw[->] (one) -- (two);
  \draw[->] (one) -- (three);
  \draw[->] (two) -- (four);
\end{tikzpicture}\\
\small{$c_5$} & \small{$c_6$} & \small{$U_3$}\\
\hline
\end{tabular}
\caption{\small User preferences with respect to \smallattrname{brand}. $U_1$=$\{c_1$,$c_2\}$, $U_2$=$\{c_3$,$c_4\}$, $U_3$=$\{c_5$,$c_6\}$.}\vspace{-8mm}
\label{tab:customer4}
\end{table}
\textbf{Approximation}\hspace{2mm} The clustering algorithm may produce clusters that comprise few users, due to diverse preferences. With small clusters, the shared computation mentioned above may not pay off its overhead. Our response to this challenge is to use approximation. As in many data retrieval scenarios, insisting on exact answers is unnecessary and answers in close vicinity of the exact ones can be just good enough. Specifically, given a set of users, if a sizable subset of the users agree with a preference, the preference can be considered an approximate common preference. This relaxation eases the aforementioned concern regarding small clusters as more approximate common preferences lead to larger clusters. As an example, in Table~\ref{tab:customer}, while $c_2$ does not share with $c_1$ the preference of \emph{Apple} over \emph{Samsung}, its preference does not oppose it either. We can consider ``\emph{Apple} over \emph{Samsung}'' as an \emph{approximate common preference}. A possible set of approximate common preferences of $c_1$ and $c_2$ form the strict partial orders in the row for virtual user $\widehat{U}$. 

\textbf{Alive objects}\hspace{2mm} Objects can have limited lifetime. The trends in social networks and news media change rapidly. Similarly, in any inventory, products become unavailable over time. In these scenarios users look for \emph{alive} objects only. To meet this real-world requirement, we further extend our algorithms to operate under the semantics of a \emph{sliding window} and thus to disseminate an object only during its lifespan.

In summary, the contributions of this paper are as follows:\vspace{-1mm} 
\begin{list}{$\bullet$}
{ \setlength{\leftmargin}{1em} \setlength{\itemsep}{0em}}
\item We study the problem of continuous object dissemination and formalize it as finding Pareto-optimal objects regarding partial orders. Given a large number of users and continuously arriving objects, our goal is to swiftly disseminate a newly arrived object to a user if the user's preferences---modeled as strict partial orders on individual attributes---approve the object as Pareto-optimal.% (Section~\ref{sec:model}).
\item We devise efficient solutions exploiting shared computation across similar preferences of different users.% (Section~\ref{sec:method}).
\item We study the novel challenge of clustering user preferences represented as strict partial orders. Particularly we design similarity measures for such preferences.% (Section~\ref{sec:cluster}).
\item To address performance degradation due to small clusters, we present an approximate similarity measure that achieves high efficiency and accuracy of answers.% (Section~\ref{sec:approx}).
\item We extend our proposed solutions to deal with Pareto frontier maintenance under sliding window. %(Section~\\label{sec:window}). 
\item We conduct extensive experiments using simulations on two real datasets (a movie dataset and a publication dataset). The results demonstrate clear strengths of our solutions in comparison with baselines, in terms of execution time and efficacy.\vspace{-2mm}
\end{list}
\section{Related Work}\label{sec:related}
Pareto-optimality is a subject of extensive investigation. Its study in the computing fields can be dated back to \emph{admissible points}~\cite{barndorff1966distribution} and \emph{maximal vectors}~\cite{paretoopt}. B\"{o}rzs\"{o}nyi et al.~\cite{656550} introduced the concept of skyline---a special case of Pareto frontier---in which all attributes are numeric and amenable to total orders. Kie{\ss}ling~\cite{kiessling2002foundations} defined preferences as strict partial orders on which preference queries operate. After that, several studies specialized on skyline query evaluation over categorical attributes~\cite{chan2005stratified, sarkas2008categorical, sacharidis2009topologically, zhang2010efficient}, among which~\cite{sarkas2008categorical, sacharidis2009topologically, zhang2010efficient} particularly considered query answer maintenance and only~\cite{sacharidis2009topologically, zhang2010efficient} allow partial orders on attribute values. Nevertheless, they all consider only one user and none utilizes shared computation across multiple users' partial orders. 

Given a set of objects, Wong et al.~\cite{wong2007mining, wong2008efficient, wong2009online} identify the minimum set of preference relations that preclude an object from being in the Pareto frontier. This minimum set is the combination of each possible preference relation with regard to the values of all unique objects in the set. In case of any update in the object set, the minimum disqualifying condition must be recomputed. Hence, it is not designed for continuously arriving objects.

Vlachou et al.~\cite{vlachou2010reverse, vlachou2013branch} and Yu et al.~\cite{yu2012processing} aimed at finding all users who view a given object as one of their top-$k$ favourites, i.e., the results of a reverse top-k query. Dellis et al.~\cite{dellis2007efficient} studied reverse skyline query---selecting users to whom a given object is in the skyline. These works consider only numeric attributes. There is no clear way to extend them for categorical attributes or even partial orders.

All these studies, while about object dissemination, focused on different aspects of the problem than ours. %%%In this paper, we study continuous monitoring of Pareto frontiers for many users and we propose to tackle the associated efficiency and scalability challenges through shared computation across users. 
Particularly, no previous studies on Pareto frontier maintenance have exploited shared computation across users' preferences. Besides, as Sec.~\ref{sec:cluster} shall explain, no prior work studied similarity measures for partial orders or how to cluster partial orders.

\section{Problem Statement}\label{sec:model}

\begin{table}[h!]
\scriptsize
\centering
\begin{tabular}{|@{\hspace{0.3em}}c@{\hspace{0.3em}}|@{\hspace{0.3em}}l@{\hspace{0.3em}}|}
\hline
$\mathcal{O}$ & set of objects\\
\hline
$d \in \mathcal{D}$ & attribute\\
\hline
$c \in \mathcal{C}$ & user\\
\hline
$\succ_c^d$ & binary relation over $dom(d)$ with regard to $c$'s preference\\
\hline
$o' \succ_c o$ & $c$ prefers $o'$ to $o$\\
\hline
$\mathcal{P}_c$ & the Pareto frontier with regard to $c$\\
\hline
$\mathcal{C}_o$ & the target users of $o$\\
\hline
$U \subseteq \mathcal{C}$ & set of users\\
\hline
$sim(U_1,U_2)$ & the similarity measure between two clusters $U_1$ and $U_2$\\
\hline
$S_U^d$ & the maximal values of $\succ_{U}^d$\\
\hline
$h$ & branch cut in dendrogram\\
\hline
\end{tabular}
\caption{\small Notations}\label{tab:notation}
\end{table}

This section provides a formal description of our data model and problem statement. Table~\ref{tab:notation} lists the major notations. Consider a set of users $\mathcal{C}$ and a table of objects $\mathcal{O}$ that are described by a set of attributes $\mathcal{D}$.
For each user $c \in \mathcal{C}$, their preference regarding $\mathcal{O}$ is represented by strict partial orders.
For each attribute $d \in \mathcal{D}$, the strict partial order corresponding to $c$'s preference on $d$ is a binary relation over $dom(d)$---the domain of $d$, as follows.\vspace{-1mm}

\begin{definition}[Preference Relation and Tuple]\label{def:preference_tuple}
Given a user $c \in \mathcal{C}$ and an attribute $d \in \mathcal{D}$, the corresponding \emph{preference relation} is denoted $\succ_c^d$.
For two attribute values $x, y \in dom(d)$, if $(x,y)$ belongs to $\succ_c^d$ (i.e., $(x,y) \in \succ_c^d$, also denoted $x \succ_c^d y$), it is called a \emph{preference tuple}.  It is interpreted as ``user $c$ prefers $x$ to $y$ on attribute $d$''.
A preference relation is irreflexive $((x,x) \notin \succ_c^d$) and transitive ($(x,y) \in \succ_c^d \wedge (y,z) \in \succ_c^d \Rightarrow (x,z) \in \succ_c^d$), which together also imply asymmetry ($(x,y) \in \succ_c^d \Rightarrow (y,x) \notin \succ_c^d$).\closedef\vspace{-1mm}
\end{definition}

\begin{definition}[Object Dominance]
A user $c$'s preferences regarding all attributes induce another strict partial order $\succ_c$ that represents $c$'s preferences on objects. Given two objects $o, o' \in \mathcal{O}$, $c$ prefers $o'$ to $o$ if $o'$ is identical or preferred to $o$ on all attributes and $o'$ is preferred to $o$ on at least one attribute.
More formally, $o' \succ_c o$ (called $o'$ \emph{dominates} $o$), if and only if $(\forall d \in \mathcal{D} : o.d=o'.d \vee o'.d \succ_c^d o.d) \wedge (\exists d \in \mathcal{D} : o'.d \succ_c^d o.d)$. If $(\forall d \in \mathcal{D} : o.d=o'.d)$, we say that $o$ and $o'$ are \emph{identical}, denoted as $o$ $=$ $o'$.\closedef\vspace{-1mm}
\end{definition}

\begin{definition}[Pareto Frontier]\label{def:pareto_frontier}
An object $o$ is \emph{Pareto-optimal} with respect to $c$, if no other object in $\mathcal{O}$ dominates it. The set of \emph{Pareto-optimal objects} (i.e., the \emph{Pareto frontier}) in $\mathcal{O}$ for $c$ is denoted $\mathcal{P}_c$, i.e., $\mathcal{P}_c$ $=$ $\{o\in \mathcal{O}|\nexists o'\in \mathcal{O}\ \text{s.t.}\ o' \succ_c o\}$. Note that the concept of skyline points~\cite{656550} is a specialization of the more general Pareto frontier, in that the preference relations for skyline points are defined as total orders (with ties) instead of general strict partial orders.\closedef\vspace{-1mm}
\end{definition}

\begin{definition}[Target Users]\label{def:target_user}
Given an object $o$, the set of all users for whom $o$ belongs to their Pareto frontiers are called the \emph{target users}.
The target user set is denoted $\mathcal{C}_o$, i.e., $\mathcal{C}_o$ $=$ $\{c \in \mathcal{C} | o \in \mathcal{P}_c\}$.\closedef\vspace{-1mm}
\end{definition}

\begin{example}\label{exp:model}
Consider Table~\ref{tab:product} and Table~\ref{tab:customer}. $\mathcal{O}$ $=$ $\{o_1$, $o_2$, $\ldots$, $o_{15}\}$ (ignore $o_{16}$ for now), $\mathcal{C}$ $=$ $\{c_1, c_2\}$, and $\mathcal{D}$ $=$ \{\smallattrname{display}, \smallattrname{brand}, \smallattrname{CPU}\}.
With respect to $c_1$, $(10$$-$$12.9$, $16$$-$$18.9)$, $($\emph{Apple}, \emph{Samsung}$)$ and $($\emph{dual}, \emph{triple}$)$ are some of the preference tuples on attributes \smallattrname{display}, \smallattrname{brand} and \smallattrname{CPU}, respectively. Similarly, for $c_2$, $(16$$-$$18.9$, $19$$-$$up)$, $($\emph{Toshiba}, \emph{Sony}$)$ and $($\emph{triple}, \emph{dual}$)$ are some sample preference tuples. 

$\mathcal{P}_{c_1}$ $=$ $\{o_2\}$, since all other objects are dominated by $o_2$ with respect to $c_1$.  $\mathcal{P}_{c_2}$ $=$ $\{o_2$, $o_3$, $o_{15}\}$, as $o_2$, $o_3$ and $o_{15}$ dominate $\{o_1$, $o_4$, $o_6$, $o_8$, $o_9$, $o_{13}\}$, $\{o_4$, $o_6$, $o_8$, $o_{13}\}$ and $\{o_4$, $o_5$, $o_7$, $o_{10}$, $o_{11}$, $o_{12}$, $o_{14}\}$, respectively.  Therefore, $\mathcal{C}_{o_2}$ $=$ $\{c_1, c_2\}$ and $\mathcal{C}_{o_3}$ $=$ $\mathcal{C}_{o_{15}}$ $=$ $\{c_2\}$. Objects other than $o_2$, $o_3$, $o_{15}$ do not have target users in $\mathcal{C}$, i.e., $\mathcal{C}_{o}$ $=$ $\phi$, $\forall o \in \mathcal{O} - \{o_2, o_3, o_{15}\}$.\closedef\vspace{-1mm}
\end{example}

\textbf{Problem Statement}\hspace{2mm} The problem of \emph{continuous monitoring of Pareto frontiers} is, given a set of users $\mathcal{C}$, their preference relations on attributes $\mathcal{D}$, and a set of continuously growing objects $\mathcal{O}$ with the latest object $o$, find $\mathcal{C}_o$---the target users of $o$.

\section{Sharing Computation across Users}\label{sec:method}
\textbf{Algorithm \algname{Baseline}}\hspace{2mm} A simple method to our problem will check, for every user, whether a new object belongs to the corresponding Pareto frontier.  The pseudo code of this approach, named \algname{Baseline}, is shown in Alg. \ref{alg:Baseline}. Upon the arrival of a new object $o$, for every user $c$, it sequentially compares $o$ with the current Pareto-optimal objects in $\mathcal{P}_{c}$. 1) If $o$ is dominated by any $o'$ or $o$ is identical to $o'$, further comparison with the remaining objects in $\mathcal{P}_{c}$ is skipped. In the case of $o$ being dominated by $o'$, $o$ is disqualified from being a Pareto-optimal object; if $o$ is identical to $o'$, then $o$ is Pareto-optimal, i.e., it is inserted into $\mathcal{P}_{c}$. 2) If $o$ dominates any $o'$, $o'$ is discarded from $\mathcal{P}_{c}$. It can be concluded already that $o$ belongs to $\mathcal{P}_{c}$, but the comparisons should continue since $o$ may dominate other existing objects in $\mathcal{P}_{c}$.  3) If $o$ is not dominated by any object in $\mathcal{P}_{c}$, it becomes an element of $\mathcal{P}_{c}$.  Readers familiar with the literature on skyline queries may have realized that the gist of the algorithm is essentially the basic skyline query algorithm~\cite{656550}. The crux of its operation is based on an important property, that it suffices to compare new objects with only the Pareto-optimal objects, since any new object dominated by a non Pareto-optimal object must be dominated by some Pareto-optimal objects too.

\begin{algorithm}[h!]
\LinesNumbered
\small

\SetKwFunction{updateParetoFrontier}{\algname{updateParetoFrontier1}}

\KwIn{$\mathcal{C}$: all users; $\mathcal{O}$: existing objects; $o$: a new object}
\KwOut{$\mathcal{C}_o$: target users of $o$}

\BlankLine

$\mathcal{C}_o \leftarrow \emptyset$;

\ForEach{$c \in \mathcal{C}$}
{
    \algname{updateParetoFrontier}$(c,o)$;\\
}

\Return{$\mathcal{C}_o$};\\
\BlankLine
\SetKwInput{KwProc}{Procedure}
\setcounter{AlgoLine}{0}
\SetKwFunction{camember}{\algname{updateParetoFrontier}}\KwProc{\camember$(c,o)$}

$\mathit{isPareto} \leftarrow \KwSty{true}$;\\

\ForEach{$o' \in \mathcal{P}_{c} $}
{
    \uIf{$o$ $\succ_{c}$ $o'$}
    {
        $\mathcal{P}_{c} \leftarrow \mathcal{P}_{c} - \{o'\}$;\label{line:remove-member1}
        $\mathcal{C}_{o'} \leftarrow \mathcal{C}_{o'} - \{c\}$;\label{line:user-remove1}
    }
    \lElseIf{$o'$ $\succ_{c}$ $o$}
    {
        $\mathit{isPareto} \leftarrow \KwSty{false}$;$\KwSty{break}$\label{line:skip1-member}
    }
    \lElseIf{$o'$.$\mathcal{D}$ $=$ $o$.$\mathcal{D}$}
    {
        $\mathit{isPareto} \leftarrow \KwSty{true}$;\KwSty{break}\label{line:skip2-member}
    }
}
\uIf{$\mathit{isPareto}$}
{
    $\mathcal{P}_{c} \leftarrow \mathcal{P}_{c} \cup \{o\}$;\label{line:insert-member}
    $\mathcal{C}_o \leftarrow \mathcal{C}_o \cup \{c\}$;\label{line:user-add}
}\label{line:check-isPareto}
\caption{\small \algname{Baseline}}
\label{alg:Baseline}
\end{algorithm}

For a user $c$, suppose the aforementioned baseline approach  on average takes time $t$ to maintain the Pareto frontier upon the entrance of an object.  Maintaining $n$ objects for all users in $\mathcal{C}$ needs $O(n\cdot |\mathcal{C}| \cdot t)$ time.  The drawback of \algname{Baseline} is it repeatedly applies the same procedure for every user. In terms of computation efficiency, the approach may become particularly unappealing when there are a large number of users and new objects constantly arrive.
To counter this drawback, our idea is to share computations across the users that exhibit similar preferences.
To this end, our method is simple and intuitive.
If several users share a set of preference tuples, it is only necessary to compare two objects once, if they attain the attribute values in the preference tuples.  If an object is dominated by another object according to these common preference tuples, it is dominated with respect to all users sharing the same preferences. This idea guarantees to filter out only ``true negatives'' for these users, and it only needs to further discern ``false positives'' for each individual user.

\begin{definition}[Common Preference Tuple and Relation]\label{def:common_preference_tuple}
Given a set of users $U$ $\subseteq$ $\mathcal{C}$, an attribute $d \in \mathcal{D}$, and two values $x, y \in dom(d)$, if $(x,y)$ belongs to preference relation $\succ_c^d$ for all $c \in U$, then it is called a \emph{common preference tuple}. The set of common preference tuples of $U$ on attribute $d$ is denoted $\succ_U^d$, i.e., $\succ_U^d=\bigcap_{c \in U}\succ_c^d$. By definition, $\succ_U^d$ also represents a strict partial order (Theorem~\ref{thm:commonpref}, proof omitted). We call it a \emph{common preference relation}. It can be viewed as the preference of a virtual user that is denoted $U$.\closedef
\end{definition}

\begin{theorem}\label{thm:commonpref}
$\succ_U^d$ is a strict partial order.\closedef
\end{theorem}

Since, for each $d$, $\succ_U^d$ is a strict partial order, the set of users' preferences (i.e., the virtual user $U$'s preferences) regarding all attributes in $\mathcal{D}$ induce another strict partial order $\succ_U$ on objects.

\begin{definition}[Pareto Frontier for $U$]\label{def:pareto_frontier_common_preference_relation}
An object $o$ is \emph{Pareto-optimal} with respect to $U$ if no other object dominates it according to $\succ_U$. The Pareto frontier of $\mathcal{O}$ for $U$ is denoted $\mathcal{P}_U$, i.e., $\mathcal{P}_U$ $=$ $\{o\in \mathcal{O}|\nexists o'\in \mathcal{O}\ \text{s.t.}\ o' \succ_U o\}$.\closedef
\end{definition}

\begin{example}\label{exp:common_preference_tuple}
From Table~\ref{tab:customer}, $\succ_{c_1}^{\attrname{CPU}} = \{($\emph{dual}, \emph{single}$)$, $($\emph{dual}, \emph{quad}$)$, $($\emph{dual}, \emph{triple}$)$, $($\emph{triple}, \emph{single}$)$, $($\emph{quad}, \emph{single}$)\}$ and $\succ_{c_2}^{\attrname{CPU}} = \{($\emph{dual}, \emph{single}$)$, $($\emph{triple}, \emph{single}$)$, $($\emph{quad}, \emph{single}$)$, $($\emph{triple}, \emph{dual}$)$, $($\emph{quad}, \emph{dual}$)$, $($\emph{quad}, \emph{triple}$)\}$. According to Def.~\ref{def:common_preference_tuple}, the common preference relation of $c_1$ and $c_2$ is $\succ_{\{c_1,c_2\}}^{\attrname{CPU}} = \{($\emph{dual}, \emph{single}$)$, $($\emph{triple}, \emph{single}$)$, $($\emph{quad}, \emph{single}$)\}$.  Similarly we can derive $\succ_{\{c_1,c_2\}}^{\attrname{display}}$ and $\succ_{\{c_1,c_2\}}^{\attrname{brand}}$. In Table~\ref{tab:customer}, the three partial orders are depicted in a row labeled as a virtual user $U$.  The Pareto frontier of $U$ is $\mathcal{P}_{U} = \{o_2$, $o_3$, $o_{10}$, $o_{15}\}$.\closedef
\end{example}

\begin{theorem}\label{thm:framework1}
Given any set of users $U$, for all $c \in U$, $\mathcal{P}_{U} \supseteq \mathcal{P}_c$ and $\overline{\mathcal{P}}_{U} \subseteq \bigcap _{c \in U}\overline{\mathcal{P}}_{c}$.\closedef
\end{theorem}

\emph{Proof:} We prove by contradiction. Suppose that there exists $c$ $\in$ $U$ such that $\mathcal{P}_{U} \nsupseteq \mathcal{P}_c$, which would mean there exists $o$ $\in$ $\mathcal{O}$ such that $o$ $\in$ $\mathcal{P}_c$ and $o$ $\notin$ $\mathcal{P}_{U}$. That implies the existence of an $o' \in \mathcal{O}$ such that $o'$ $\succ_{U}$ $o$ and $o'$ $\nsucc_{c}$ $o$. However, by Def.~\ref{def:common_preference_tuple}, $o'$ $\succ_{U}$ $o$ implies $o'$ $\succ_{c}$ $o$.
Therefore, the existence of $o'$ is impossible.  This contradiction eventually leads to that $\mathcal{P}_{U} \supseteq \mathcal{P}_c$.
Hence, $\mathcal{P}_{U} \supseteq \bigcup _{c \in U}\mathcal{P}_c$, which implies $\overline{\mathcal{P}}_{U} \subseteq \bigcap _{c \in U}\overline{\mathcal{P}}_{c}$ according to De Morgan's laws.

\begin{lemma}\label{lemma:framework1}
Given any set of users $U$, for all $c \in U$, $\mathcal{P}_c$ $=$ $\{o\in \mathcal{P}_{U}|\nexists o'\in \mathcal{P}_{U}\ \text{s.t.}\ o' \succ_c o\}$.\closedef
\end{lemma}

\begin{example}
In Table~\ref{tab:customer}, $\mathcal{P}_{U}$ $=$ $\{o_2$, $o_3$, $o_{10}$, $o_{15}\}$ and $\mathcal{P}_{c_1}$ $\cup$ $\mathcal{P}_{c_2}$ $=$ $\{o_2$, $o_3,o_{15}\}$.  $\mathcal{P}_{U} \supseteq \mathcal{P}_{c_1}$ $\cup$ $\mathcal{P}_{c_2}$. Moreover, $\overline{\mathcal{P}}_{U}$ = $\{o_1$, $o_4$, $o_5$, $o_6$, $o_7$, $o_8$, $o_9$, $o_{11}$, $o_{12}$, $o_{13}$, $o_{14}\}$ and $\overline{\mathcal{P}}_{c_1}$ $\cap$ $\overline{\mathcal{P}}_{c_2}$ $=$ $\{o_1$, $o_4$, $o_5$, $o_6$, $o_7$, $o_8$, $o_9$, $o_{10}$, $o_{11}$, $o_{12}$, $o_{13}$, $o_{14}$, $o_{15}\}$. $\overline{\mathcal{P}}_{U} \subseteq \overline{\mathcal{P}}_{c_1} \cap \overline{\mathcal{P}}_{c_2}$. 
\closedef
\end{example}

Theorem~\ref{thm:framework1} suggests an appealing quality of the common preference relations of $U$.  By $\mathcal{P}_{U} \supseteq \mathcal{P}_c$, the Pareto frontier of $U$ subsumes the Pareto frontier of every user member in $U$.  What it means is that, if we simply compute the Pareto frontier of $U$, we get to retain all the objects that we eventually look for. Consider $\mathcal{P}_c$ as the ground truth and $\mathcal{P}_{U}$ as the predictions. The objects that are filtered out ($\overline{\mathcal{P}}_{U}$) are all ``true negatives'' and there are no ``false negatives''.  The set $\mathcal{P}_{U}$ may contain ``false positives'', which we just need to throw out after further verification, as Lemma~\ref{lemma:framework1} suggests. 

This approach's merit is the potential saving on object comparisons.
%%In any application that selectively recommends objects, most objects are expected to be non Pareto-optimal.
For a cluster of users, many non Pareto-optimal objects may be filtered out altogether for all the users,
without incurring the same comparisons repeatedly for each user.

To capitalize on the above ideas, our method must answer three questions. 
(1) How to find users sharing similar preferences? 
(2) For a set of similar users $U$, how to maintain the corresponding Pareto frontier $\mathcal{P}_U$ based on their common preference relations $\succ_{U}^d$ for different attributes $d$? 
(3) For each user $c$ in $U$, how to discern the ``false positives'' in $\mathcal{P}_U$ and thus find $\mathcal{P}_c$.
Note that the second and the last challenges need to be addressed for constantly arriving new objects.

For (1), our method is to cluster users based on the similarity between their preference relations. While many clustering methods have been developed for various types of data, none is specialized in clustering partial orders.  Our clustering method is discussed in Sec.~\ref{sec:cluster}.
For (2) and (3), our algorithm takes a \emph{filter-then-verify} approach and is thus named \algname{FilterThenVerify}, of which the pseudo code is displayed in Alg \ref{alg:filterthenverify}.

\textbf{Alg. \algname{FilterThenVerify}}\hspace{2mm} Upon the arrival of a new object $o$, for every cluster $U$, \algname{FilterThenVerify} compares $o$ with the current members of $\mathcal{P}_{U}$ based on the preference relations of the virtual user $U$.
Various actions are taken, depending on the comparison outcomes, as follows: 

\textbf{I)} If $o$ dominates any $o'$ in $\mathcal{P}_{U}$ according to $\succ_U^d$ of all relevant $d$, $o'$ is removed from $\mathcal{P}_{U}$ (Line~\ref{line:remove-head} of Procedure~\algname{updateParetoFrontierU} in Alg. \ref{alg:filterthenverify}).  For every $c \in \mathcal{C}$ such that $o' \in \mathcal{P}_{c}$, $o'$ is also discarded from $\mathcal{P}_{c}$ (Line~\ref{line:remove-member2} of Procedure \algname{updateParetoFrontierU}).

\textbf{II)} If $o$ is dominated by any $o'$ in $\mathcal{P}_{U}$, then $o$ does not occupy the Pareto frontier of any user in $U$ (Theorem~\ref{thm:framework1}). Further operations involving $o$ are unnecessary (Line~\ref{line:skip1-head} of Procedure \algname{updateParetoFrontierU}).

\textbf{III)} After comparing $o$ with all current objects in $\mathcal{P}_{U}$, if it is realized that $o$ is not dominated by any $o'$, then $o$ becomes a member of $\mathcal{P}_{U}$ (Line~\ref{line:insert-head} of \algname{updateParetoFrontierU}). Furthermore, for each $c \in U$, $o$ is further compared with the members of $\mathcal{P}_{c}$ based on the preference relations of $c$, by using Procedure \algname{updateParetoFrontier} of Alg.\ref{alg:Baseline} (Line~\ref{line:individual} of Alg.\ref{alg:filterthenverify}).

\begin{algorithm}[t]
\LinesNumbered
\small

\KwIn{$U_1$, $U_2$,..., $U_n$: clusters of users; $\mathcal{O}$: existing objects; $o$: a new object}

\KwOut{$\mathcal{C}_o$: target users of $o$}

\BlankLine

$\mathcal{C}_o \leftarrow \emptyset$;

\ForEach{$U \in \{U_1, U_2, ..., U_n\}$}
{
    $isPareto \leftarrow$ \algname{updateParetoFrontierU}$(U,o)$;\\

    \uIf{$isPareto$}
    {
        \ForEach{$c \in U$}
        {
            \algname{updateParetoFrontier}$(c,o)$; //Algorithm~\ref{alg:Baseline}\label{line:individual}\\
        }
    }\label{line:check-isPareto2}
}

\Return{$\mathcal{C}_o$};\\
\BlankLine
\SetKwInput{KwProc}{Procedure}
\setcounter{AlgoLine}{0}
\SetKwFunction{cahead}{\algname{updateParetoFrontierU}}\KwProc{\cahead$(U,o)$}

$isPareto \leftarrow \KwSty{true}$;\\

\ForEach{$o' \in \mathcal{P}_{U}$}
{
    \uIf{$o$ $\succ_{U}$ $o'$}
    {
        \ForEach{$c \in U$}
        {
            \uIf{$o'$ $\in$ $\mathcal{P}_{c}$}
            {
                $\mathcal{P}_{c} \leftarrow \mathcal{P}_{c} - \{o'\}$;\label{line:remove-member2}
                $\mathcal{C}_{o'} \leftarrow \mathcal{C}_{o'} - \{c\}$;\label{line:user-remove2}
            }
        }
        $\mathcal{P}_{U} \leftarrow \mathcal{P}_{U} - \{o'\}$;\label{line:remove-head}\\

    }
    \lElseIf{$o'$ $\succ_{U}$ $o$}
    {
        $\mathit{isPareto} \leftarrow \KwSty{false}$;\label{line:skip1-head}
        \KwSty{break}
    }
}

\lIf{$isPareto$}
{
    $\mathcal{P}_{U} \leftarrow \mathcal{P}_{U} \cup \{o\}$\label{line:insert-head}
}

\Return{$isPareto$};\\
\caption{\small \algname{FilterThenVerify}}
\label{alg:filterthenverify}
\end{algorithm}

\begin{example}\label{exp:alg_ca}
In this example we explain the execution of \algname{FilterThenVerify} on Table~\ref{tab:product} and Table~\ref{tab:customer}.
Suppose users $c_1$ and $c_2$ form a cluster $U$, of which the preference relations are depicted in Table~\ref{tab:customer}. 
The existing objects are $o_1$ to $o_{14}$, and $o_{15}$ $=$ $\langle 16.5''$, \emph{Lenovo}, \emph{quad}$\rangle$ is the object that just becomes available.
Before $o_{15}$ arrives, the Pareto frontier of $U$ is $\mathcal{P}_{U}$ $=$ $\{o_2$, $o_3$, $o_7$, $o_{10}\}$.
The algorithm starts by comparing $o_{15}$ with each element in $\mathcal{P}_{U}$.
As $o_{15}$ dominates $o_7$ $=$ $\langle 9.5''$, \emph{Lenovo}, \emph{quad}$\rangle$ according to $U$'s preference relations,
$o_7$ is discarded from $\mathcal{P}_{U}$. 
Before $o_{15}$ arrives, $o_7$ also belongs to $\mathcal{P}_{c_2}$.
Therefore, $o_7$ is removed from $\mathcal{P}_{c_2}$ as well. 
$o_{15}$ does not dominate any other object in $\mathcal{P}_{U}$. It is not dominated by any either. 
Hence, it is inserted into $\mathcal{P}_{U}$. 
 
$o_{15}$ is further compared with the existing members of $\mathcal{P}_{c_1}$ and $\mathcal{P}_{c_2}$.
It is dominated by $o_2$$=$$\langle 14''$, \emph{Apple}, \emph{dual}$\rangle$ according to $c_1$'s preference relations.
Thus it is not part of $\mathcal{P}_{c_1}$. 
According to $c_2$'s preferences, $o_{15}$ does not dominate any existing Pareto optional object (except the aforementioned $o_7$ which by now is already discarded).
Therefore $\mathcal{P}_{c_2}$ is not further changed and $o_{15}$ becomes part of $\mathcal{P}_{c_2}$.
Overall, $\mathcal{C}_{o_{15}}$$=$$\{c_2\}$.

Moreover, consider the arrival of $o_{16}$ $=$ $\langle 16''$, \emph{Toshiba}, \emph{single}$\rangle$ after $o_{15}$.
In the process of comparing $o_{16}$ with $\mathcal{P}_{U}$ $=$ $\{o_2$, $o_3$, $o_{10}$, $o_{15}\}$,
it is realized that $o_{16}$ is dominated by $o_2$ according to $U$'s preference relations.
Therefore, it does not belong to $\mathcal{P}_{U}$.
It is thus unnecessary to further compare $o_{16}$ with $\mathcal{P}_{c_1}$ or $\mathcal{P}_{c_2}$.
$\mathcal{C}_{o_{16}}$$=$$\emptyset$. %%As discussed earlier, in any real-world application, most of the objects fail to occupy the Pareto frontier. 
Thereby, \algname{updateParetoFrontierU} acts as a sieve to filter out non Pareto-optimal objects such as $o_{16}$. 
In this way \algname{FilterThenVerify} reduces computation cost by avoiding repeated comparisons with such objects.\closedef 
\end{example}

\textbf{Complexity Analysis of Algorithm \ref{alg:filterthenverify}}\hspace{2mm} According to previous study, given the number of attributes, the computation of Pareto frontier is polynomial with respect to the number of objects. Therefore, given a set of attributes $\mathcal{D}$, the complexity of Pareto frontier maintenance for a user $c$ or a cluster $U$ is polynomial with regard to the corresponding candidate Pareto-optimal objects.

For a cluster $U$, suppose the aforementioned filter-then-verify approach takes $t'$ time on average to maintain the Pareto frontier upon the entrance of an object. Consider $k$ as the number of clusters. Therefore, maintaining $n$ objects for all clusters in $\mathcal{C}$ needs $n \cdot k \cdot t'$ time. With regard to each cluster, consider the Pareto frontier includes $m$ object on average, i.e., $m$ objects occupy the candidate Pareto-optimal object for each users in the corresponding cluster. Now with regard to a particular user $c$, suppose \algname{FilterThenVerify} takes $t''$ time on average to maintain the Pareto frontier for each objects in $m$. Hence, all users in $\mathcal{C}$ needs $O(m \cdot |\mathcal{C}| \cdot t'')$ time to maintain $m$ objects. In summation, \algname{FilterThenVerify} needs $O((n \cdot k \cdot t')+(m \cdot |\mathcal{C}| \cdot t''))$ time to find the target users for all objects. Now we compare \algname{FilterThenVerify} and \algname{Baseline} in terms of time complexity. We can assume that $t \approx t'$ and $k < |\mathcal{C}|$. Therefore, $n \cdot k \cdot t' \leq n \cdot |\mathcal{C}| \cdot t$. Besides, it is intuitive that $m < n$ and thus $t'' < t$. Therefore, $m \cdot |\mathcal{C}| \cdot t'' \leq n \cdot |\mathcal{C}| \cdot t$. In summation, $(n \cdot k \cdot t')+(m \cdot |\mathcal{C}| \cdot t'') < n \cdot |\mathcal{C}| \cdot t$.

Now the question is: how to maximize ${n \cdot |\mathcal{C}| \cdot t} \over {(n \cdot k \cdot t')+(m \cdot |\mathcal{C}| \cdot t'')}$? The key is to minimize the number of clusters $k$ as well as the filtered objects $m$. As we discussed earlier, larger clusters tend to share fewer preference tuples which may leave a number of objects as candidate Pareto-optimal objects, i.e., $m \approx n$. On the contrary, the presence of lots of tiny clusters, i.e., $k \approx |\mathcal{C}|$ makes the filter-then-verify ineffective. Apparently, there exists a tradeoff between $k$ and $m$. In conclusion, in order to optimize $k$ and $m$, the clustering should be effective.
\section{Similarity Measures for Clustering User Preferences}\label{sec:cluster}
This section discusses how to cluster users based on their preference relations.
Our focus is on the similarity measures rather than the clustering method.
The method we adopt is the conventional hierarchical agglomerative clustering algorithm~\cite{dmbook}.
At every iteration, the method merges the two most similar clusters.
The common preference relation of the merged cluster $U$ on each attribute $d$, i.e., $\succ^d_{U}$, is computed.
It then calculates the similarity between $U$ and each remaining cluster.
Given two clusters $U_1$ and $U_2$, their similarity $sim(U_1$, $U_2)$ is defined as the summation of the similarities between their preference relations on individual attributes, as follows.
This resembles the high-level idea of using $L_1$ norm distance between centroids for measuring inter-cluster similarity in conventional hierarchial clustering.
\begin{flalign}
&sim(U_1,U_2)=\sum_{d \in \mathcal{D}}sim^d(U_1,U_2)&
\end{flalign}

Individual users' and clusters' preference relations on attributes are strict partial orders. No prior work studied clustering approaches or similarity measures for partial orders.  Similarity measures commonly used in clustering algorithms assume numeric or categorical attributes. Kamishima et al.~\cite{kamishima2003clustering, kamishima2009efficient} and Ukkonen et al.~\cite{ukkonen2011clustering} cluster total orders but not partial orders. Given two totally ordered attributes, these works use the comparative ranks of the corresponding values to measure similarity. Clearly, such similarity measures are not applicable for partially ordered attributes.  

In this section we propose four different similarity functions for defining $sim^d(U_1,U_2)$.

\textbf{1) Intersection size}\hspace{2mm} This is simply the size of the intersection of $\succ^d_{U_1}$ and $\succ^d_{U_2}$, i.e., the number of common preference tuples of all users in the two clusters $U_1$ and $U_2$. It is defined as
\begin{flalign}
&sim^d_i(U_1,U_2)=|\succ^d_{U_{1}}\cap \succ^d_{U_{2}}|&
\end{flalign}

\begin{example}
\label{example:sim_i}
Table~\ref{tab:customer4} shows three clusters $U_1$ ($\{c_1$, $c_2\}$), $U_2$ ($\{c_3$, $c_4\}$), and $U_3$ ($\{c_5$, $c_6\}$) and the common preference relation associated with each cluster on attribute $\smallattrname{brand}$. $U_1$ and $U_2$ do not share any preference tuple and thus $sim^\attrname{brand}_{i}(U_1,U_2) = 0$.  $U_1$ and $U_3$ have $($\emph{Apple}, \emph{Samsung}$)$ and $($\emph{Lenovo}, \emph{Samsung}$)$ as common preference tuples, i.e., $sim^\attrname{brand}_{i}(U_1,U_3) = 2$. Similarly, $U_2$ and $U_3$ share $($\emph{Lenovo}, \emph{Apple}$)$ and $($\emph{Lenovo}, \emph{Toshiba}$)$, i.e., $sim^\attrname{brand}_{i}(U_2,U_3) = 2$. \closedef
\end{example}

\textbf{2) Jaccard similarity}\hspace{2mm} The measure $sim_i$ captures the absolute size of the intersection of two preference relations.
It does not take into account their differences.
Consider three clusters $U_1$, $U_2$ and $U_3$ such that $sim^d_i(U_1,U_2)$ $=$ $sim^d_i(U_1,U_3)$ (i.e.,  $|\succ^d_{U_{1}}\cap \succ^d_{U_{2}}|$ $=$ $|\succ^d_{U_{1}}\cap \succ^d_{U_{3}}|$) and $|\succ^d_{U_{1}}\cup \succ^d_{U_{2}}|$ $<$ $|\succ^d_{U_{1}}\cup \succ^d_{U_{3}}|$.
We can argue that the similarity between $U_1$ and $U_2$ should be higher than (instead of equal to) that between $U_1$ and $U_3$, because $U_1$ and $U_2$ have a larger percentage of common preference tuples than $U_1$ and $U_3$.
To address this limitation of $sim_i$, we define the \emph{Jaccard similarity} between two preference relations as their intersection size over their union size, i.e., the ratio of common preference tuples to all preference tuples in the two preference relations. Formally,
\begin{flalign}
&sim^d_j(U_1,U_2) = {|\succ^d_{U_{1}}\cap \succ^d_{U_{2}}| \over |\succ^d_{U_{1}}\cup \succ^d_{U_{2}}|} = {sim^d_i(U_1,U_2) \over |\succ^d_{U_{1}}\cup \succ^d_{U_{2}}|}&
\end{flalign}

\begin{example}
\label{example:sim_j}
Continue Example~\ref{example:sim_i}. $\succ^\attrname{brand}_{U_1}$ and $\succ^\attrname{brand}_{U_3}$ have $6$ preference tuples in total while $\succ^\attrname{brand}_{U_2}$ and $\succ^\attrname{brand}_{U_3}$ have $7$. Thus, $sim^\attrname{brand}_{j}(U_1,U_3)$$=$$2/6$ and $sim^\attrname{brand}_{j}(U_2,U_3)$$=$$2/7$.
%%Note that, although $sim^\attrname{brand}_{i}(U_1,U_3) = sim^\attrname{brand}_{i}(U_2,U_3)$, $sim^\attrname{brand}_{j}(U_1$,$U_3)$ $>$ $sim^\attrname{brand}_{j}(U_2$,$U_3)$ since the union sizes differ in these two pairs.
\closedef
\end{example}

\textbf{3) Weighted intersection size}\hspace{2mm} Intersection size and Jaccard similarity are based on the cardinalities of intersection and union sets of preference relations. In counting the cardinalities, they both treat all preference tuples equal. We argue that this is counter-intuitive.
Values at the top of a partial order matter more than those at the bottom, in terms of their impact on which objects belong to the Pareto frontier. Accordingly we introduce \emph{weighted intersection size}, a modified version of intersection size $sim_i$. In counting the common preference tuples of two preference relations, it assigns a weight to each preference tuple. Formally,\vspace{-1mm}

\begin{small}
\begin{flalign}
&sim^d_{wi}(U_1,U_2) = \hspace{-9mm} \sum_{(v, v') \in \succ^d_{U_{1}}\cap \succ^d_{U_{2}}}\hspace{-5mm} {1 \over 2} \times({1 \over \displaystyle \min_{s \in S_{U_1}^d}\hspace{-1mm}D(s,v)\hspace{-1mm}+\hspace{-1mm}1} + {1 \over \displaystyle \min_{s \in S_{U_2}^d}\hspace{-1mm}D(s,v)\hspace{-1mm}+\hspace{-1mm}1})&\hspace{-2mm}
\end{flalign}
\end{small}

In the above equation, with regard to an attribute $d$, the similarity between two clusters' preference relations is a summation over their common preference tuples.
For each common preference tuple $(v,v')$, it computes the average weight of the better value $v$ with respect to $U_1$ and $U_2$, respectively.
Given a cluster $U$, $S^d_{U}$ is the set of \emph{maximal values} in the partial order $\succ^d_{U}$ and $D(s,v)$ for each $s \in S^d_{U}$ is the shortest distance from $s$ to $v$ in $\succ^d_{U}$.
The weight of $v$ in $U$ is the inverse of the minimal distance from any maximal value to $v$ (plus 1, to avoid division by zero).
The concept of maximal value is defined as follows.

\begin{definition}[Maximal Value]
With regard to $\succ_{U}^d$, value $x\in dom(d)$ is a \emph{maximal value} if no other value in $dom(d)$ is preferred over $x$. The set of maximal values for $\succ_{U}^d$ is denoted $S_U^d$. Formally, $S_U^d = \{x \in dom(d)\ |\ \nexists y \in dom(d)\ s.t.\ (y,x) \in$ $\succ_{U}^d\}$.\closedef
\end{definition}

\begin{example}
\label{example:sim_wi}
Continue Example~\ref{example:sim_i}. The maximal values in $\succ^\attrname{brand}_{U_1}$, $\succ^\attrname{brand}_{U_2}$ and $\succ^\attrname{brand}_{U_3}$ are $S_{U_1}^{\attrname{brand}}$$=$$\{$\emph{Apple}, \emph{Toshiba}$\}$, $S_{U_2}^{\attrname{brand}}=\{$\emph{Samsung}$\}$ and $S_{U_3}^{\attrname{brand}}$$=$$\{$\emph{Lenovo}$\}$, respectively.
In the partial order corresponding to $\succ_{U_1}^{\attrname{brand}}$, the minimal shortest distances to \emph{Apple}, \emph{Lenovo}, \emph{Samsung}, and \emph{Toshiba} from the maximal values $\{$\emph{Apple}, \emph{Toshiba}$\}$ are $0$, $1$, $1$ and $0$, respectively.  The corresponding weights are $1$, $1/2$, $1/2$ and $1$. Similarly, in $\succ_{U_2}^{\attrname{brand}}$, the weights of \emph{Apple}, \emph{Lenovo}, \emph{Samsung} and \emph{Toshiba} are $1/3$, $1/2$, $1$ and $1/3$, respectively.
In $\succ_{U_3}^{\attrname{brand}}$, the corresponding weights are $1/2$, $1$, $1/3$ and $1/2$, respectively.

$U_1$ and $U_3$ have (\emph{Apple}, \emph{Samsung}) and (\emph{Lenovo}, \emph{Samsung}) as common preference tuples. For the two better-values in these preference tuples---\emph{Apple} and \emph{Lenovo}, the average weights are both $3/4$.
The similarity $sim^\attrname{brand}_{wi}(U_1,U_3)$$=$${1+{1 \over 2} \over 2} + {{1 \over 2}+1 \over 2}$$=$${3 \over 2}$. Similarly, $U_2$ and $U_3$ have $($\emph{Lenovo}, \emph{Apple}$)$ and $($\emph{Lenovo}, \emph{Toshiba}$)$ as common preference tuples. In $U_2$ and $U_3$, the average weight of \emph{Lenovo}---the better-value in both common preference tuples---is $3/4$. The similarity $sim^\attrname{brand}_{wi}(U_2,U_3)$$=$${{1 \over 2}+1 \over 2} + {{1 \over 2}+1 \over 2}$$=$${3 \over 2}$. \closedef
\end{example}

\textbf{4) Weighted Jaccard similarity}\hspace{2mm}  This measure is a combination of the last two ideas---Jaccard similarity and weighted intersection size.
As in Jaccard similarity, \emph{weighted Jaccard similarity} computes the ratio of intersection size to union size.
Similar to weighted intersection size, the values in a preference relation are assigned weights corresponding to their minimal shortest distances to the preference relation's maximal values.
The measure's definition is as follows.\vspace{-1mm}

\begin{small}
\begin{flalign}
sim^d_{wj}(U_1,U_2) = \hspace{-9mm} \sum_{(v, v') \in \succ^d_{U_{1}}\cap \succ^d_{U_{2}}}\hspace{-4mm} {1 \over 2} \times ({1 \over \displaystyle \min_{s \in S_{U_1}^d}\hspace{-1mm}D(s,v)\hspace{-1mm}+\hspace{-1mm}1} + {1 \over \displaystyle \min_{s \in S_{U_2}^d}\hspace{-1mm}D(s,v)\hspace{-1mm}+\hspace{-1mm}1})  \nonumber \\
\Big/\ \sum_{(v, v') \in \succ^d_{U_{1}}\cup \succ^d_{U_{2}}}\hspace{-4mm} {1 \over 2} \times ({1 \over \displaystyle \min_{s \in S_{U_1}^d}\hspace{-1mm}D(s,v)\hspace{-1mm}+\hspace{-1mm}1} + {1 \over \displaystyle \min_{s \in S_{U_2}^d}\hspace{-1mm}D(s,v)\hspace{-1mm}+\hspace{-1mm}1}) \nonumber \\ 
 = sim^d_{wi}(U_1,U_2) \Big/\ \Big[ sim^d_{wi}(U_1,U_2) + \hspace{-8mm} {\displaystyle \sum_{(v, v') \in \succ_{U_1}^d - \succ_{U_2}^d} {1 \over \displaystyle  \min_{s \in S_{U_1}^d} D(s,v)+1}} \nonumber \\
+ \hspace{-4mm}{\displaystyle \sum_{(v, v') \in \succ_{U_2}^d - \succ_{U_1}^d} {1 \over \displaystyle  \min_{s \in S_{U_2}^d} D(s,v)+1}} \Big] 
\end{flalign}
\end{small}

\begin{example}
\label{example:sim_wj}
Continue Example~\ref{example:sim_wi}.  Now $sim^\attrname{brand}_{wj}(U_1,U_3)$=\\${{3 \over 2} \over {(1+1)+(1+1)+{3 \over 2}}}$=${3 \over 11}$, since $\succ_{U_1}^d$$-$$\succ_{U_3}^d$=$\{($\emph{Apple}, \emph{Lenovo}$)$, $($\emph{Toshiba}, \emph{Samsung}$)\}$ and $\succ_{U_3}^d-\succ_{U_1}^d$=$\{($\emph{Lenovo}, \emph{Apple}$)$, $($\emph{Lenovo}, \emph{Toshiba}$)\}$.  Similarly,  $sim^\attrname{brand}_{wj}(U_2,U_3$)=${{3 \over 2} \over {(1+1+1)+(1+{1 \over 2})+{3 \over 2}}}$=${3 \over 12}$, as $\succ_{U_2}^d-\succ_{U_3}^d$ =$\{($\emph{Samsung}, \emph{Lenovo}), (\emph{Samsung}, \emph{Apple}), (\emph{Samsung}, \emph{Toshiba})$\}$ and $\succ_{U_3}^d$$-$$\succ_{U_2}^d$$=$$\{($\emph{Lenovo}, \emph{Samsung}$)$, $($\emph{Apple}, \emph{Samsung}$)\}$. Note that $sim^\attrname{brand}_{wj}(U_1,U_3)>sim^\attrname{brand}_{wj}(U_2,U_3)$ although $sim^\attrname{brand}_{wi}(U_1,U_3$) =$sim^\attrname{brand}_{wi}(U_2,U_3)$.\closedef
\end{example}
\section{Approximate User Preferences}\label{sec:approx}
%%After clustering users based on their preferences, \algname{FilterThenVerify} uses a cluster's common preference relations to filter out new objects that do not qualify in Pareto-optimality for any member of the cluster.
Two conflicting factors have crucial impacts on the effectiveness of \algname{FilterThenVerify}. 
One is the size of the common preference relations.
The other is the size of the clusters.
Specifically, the more preference tuples a cluster's users share, the more objects can be filtered out and thus the less verifications need to be done for individual users.
On the contrary, the more users a cluster contains, the more repeated comparisons are avoided for these individual users.
There is a clear tradeoff between these two factors, since larger clusters (i.e., more users in each cluster) naturally leads to smaller common preference relations. 

Our approach to this challenge is \emph{approximation}. As discussed in Sec.~\ref{sec:intro}, it suffices for many applications to approximately identify target users. In this section, we show that we can find such approximation through a relaxed notion of common preference tuple, namely \emph{approximate common preference tuple}. For a set of users, it allows a preference tuple to be absent from a tolerably small subset. 
If a sizable subset of the users agree with the preference tuple, it is considered an approximate common preference tuple. This relaxation addresses the aforementioned concern, since more approximate common preferences lead to larger clusters.\vspace{-2mm}

\subsection{Approximate Common Preference Tuples and Relations}\label{sec:approx-computing}
\begin{definition}[Approximate Common Preference Tuple and Relation]\label{def:approx_common_preference_tuple}
Given a set of users $U$$\subseteq$$\mathcal{C}$, an attribute $d$$\in$$\mathcal{D}$ of which $|dom(d)|$$=$$m$, consider $A_{1 \ldots \Perm{m}{2}}$ which is an ordered permutation of all possible preference tuples $\{(x,y) \in {dom(d) \times dom(d)}\ |\ x \neq y\}$ such that $freq({A_{i}})$$\geq$$freq({A_{i+1}})$ for $i \in [1, \Perm{m}{2}-1]$, in which $freq(A_{i})$ denotes the percentage of users in $U$ whose preference relations contain preference tuple $A_{i}$. The \emph{approximate common preference relation} $\widehat{\succ}_U^d$ is defined as ç $=$ $R_j$ in which $j$ is the largest index $i \in [1, \Perm{m}{2}]$ that satisfies the condition $(|R_i|$ $<$ $\theta_1$ $\land$ $freq({A_{i}}) > \theta_2)$ $\lor$ $freq({A_{i}}) = 1$ where $R_i$ is defined as\vspace{-2mm}

\begin{small}
\[
 R_i =
  \begin{cases}
  \{{A_{1}}\} & \text{if } {i = 1} \\
   (R_{i-1} \cup \{{A_{i}}\})^+ & \text{if } R_{i-1} \cup \{{A_{i}}\}\ \text{is a strict partial order} \\
   R_{i-1}       & \text{otherwise}
  \end{cases}
\]
\end{small}\\
and $\theta_1$ and $\theta_2$ are two given thresholds. $\theta_1$ limits the size of the resulting $\widehat{\succ}_U^d$ while $\theta_2$ excludes infrequent preference tuples from $\widehat{\succ}_U^d$.\closedef
\end{definition}

By this definition, the resulting approximate preference relation always includes the common preference tuples. The remaining possible preference tuples are considered in descending order of their frequencies, since preference tuples with higher frequencies are shared by more users. A preference tuple is included into $\widehat{\succ}_U^d$ only if its reverse tuple is not included. This guarantees asymmetry. Furthermore, when a preference tuple is included into $\widehat{\succ}_U^d$, the transitive closure of the updated $\widehat{\succ}_U^d$ is also included. This guarantees transitivity. Irreflexivity is guaranteed too since $A_{1 \ldots \Perm{m}{2}}$ does not include preference tuples in the form of $(x,x)$. These altogether assure $\widehat{\succ}_U^d$ is a strict partial order. Given an append-only database of objects, a strict partial order ensures that the preference query results are independent of the order by which objects are appended to the database. Therefore, $\widehat{\succ}_U^d$ can be viewed as the preference of a virtual user (denoted $\widehat{U}$) on attribute $d$. Moreover, we denote the Pareto frontier of $\mathcal{O}$ for $\widehat{U}$ as $\ParetoApprox{U}$. 

$\theta_1$ and $\theta_2$ regulate the size of $\widehat{\succ}_U^d$. A pair of large $\theta_1$ and small $\theta_2$ allows $\widehat{\succ}_U^d$ to include infrequent preference tuples. In such a case the approximate common preference relation becomes ineffective, since Procedure~\algname{updateParetoFrontierU} in Alg.\ref{alg:filterthenverify} may retain a large number of candidates that must be verified for each $c \in U$. On the other hand, a pair of small $\theta_1$ and large $\theta_2$ may limit $\widehat{\succ}_U^d$ to contain only ${\succ}_U^d$, in which case the concern regarding small common preference relation remains. 

As Def.~\ref{def:approx_common_preference_tuple} itself is procedural, it naturally corresponds to a greedy algorithm for constructing approximate preference relation $\widehat{\succ}_U^d$. The  pseudo code \algname{GetApproxPreferenceTuples} is in Alg.~\ref{alg:approx}. %%%It starts by sorting the preference tuples' frequencies in descending order (Line \ref{line:approx-sort}). 
First, all the common preference tuples are included  (Lines \ref{line:approx-common-extract_start}-\ref{line:approx-common-extract_end}). After that, preference tuples are considered in the order of frequency, as long as the two thresholds are satisfied (Line \ref{line:approx-threshold-start}). For each preference tuple in consideration, if it together with all chosen tuples hitherto do not violate the properties of a strict partial order, their transitive closure is included into the approximate preference relation (Lines \ref{line:approx-spo}-\ref{line:approx-extract_end}).

\begin{algorithm}[h!]
\LinesNumbered
\small

\KwIn{$A_i$: ordered permutation of all possible preference tuples, defined on $dom(d)$, in descending order of their frequencies among users $U$, $\theta_1$ and $\theta_2$: thresholds}

\KwOut{$\widehat{\succ}_U^d$: approximate common preference relation of $U$ on attribute $d$}

\BlankLine

\For{$i = 1$ \emph{\textbf{to}} $\Perm{|dom(d)|}{2}$} 
{
    \uIf{freq(${A_{i}}) = 1$}
    {\label{line:approx-common-extract_start}
        $\widehat{\succ}_U^d \leftarrow \widehat{\succ}_U^d \cup \{A_{i}\}$;
        \KwSty{continue};
    }\label{line:approx-common-extract_end}
    \uIf{$|{\widehat{\succ}_U^d}| \ge \theta_1$ \KwSty{or} freq(${A_{i}}) \le \theta_2$}
    {\label{line:approx-threshold-start}
        \KwSty{break};
    }\label{line:approx-threshold-end}
    \uIf{$(\widehat{\succ}_U^d \cup \{A_{i}\})$ \emph{is a strict partial order}}
    {\label{line:approx-spo}
        $\widehat{\succ}_U^d \leftarrow (\widehat{\succ}_U^d \cup \{A_{i}\})^+$;
    }\label{line:approx-extract_end}
}

\Return{$\widehat{\succ}_U^d$};\\
\caption{\small \algname{GetApproxPreferenceTuples}}
\label{alg:approx}
\end{algorithm}

\vspace{-4mm}
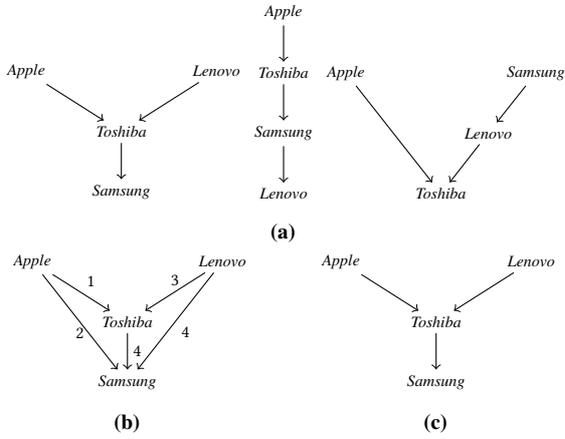
\begin{figure}[h!]
\scriptsize
\noindent
\begin{subfigure}[t]{\linewidth}
\centering
\begin{tikzpicture}
  \node (one) at (-1.25,0) {\emph{Apple}};
  \node (two) at (1.25,0) {\emph{Lenovo}};
  \node (three) at (0,-0.8) {\emph{Toshiba}};
  \node (four) at (0,-1.6) {\emph{Samsung}};
  \draw[->] (one) --   (three);
  \draw[->] (two) -- (three);
  \draw[->] (three) -- (four);
\end{tikzpicture}
\begin{tikzpicture}
  \node (one) at (0,0) {\emph{Apple}};
  \node (two) at (0,-2.4) {\emph{Lenovo}};
  \node (three) at (0,-0.8) {\emph{Toshiba}};
  \node (four) at (0,-1.6) {\emph{Samsung}};
  \draw[->] (one) --   (three);
  \draw[->] (three) -- (four);
  \draw[->] (four) -- (two);
\end{tikzpicture}
\begin{tikzpicture}
  \node (one) at (-1.25,0) {\emph{Apple}};
  \node (two) at (0.62,-0.8) {\emph{Lenovo}};
  \node (three) at (0,-1.6) {\emph{Toshiba}};
  \node (four) at (1.25,0) {\emph{Samsung}};
  \draw[->] (one) --   (three);
  \draw[->] (four) -- (two);
  \draw[->] (two) -- (three);
\end{tikzpicture}
\caption{\small}
\label{fig:approx_a}
\end{subfigure}
\begin{subfigure}[t]{0.49\linewidth}
\centering
\begin{tikzpicture}
  \node (one) at (-1.25,0) {\emph{Apple}};
  \node (two) at (1.25,0) {\emph{Lenovo}};
  \node (three) at (0,-0.8) {\emph{Toshiba}};
  \node (four) at (0,-1.6) {\emph{Samsung}};
  \draw[->] (one) --   node[auto] {$1$} (three);
  \draw[->] (one) --  node[below] {$2$} (four);
  \draw[->] (two) -- node[above] {$3$} (three);
  \draw[->] (two) -- node[auto] {$4$} (four);
  \draw[->] (three) -- node[auto] {$4$} (four);
\end{tikzpicture}
\caption{\small}\vspace{-1mm}
\label{fig:approx_b}
\end{subfigure}
\begin{subfigure}[t]{0.49\linewidth}
\centering
\begin{tikzpicture}
  \node (one) at (-1.25,0) {\emph{Apple}};
  \node (two) at (1.25,0) {\emph{Lenovo}};
  \node (three) at (0,-0.8) {\emph{Toshiba}};
  \node (four) at (0,-1.6) {\emph{Samsung}};
  \draw[->] (one) --   (three);
  \draw[->] (two) -- (three);
  \draw[->] (three) -- (four);
\end{tikzpicture}
\caption{\small}
\label{fig:approx_c}
\end{subfigure}
\vspace{-2mm}
\caption{\small Execution of \algname{GetApproxPreferenceTuples}. \textbf{a)} Input: the preferences of $3$ users w.r.t. \smallattrname{brand}. \textbf{b)} The sequence of included approximate preference tuples. \textbf{c)} Output: the final Hasse diagram representation of the partial order.}\vspace{-4mm}
\label{fig:approx}
\end{figure}

\begin{table}[h!]
\scriptsize
\centering
\begin{tabular}{|@{\hskip+0.8mm}c@{\hskip+0.8mm}|*{12}{@{\hskip+0.8mm}c@{\hskip+0.8mm}|}}
\hline
(\emph{A}, \emph{T})& (\emph{A}, \emph{S})& (\emph{L}, \emph{T}) &(\emph{T}, \emph{S}) & (\emph{S}, \emph{L}) & (\emph{A}, \emph{L}) & (\emph{L}, \emph{S}) & (\emph{T}, \emph{L}) & (\emph{S}, \emph{T}) & (\emph{L}, \emph{A}) & (\emph{T}, \emph{A}) & (\emph{S}, \emph{A})  \\
\hline
3/3& 2/3 & 2/3 & 2/3 & 2/3 & 1/3 & 1/3 & 1/3 & 1/3 & 0/3 & 0/3 & 0/3 \\
\hline
\end{tabular}\\
\vspace{2mm}
\caption{\small All possible preference tuples in order of frequency. (\emph{A}, \emph{L}, \emph{S} and \emph{T} stand for \emph{Apple}, \emph{Lenovo}, \emph{Samsung} and \emph{Toshiba}.)}\vspace{-6mm}
\label{tab:approx_frequency_sort}
\end{table}

\begin{example}
We use Figure~\ref{fig:approx} to explain the execution of \algname{GetApproxPreferenceTuples}. Figure~\ref{fig:approx_a} depicts three users' preference relations on \smallattrname{brand}. Suppose together these three users form a cluster. Assume $\theta_1=7$ and $\theta_2=60\%$. 

Table~\ref{tab:approx_frequency_sort} shows the frequencies of all possible preference tuples after sorting. 
For instance, since all users prefer \emph{Apple} to \emph{Toshiba}, the corresponding frequency is $3/3$; the frequency of $(\emph{Apple}$, $\emph{Samsung})$ is $2/3$ as two of these three users prefer \emph{Apple} to \emph{Samsung}. At first \algname{GetApproxPreferenceTuples} includes the common preference tuple $(\emph{Apple}$, $\emph{Toshiba})$  into $\widehat{\succ}_U^d$. It then includes $(\emph{Apple}$, $\emph{Samsung})$, $(\emph{Lenovo}$, $\emph{Toshiba})$, and (\emph{Toshiba}, \emph{Samsung}) as approximate preference tuples too. Furthermore, upon the addition of (\emph{Toshiba}, \emph{Samsung}), \algname{GetApproxPreferenceTuples} includes (\emph{Lenovo}, \emph{Samsung}) as well since $(\emph{Lenovo}$, $\emph{Toshiba})$ and $(\emph{Toshiba}, \emph{Samsung})$ transitively induce it.  The algorithm then considers (\emph{Samsung}, \emph{Lenovo}), which is disqualified since its reverse tuple $(\emph{Lenovo}$, $\emph{Samsung})$ is already included. Otherwise the tuples will not form a strict partial order. The algorithm stops at $(\emph{Apple}$, $\emph{Lenovo})$ because its frequency is below the threshold $60\%$. Fig.\ref{fig:approx_b} illustrates the sequence of the included tuples and Fig.\ref{fig:approx_c} depicts the output approximate preference relation in the form of a Hasse diagram.\closedef\vspace{-1mm}
\end{example}

\subsection{False Positives and False Negatives due to Approximation}\label{sec:approx-impact}

\algname{FilterThenVerify} (Alg.\ref{alg:filterthenverify}) is extended to use approximate preference tuples and thus we rename it \algname{FilterThenVerifyApprox}. The algorithm itself remains the same. Procedure \algname{updateParetoFrontierU} maintains $\ParetoApprox{U}$ as the candidate Pareto frontier. The algorithm eventually returns $\ParetoApprox{c}$ for each user $c \in U$, in which $\ParetoApprox{c}$ $=$ $\{o\in \ParetoApprox{U}|\nexists o'\in \ParetoApprox{U}\ \text{s.t.}\ o' \succ_c o\}$, i.e., $\ParetoApprox{U} \supseteq \ParetoApprox{c}$. Thus, $\widehat{\mathcal{C}}_o$ $=$ $\{c \in \mathcal{C} | o \in \ParetoApprox{c}\}$. We use the example below to explain its execution over approximate preference relations. 

\begin{example}
Reconsider Example~\ref{exp:alg_ca}, but use the approximate preference relations associated with virtual user $\widehat{U}$ in Table \ref{tab:customer}. Upon the arrival of $o_{15}$, it is compared with the elements in $\ParetoApprox{U} = \{o_2$, $o_7\}$. $\ParetoApprox{U}$ becomes $\{o_2$, $o_{15}\}$ since $o_{15}$ dominates $o_7$. $o_7$ is then also removed from $\ParetoApprox{c_2}$. $o_{15}$ is further compared with $\ParetoApprox{c_1} = \{o_2\}$ and $\ParetoApprox{c_2} = \{o_2\}$, which does not lead to any further change. Overall, $\widehat{\mathcal{C}}_{o_{15}}$ $=$ $\{c_2\}$. The target users using approximate preference relations remain identical to the exact ones, i.e., no loss of accuracy in this case.\closedef
\end{example}

The rest of this section focuses on the accuracy of \algname{FilterThenVerifyApprox}. 
It produces \emph{false positives} if there exists such an $o$ that $o \in \ParetoApprox{c}$ but $o \notin \mathcal{P}_c$. 
It produces \emph{false negatives} if there exists such an $o$ that $o \notin \ParetoApprox{c}$ but $o \in \mathcal{P}_c$. 
Below we present Theorems~\ref{thm:approx1} and \ref{thm:approx2} to analyze how $\ParetoApprox{U}$ and $\ParetoApprox{c}$ relate to $\mathcal{P}_{U}$ and $\mathcal{P}_c$. 

\begin{lemma}\label{lemma:approx1}
Given a set of users $U$ and an attribute $d$, the common preference relation $\succ_U^d$ and an approximate common preference relation $\widehat{\succ}_U^d$ satisfy the following properties:

\textbf{1)} The approximate preference tuples are a superset of the common preference tuples, i.e., $\widehat{\succ}_U^d \supseteq \succ_{U}^d$.

\textbf{2)} If any preference tuple along with its reverse tuple do not belong to the approximate common preference relation, neither of them belongs to the common preference relation either, i.e., $(x,y)$$\notin \widehat{\succ}_U^d$ $\wedge$ $(y,x)$$\notin \widehat{\succ}_U^d$ $\Rightarrow$ $(x,y)$ $\notin$ $\succ_{U}^d$ $\wedge$ $(y,x)$ $\notin$ $\succ_{U}^d$.\closedef
\end{lemma}

\begin{table}[t]
\noindent
\begin{minipage}[b]{0.48\linewidth}
\centering
\pagestyle{empty}
\def\firstcircle{(0,0) circle (0.5cm)}
\def\secondcircle{(0:0.4cm) circle (1cm)}
\def\firstellipse{(0.55,0) ellipse (0.65cm and 0.9cm)}
\def\secondellipse{(0.25,0) ellipse (0.35cm and 0.71cm)}
\centering
\begin{tikzpicture}
\begin{scope}[transparency group]
    \begin{scope}[blend mode=multiply]
        \firstcircle;
        \secondcircle;
        \firstellipse
        \secondellipse;
        \draw[blue] \firstcircle;
        \draw[blue] \secondcircle;
        \draw[blue] (-1.25,-1.25) rectangle (1.73,1.25);
        \draw[red] \firstellipse;
        \draw[red] \secondellipse;
    \end{scope}
    \end{scope}
      \node at ( 0:-0.3cm)    {\scriptsize \Romannum{3}};
      \node at ( 0:0.3cm)    {\scriptsize \Romannum{4}};
      \node at ( 65:0.6cm)    {\scriptsize \Romannum{5}};
      \node at ( 0:0.9cm)    {\scriptsize \Romannum{6}};
      \node at ( 40:1.5cm)    {\scriptsize \Romannum{1}};
      \node at ( 96:0.7cm)    {\scriptsize \Romannum{2}};
\end{tikzpicture}
   \captionof{figure}{\small Venn diagram depicting $\mathcal{O}$, $\mathcal{P}_{U}$, $\ParetoApprox{U}$, $\mathcal{P}_c$ and $\ParetoApprox{c}$.}\vspace{-4mm}
   \label{fig:venn_approx}
\end{minipage}
\hspace{1mm}
\noindent
\begin{minipage}[b]{0.48\linewidth}
\centering
\small
\begin{tabular}{|c|c|}
\hline
Set & Area Covered\\
\hline
\hline
$\mathcal{O}$ & \Romannum{1},\Romannum{2},\Romannum{3},\Romannum{4},\Romannum{5},\Romannum{6}\\
\hline
$\mathcal{P}_{U}$ & \Romannum{2},\Romannum{3},\Romannum{4},\Romannum{5},\Romannum{6}\\
\hline
$\ParetoApprox{U}$ & \Romannum{4},\Romannum{5},\Romannum{6}\\
\hline
$\mathcal{P}_c$ & \Romannum{3},\Romannum{4}\\
\hline
$\ParetoApprox{c}$ & \Romannum{4},\Romannum{5}\\
\hline
\end{tabular}
\caption{\small Areas covered by $\mathcal{O}$, $\mathcal{P}_{U}$, $\ParetoApprox{U}$, $\mathcal{P}_c$ and $\ParetoApprox{c}$ in Fig.\ref{fig:venn_approx}.}\vspace{-4mm}
\label{tab:venn_approx}
\end{minipage}
\end{table}

\begin{table}[t]
\centering
\scriptsize
\begin{tabular}{|c|c|c|c|}
\hline
                  & \multicolumn{3}{c|}{\textbf{Exact}} \\ \hline
\multirow{3}{*}{\textbf{Approx.}} & & \textbf{Pareto frontier} &  \textbf{Non Pareto frontier} \\ \cline{2-4}
                  & \textbf{Pareto frontier} & \Romannum{4} & \Romannum{5} \\ \cline{2-4}
                  & \textbf{Non Pareto frontier} & \Romannum{3} & \Romannum{1},\Romannum{2},\Romannum{6} \\ \hline
\end{tabular}
\caption{\small Confusion matrix w.r.t. $c$.}\vspace{-8mm}
\label{tab:matrix}
\end{table}

\begin{theorem}\label{thm:approx1}
Given objects $\mathcal{O}$ and users $U$, the Pareto frontier with regard to approximate common preference relations is a subset of the Pareto frontier with regard to common preference relations, i.e., $\ParetoApprox{U} \subseteq \mathcal{P}_{U}$.\closedef
\end{theorem}

\emph{Proof:} We prove by contradiction. Suppose $\ParetoApprox{U} \nsubseteq \mathcal{P}_{U}$, which would mean there exists $o \in \mathcal{O}$ such that $o \in \ParetoApprox{U}$ and $o \notin \mathcal{P}_{U}$.  That leads to the existence of an $o'$ such that $o' \succ_{U} o$ and $o' \nsucc_{\widehat{U}} o$. However, $o' \succ_{U} o$ implies $o' \succ_{\widehat{U}} o$ because $\widehat{\succ}_U^d \supseteq \succ_{U}^d$ for every $d$ (Lemma~\ref{lemma:approx1}). 
Therefore, the existence of $o'$ is impossible. This contradiction proves that $\ParetoApprox{U} \subseteq \mathcal{P}_{U}$.\qed

\begin{lemma}\label{lemma:approx2}
Given any set of users $U$, for all user $c \in U$, $\ParetoApprox{U} \supseteq \ParetoApprox{c}$.\closedef\vspace{-0.5mm}
\end{lemma}

\begin{theorem}\label{thm:approx2}
Given any set of users $U$, for all user $c \in U$, $\ParetoApprox{U} \cap \mathcal{P}_c \subseteq \ParetoApprox{c}$.\closedef
\end{theorem}

\emph{Proof:} We prove by contradiction. Suppose $\ParetoApprox{U} \cap \mathcal{P}_c \nsubseteq \ParetoApprox{c}$, which would mean there exists $o \in \mathcal{O}$ such that $o$ $\in$ $\ParetoApprox{U} \cap \mathcal{P}_c$ and $o$ $\notin$ $\ParetoApprox{c}$. $o$ $\notin$ $\ParetoApprox{c}$ implies the existence of an $o' \in \mathcal{O}$ such that $o' \in \ParetoApprox{c}$ and $o'$ $\succ_{c}$ $o$ (since  $o$ $\in$ $\ParetoApprox{U} \cap \mathcal{P}_c$ and thus $o$ $\in$ $\ParetoApprox{U}$ which means $o' \nsucc_{\widehat{U}} o$). Since $o'$ $\succ_{c}$ $o$, $o$ $\notin$ $\mathcal{P}_c$ (Def.~\ref{def:pareto_frontier}) and thus $o$ $\notin$ $\ParetoApprox{U} \cap \mathcal{P}_c$. In other words, the existence of $o'$ is impossible. This contradiction proves that $\ParetoApprox{U} \cap \mathcal{P}_c \subseteq \ParetoApprox{c}$.\qed

%%%\textcolor{red}{Theorems~\ref{thm:approx1} and \ref{thm:approx2} provide useful insights into the impact of approximate preference relations on algorithm \algname{FilterThenVerifyApprox}'s accuracy in finding Pareto frontier. Theorem~\ref{thm:approx1} ensures $\ParetoApprox{U}$ will never include objects from $\mathcal{O}-\mathcal{P}_{U}$. As a result, $\ParetoApprox{c}$ may include $o$ while $o \notin \mathcal{P}_c$ such that there exists an object $o' \succ_c o$ and $o' \notin \ParetoApprox{U}$, i.e., $o$ becomes a \emph{false positive}. Moreover, we use Theorem~\ref{thm:approx1} to prove Theorem~\ref{thm:approx2}. Theorem~\ref{thm:approx2} says that given a cluster $U$, if an object $o$ belongs to both $\ParetoApprox{U}$ and $\mathcal{P}_c$, then $\ParetoApprox{c}$ includes $o$ as well. Violation of Theorem~\ref{thm:approx2} allows $o \notin \ParetoApprox{c}$ such that $o \in \ParetoApprox{U} \wedge o \in \mathcal{P}_c$, i.e., $o$ is a \emph{false negative}.}

Consider a cluster ${U}$  and a user $c \in U$. The Venn diagram in Fig.~\ref{fig:venn_approx} shows the effect of approximation through depicting $\mathcal{O}$ (rectangle), $\mathcal{P}_{U}$ (outer blue circle), $\ParetoApprox{U}$ (outer red ellipse), $\mathcal{P}_c$ (inner blue circle), and $\ParetoApprox{c}$ (inner red ellipse). Besides, Table~\ref{tab:venn_approx} elaborates the area covered by these sets while Table~\ref{tab:matrix} shows the confusion matrix for $c$. Note that using approximate common preference relations results in false negatives (\Romannum{3}).  Mistakenly declaring  \Romannum{3} as not Pareto-optimal further allows false positives (\Romannum{5}) to sneak in. 

With these notations in place, we are ready to quantify the accuracy of \algname{FilterThenVerifyApprox} using standard evaluation measures in information retrieval.   Specifically, \emph{precision} is the fraction of objects found by \algname{FilterThenVerifyApprox} that are truly Pareto-optimal, i.e., $\sum_{c \in \mathcal{C}}{{\ParetoApprox{c} \cap \mathcal{P}_c}} \over \sum_{c \in \mathcal{C}}{\ParetoApprox{c}}$. \emph{Recall} is the fraction of Pareto-optimal objects that are correctly found by \algname{FilterThenVerifyApprox}, i.e., $\sum_{c \in \mathcal{C}}{{\ParetoApprox{c} \cap \mathcal{P}_c}} \over \sum_{c \in \mathcal{C}}{{\mathcal{P}_c}}$. With regard to a specific user $c$, the algorithm's precision, recall and accuracy can be represented using the areas in Fig.~\ref{fig:venn_approx}, as follows.\vspace{-3mm}

\begin{small}
\begin{flalign}
&precision = {|\ \text{\Romannum{4}}\ | \over |\ \text{\Romannum{4} $\cup$ \Romannum{5}}\ |}&
\end{flalign}\vspace{-3mm}
\begin{flalign}
&recall = {|\ \text{\Romannum{4}}| \over |\ \text{\Romannum{3} $\cup$ \Romannum{4}}\ |}&
\end{flalign}\vspace{-3mm}
\begin{flalign}
&accuracy = {|\ \text{\Romannum{1} $\cup$ \Romannum{2} $\cup$ \Romannum{4} $\cup$ \Romannum{6}}\ | \over |\ \text{\Romannum{1} $\cup$ \Romannum{2} $\cup$ \Romannum{3} $\cup$ \Romannum{4} $\cup$ \Romannum{5} $\cup$ \Romannum{6}}\ |}&
\end{flalign}\vspace{-3mm}
\end{small}

\subsection{Similarity Functions}\label{sec:approx-similarity}
To make the clustering solution in Sec.~\ref{sec:cluster} compatible with approximate preference relations, we extend the similarity measures, using ideas inspired by the Jaccard similarity for non-negative multidimensional real vectors~\cite{ChierichettiKPV10}. 

\textbf{1) Jaccard Similarity}\hspace{2mm} Consider an attribute $d$ with $|dom(d)| = m$. For each cluster $U$, construct a vector $\mathbf{U} = (\mathbf{U}(1)$, $\mathbf{U}(2)$, $\ldots$, $\mathbf{U}(\Perm{m}{2}))$. For $i \in [1, \Perm{m}{2}]$, $\mathbf{U}(i)$  represents the frequency of $A_i$ (Definition~\ref{def:approx_common_preference_tuple}) in $U$. Given two clusters $U$ and $V$, their \emph{Jaccard similarity} on attribute $d$ is \vspace{-3mm}

\begin{flalign}\label{fn:sim_j}
sim^d_{j}(U,V) = {\sum_i {\min(\mathbf{U}(i), \mathbf{V}(i))} \over \sum_i {\max(\mathbf{U}(i), \mathbf{V}(i))}}
\end{flalign}

\begin{example}
Consider $U_1$ and $U_3$ in Table~\ref{tab:customer4}. Suppose $A(i)$ for $i \in [1, \Perm{m}{2}]$ are ((\emph{Apple}, \emph{Lenovo}), (\emph{Apple}, \emph{Samsumg}), (\emph{Apple}, \emph{Toshiba}), (\emph{Lenovo}, \emph{Apple}), (\emph{Lenovo}, \emph{Samsung}), (\emph{Lenovo}, \emph{Toshiba}), (\emph{Toshiba}, \emph{Apple}), (\emph{Toshiba}, \emph{Lenovo}), (\emph{Toshiba}, $\emph{Samsung})$, (\emph{Samsung}, \emph{Apple}), $(\emph{Samsung}$, $\emph{Lenovo})$, $(\emph{Samsung}$, $\emph{Toshiba}))$. The two vectors are $\mathbf{U_1} = (2/2$, $2/2$, $0/2$, $0/2$, $2/2$, $0/2$, $0/2$, $1/2$, $2/2$, $0/2$, $0/2$, $0/2)$ and $\mathbf{U_3} = (0/2$, $2/2$, $1/2$, $2/2$, $2/2$, $2/2$, $0/2$, $0/2$, $1/2$, $0/2$, $0/2$, $0/2)$. For instance, $\mathbf{U_1}$ has $1/2$ on the $8^{th}$-dimension since only one of the two users' preference relations contains $(\emph{Toshiba}$, $\emph{Lenovo})$. Hence, $sim^{\attrname{brand}}_{j}(U_1$,$U_3)$ $=$ $0.36$.\closedef
\end{example}

\textbf{2) Weighted Jaccard Similarity}\hspace{2mm} This measure, denoted as $sim^d_{wj}$, extends the namesake measure in Sec.~\ref{sec:cluster} with the idea above. Its definition is the same as Eq.~\ref{fn:sim_j} except that a value $\mathbf{U}(i)$ in a vector represents the frequency of $A_i$ in $U$ that takes into consideration the weights explained in Sec.~\ref{sec:cluster}. Consider $A_i$ as the preference relation $(A_i(x), A_i(y))$. Its definition is as follows.\vspace{-3mm}

\begin{small}
\begin{flalign}
sim^d_{wj}(U,V) = \sum_i (\min({1 \over |U|} \times \sum_{c \in U}{1 \over \displaystyle \min_{s \in S_{c}^d}\hspace{-1mm}D(s,A_i(x))\hspace{-1mm}+\hspace{-1mm}1}, \nonumber \\
{1 \over |V|} \times \sum_{c \in V}{1 \over \displaystyle \min_{s \in S_{c}^d}\hspace{-1mm}D(s,A_i(x))\hspace{-1mm}+\hspace{-1mm}1}))  \nonumber \\
\Big/\ \sum_i (\max({1 \over |U|} \times \sum_{c \in U}{1 \over \displaystyle \min_{s \in S_{c}^d}\hspace{-1mm}D(s,A_i(x))\hspace{-1mm}+\hspace{-1mm}1}, \nonumber \\
{1 \over |V|} \times \sum_{c \in V}{1 \over \displaystyle \min_{s \in S_{c}^d}\hspace{-1mm}D(s,A_i(x))\hspace{-1mm}+\hspace{-1mm}1}))
\end{flalign}
\end{small}

\begin{example}
In Table~\ref{tab:customer4}, in the partial order depicting $\succ^\attrname{brand}_{c_6}$, the distance to \emph{Apple} from the maximal value $\emph{Lenovo}$ is $1$, i.e., the weight of $\emph{Apple}$ is $1/2$. Since only one of the two users in $U_3$ has $(\emph{Apple}$, $\emph{Toshiba})$ in their preference relation, $\mathbf{U_3}$ has ${{1 \over 2}+0 \over 2} = {1 \over 4}$ on the $3^{rd}$-dimension. In this way, we get $\mathbf{U_1} = (2/2$, $2/2$, $0/2$, $0/2$, $1/2$, $0/2$, $0/2$, $1/2$, $2/2$, $0/2$, $0/2$, $0/2)$ and $\mathbf{U_3} = (0/2$, $1/2$, $1/4$, $2/2$, $2/2$, $2/2$, $0/2$, $0/2$, $1/4$, $0/2$, $0/2$, $0/2)$. Therefore, $sim^{\attrname{brand}}_{wj}(U_1$,$U_3) = 0.19$.\closedef 
\end{example}
\section{Alive Object Dissemination}\label{sec:window}
In Sec.~\ref{sec:intro}, we discussed motivating applications such as social network content dissemination, news delivery and product recommendation. The significance of a particular social network content (e.g. a post in Facebook) or a piece of news diminishes eventually. Similarly, in any inventory, products are consumed and perishable products expire over time. In other words, objects can have limited lifetime. Thus, upon the arrival of a new object, it needs to compete only with the alive objects. To meet this requirement, we extend our problem as continuous monitoring of Pareto frontiers over \emph{alive objects} for many users and formalize it as finding Pareto frontiers over \emph{sliding window}.

Suppose $\mathcal{O}$ $=$ $\{o_1$, $o_2$, $\ldots$, $o_N\}$ is a stream of objects, in which the subscript of each object is its timestamp.
We consider a sliding window as a sequence of $W$ recent objects. Upon the arrival of an incoming object $o_{in}$, an object $o_{out}$ expires if $in-out=W$. Specifically, the sliding window contains objects whose timestamps are in $(out, in]$, i.e., an object $o_i \in \mathcal{O}$ is alive during $(out, in]$ if $i \in (out, in]$. Given the concept of sliding window, we extend the definition of Pareto frontier in Def.~\ref{def:pareto_frontier} and the problem statement in Sec.~\ref{sec:model}.

\begin{definition}[Pareto Frontier]\label{def:pareto_frontier_window}
An alive object $o$ is Pareto-optimal with respect to $c$, if no other alive object dominates it. $\mathcal{P}_c$ $=$ $\{o_i\in \mathcal{O}|\nexists o_j\in \mathcal{O}\ \text{s.t.}\ o_j \succ_c o_i \wedge i,j \in (out, in]\}$. The target users of $o_{in}$ is $\mathcal{C}_{o_{in}}$ $=$ $\{c \in \mathcal{C} | o_{in} \in \mathcal{P}_c\}$ (Def.~\ref{def:target_user}).\closedef
\end{definition}

\textbf{Problem Statement}\hspace{2mm} The problem of continuous monitoring of Pareto frontiers over sliding window is, given a set of users $\mathcal{C}$, their preference relations on attributes $\mathcal{D}$, and a stream of objects $\mathcal{O}$ with the incoming object $o_{in}$ as well as the outgoing object $o_{out}$, find $\mathcal{C}_{o_{in}}$---the target users of $o_{in}$. \vspace{1mm}

\textbf{Algorithms \algname{BaselineSW} and \algname{FilterThenVerifySW}}\hspace{2mm}  We extend \algname{Baseline} and \algname{FilterThenVerify} to \algname{BaselineSW} and \algname{FilterThenVerifySW}, respectively, to accommodate sliding window. Algorithm~\ref{alg:BaselineSW} and Algorithm~\ref{alg:FilterThenVerifySW} describe the pseudo codes, respectively. We note that no prior work studied Pareto frontier maintenance with regard to strict partial orders over sliding window. \cite{lin2005stabbing, tao2006maintaining, morse2007efficient} studied skyline maintenance over sliding window, assuming numeric attributes. ~\cite{sarkas2008categorical} considered categorical attributes and focused on maintaining preference query results over sliding window.

\begin{algorithm}[t]
\LinesNumbered
\small

\KwIn{$\mathcal{C}$: all users; $\mathcal{P}$: Pareto frontier; $\mathcal{PB}$: Pareto frontier buffer; $o_{in}$: incoming object; $o_{out}$: outgoing object}

\KwOut{$\mathcal{C}_{o_{in}}$: target users of $o_{in}$}

\BlankLine

\ForEach{$c \in \mathcal{C}$}
{
    \uIf{$o_{out} \in \mathcal{P}_c$}
    {
        \ForEach{$o \in \mathcal{PB}_{c} $}
        {
            \uIf{$o_{out}$ $\succ^{c}$ $o$}
            {
                \algname{mendParetoFrontierSW}$(c, o)$;
            }
        }
        %find objects in $\mathcal{PB}_c$ dominated by $o_{out}$ and repair corresponding target users as well as $\mathcal{P}_c$ based on them;\\
    }
    $\mathcal{PB}_c \leftarrow \mathcal{PB}_c - \{o_{out}\}$;\\
    \uIf{$o_{in}$ not dominated by $\mathcal{P}_c$}
    {
        \algname{updateParetoFrontierSW}$(c, o)$;\label{line:BaselineSW-updateParetoFrontierSW}
        %delete dominated objects from $\mathcal{P}_c$ and repair corresponding target users;\\
    }
    %delete dominated objects from $\mathcal{PB}_c$;
    \algname{refreshParetoBufferSW}$(c, o_{in})$;\label{line:BaselineSW-refreshParetoBufferSW}
}

\Return{$\mathcal{C}_{o_{in}}$};\\
\BlankLine
\SetKwInput{KwProc}{Procedure}
\setcounter{AlgoLine}{0}
\SetKwFunction{camember}{\algname{mendParetoFrontierSW}}

\KwProc{\camember$(c,o)$}

$\mathit{isPareto} \leftarrow \KwSty{true}$;\\
\ForEach{$o' \in \mathcal{P}_{c} $}
{
    \lIf{$o'$ $\succ^{c}$ $o$}
    {
        $\mathit{isPareto} \leftarrow \KwSty{false}$;
        \KwSty{break}
    }
}
\lIf{$\mathit{isPareto}$}
{
    $\mathcal{P}_{c} \leftarrow \mathcal{P}_{c} \cup \{o\}$;
    $\mathcal{C}_o \leftarrow \mathcal{C}_o \cup \{c\}$
}
\BlankLine
\SetKwInput{KwProc}{Procedure}
\setcounter{AlgoLine}{0}
\SetKwFunction{camember}{\algname{updateParetoFrontierSW}}

\KwProc{\camember$(c,o_{in})$}

$\mathcal{P}_c \leftarrow \mathcal{P}_c \cup \{o_{in}\}$;
$\mathcal{C}_{o_{in}} \leftarrow \mathcal{C}_{o_{in}} \cup \{c\}$;\\
\ForEach{$o \in \mathcal{P}_{c} $}
{
    \lIf{$o_{in}$ $\succ^{c}$ $o$}
    {
        $\mathcal{P}_{c} \leftarrow \mathcal{P}_{c} - \{o\}$;
        $\mathcal{C}_{o} \leftarrow \mathcal{C}_{o} - \{c\}$
    }
}
\BlankLine
\SetKwInput{KwProc}{Procedure}
\setcounter{AlgoLine}{0}
\SetKwFunction{camember}{\algname{refreshParetoBufferSW}}

\KwProc{\camember$(c,o_{in})$}

$\mathcal{PB}_c \leftarrow \mathcal{PB}_c \cup \{o_{in}\}$;\\
\ForEach{$o \in \mathcal{PB}_{c} $}
{
    \lIf{$o_{in}$ $\succ^{c}$ $o$}
    {
        $\mathcal{PB}_{c} \leftarrow \mathcal{PB}_{c} - \{o\}$
    }
}
\caption{\small \algname{BaselineSW}}
\label{alg:BaselineSW}
\end{algorithm}

\begin{algorithm}[h!]
\LinesNumbered
\small

\KwIn{$\{U_1$, $U_2$,..., $U_n\}$: all clusters; $\mathcal{P}$: Pareto frontier; $\mathcal{PB}$: Pareto frontier buffer; $o_{in}$: incoming object; $o_{out}$: outgoing object}

\KwOut{$\mathcal{C}_{o_{in}}$: target users of $o_{in}$}

\BlankLine

\ForEach{$i=1$ to $n$}
{
    \uIf{$o_{out} \in \mathcal{P}_{U_i}$}
    {
        %find objects in $\mathcal{PB}_{C}$ dominated by $o_{out}$ and repair $\mathcal{P}_{C}$ along with $\mathcal{P}_{c}$ $\forall c \in \mathcal{C}_{C}$ based on them;\\
        \ForEach{$o \in \mathcal{PB}_{U_i} $}
        {
            \uIf{$o_{out} \succ o$}
            {
                $isPareto \leftarrow$ \algname{mendParetoFrontierUSW}$(U_i,o)$;\\
                \uIf{$\mathit{isPareto}$}
                {
                    \ForEach{$c \in U_i$}
                    {
                        \algname{mendParetoFrontierSW}$(c, o)$
                    }
                }
            }
        }
    }
    $\mathcal{PB}_{U_i} \leftarrow \mathcal{PB}_{U_i} - \{o_{out}\}$;\\
    \uIf{$o_{in}$ not dominated by $\mathcal{P}_{U_i}$}
    {
        \algname{updateParetoFrontierUSW}$(U_i, o)$;\label{line:FilterThenVerifySW-updateParetoFrontierUSW}\\
        \ForEach{$c \in U_i$}
        {
            \uIf{$o_{in}$ not dominated by $\mathcal{P}_c$}
            {
                \algname{updateParetoFrontierSW}$(c, o)$;\label{line:FilterThenVerifySW-updateParetoFrontierSW}//Algorithm~\ref{alg:BaselineSW}\\
            }
        }
    }
    \algname{refreshParetoBufferSW}$(U_i, o_{in})$;\label{line:FilterThenVerifySW-refreshParetoBufferSW}
    //Algorithm~\ref{alg:BaselineSW}\\
    %delete dominated objects from $\mathcal{PB}_{C}$;
}

\Return{$\mathcal{C}_{o_{in}}$};\\
\BlankLine
\SetKwInput{KwProc}{Procedure}
\setcounter{AlgoLine}{0}
\SetKwFunction{camember}{\algname{mendParetoFrontierUSW}}

\KwProc{\camember$(U,o)$}

$\mathit{isPareto} \leftarrow \KwSty{true}$;\\
\ForEach{$o' \in \mathcal{P}_{U} $}
{
    \lIf{$o'$ $\succ^{U}$ $o$}
    {
        \Return{\KwSty{false}}
    }
}
\lIf{$\mathit{isPareto}$}
{
    $\mathcal{P}_{U} \leftarrow \mathcal{P}_{U} \cup \{o\}$
}
\Return{$isPareto$};\\
\BlankLine
\SetKwInput{KwProc}{Procedure}
\setcounter{AlgoLine}{0}
\SetKwFunction{camember}{\algname{updateParetoFrontierUSW}}

\KwProc{\camember$(U,o_{in})$}

$\mathcal{P}_U \leftarrow \mathcal{P}_U \cup \{o_{in}\}$;\\
\ForEach{$o \in \mathcal{P}_{U} $}
{
    \lIf{$o_{in}$ $\succ^{U}$ $o$}
    {
        $\mathcal{P}_{c} \leftarrow \mathcal{P}_{c} - \{o\}$;
    }
}
\caption{\small \algname{FilterThenVerifySW}}
\label{alg:FilterThenVerifySW}
\end{algorithm}

Under the constraint of having a sliding window, an object can be excluded from Pareto frontier forever if it is dominated by any succeeding object. This observation is formalized as Theorem~\ref{thm:window2}.

\begin{theorem}\label{thm:window2}
Consider a user $c \in \mathcal{C}$ and two objects $o_i, o_j \in \mathcal{O}$ such that $o_i \prec_c o_j$ and $i<j$. After the arrival of $o_j$, $o_i$ can never be part of $\mathcal{P}_c$ in its remaining lifetime.\closedef
\end{theorem}

\emph{Proof:} Since $i<j$, $o_i$ expires before $o_j$ and the sliding window always includes $o_j$ if it includes $o_i$. Since $o_j$ dominates $o_i$, $o_i$ will never get into $\mathcal{P}_c$ after the arrival of $o_j$.\qed

\begin{example}\label{exp:pareto_frontier}
Consider Table~\ref{tab:product} and Table~\ref{tab:customer}. Consider $W$, $in$ and $out$ as 5, 10 and 5, respectively. Upon the arrival of $o_{10}$ $=$ $\langle 9.5$, \emph{Lenovo}, \emph{triple}$\rangle$ and the expiration of $o_5$ $=$ $\langle 9$, \emph{Samsung}, \emph{quad}$\rangle$, we get $\mathcal{P}_{c_1} = \{o_8\}$ and $\mathcal{P}_{c_2} = \{o_7$, $o_8\}$.\closedef\vspace{-1mm}
\end{example}

By Theorem~\ref{thm:window2}, we extend our algorithms to maintain a \emph{Pareto frontier buffer} which stores at most $W$ recent objects that are not dominated by any succeeding object. Clearly, $o_{in}$ is part of the Pareto frontier buffer.

\begin{definition}[Pareto Frontier Buffer]\label{def:pareto_frontier_buffer}
With regard to user $c$ and the sliding window $(out$, $in]$, an alive object $o$ belongs to the Pareto frontier buffer if it is not dominated by any succeeding object.  The Pareto frontier buffer is $\mathcal{PB}_c$ $=$ $\{o_i\in \mathcal{O}|\nexists o_j\in \mathcal{O}\ \text{s.t.}\ o_j \succ_c o_i \wedge i,j \in (out, in] \wedge i<j\}$. By definition, $\mathcal{PB}_{c} \supseteq \mathcal{P}_c$ (Def.~\ref{def:pareto_frontier_window}).\closedef\vspace{-1mm}
\end{definition}

\begin{theorem}\label{thm:window1}
Given a set of users $U$, for all $c \in U$, i) $\mathcal{PB}_{U} \supseteq \mathcal{P}_U$ and ii) $\mathcal{PB}_{U} \supseteq \mathcal{PB}_c$ .\closedef\vspace{-1mm}
\end{theorem}

\emph{Proof:} i) Together Def.~\ref{def:pareto_frontier_window} and~\ref{def:pareto_frontier_buffer} imply that $\mathcal{PB}_{U} \supseteq \mathcal{P}_U$.

ii) We prove by contradiction. Suppose that there exists $c$ $\in$ $U$ such that $\mathcal{PB}_{U} \nsupseteq \mathcal{PB}_c$, which would mean there exists $o$ $\in$ $\mathcal{O}$ such that $o$ $\in$ $\mathcal{PB}_c$ and $o$ $\notin$ $\mathcal{PB}_{U}$. That implies the existence of an $o' \in \mathcal{O}$ such that $o'$ $\succ_{U}$ $o$ and $o'$ $\nsucc_{c}$ $o$. However, by Def.~\ref{def:common_preference_tuple}, $o'$ $\succ_{U}$ $o$ implies $o'$ $\succ_{c}$ $o$.
Therefore, the existence of $o'$ is impossible.  In conclusion, $\mathcal{PB}_{U} \supseteq \mathcal{PB}_c$.\qed

Note that, \algname{BaselineSW} needs to maintain an exclusive Pareto frontier buffer for each user ($\mathcal{PB}_{c}$) while a Pareto frontier buffer per cluster ($\mathcal{PB}_{U}$) is sufficient for \algname{FilterThenVerifySW}.\vspace{-2mm}

\begin{example}
Continue Example~\ref{exp:pareto_frontier}. We get $\mathcal{PB}_{c_1}$=$\{o_8$, $o_9$, $o_{10}\}$. In this case, $o_8$ is the only element of $\mathcal{P}_{c_1}$. Since $o_6$ or $o_7$ could never been qualified in Pareto-optimality as they arrive before $o_8$, we do not need to store them. Nevertheless, upon the expiration of $o_8$, either $o_9$ or $o_{10}$ could attain Pareto-optimality during their lifetime if they are not dominated by any following object. Therefore, $o_9$ and $o_{10}$ are stored in $\mathcal{PB}_{c_1}$. For instance, $o_{10}$ acquires Pareto-optimality during $(8, 13]$ as it does not dominated by any following object.\closedef
\end{example}

\begin{table}[t]
\centering
\scriptsize
\begin{tabular}{|l|*{4}{c|}}\hline
$ $ & \smallattrname{{\scriptsize display}} & \smallattrname{{\scriptsize brand}} & \smallattrname{{\scriptsize CPU}}\\
\hline
\hline
{\scriptsize $o_1$} & 17 & \emph{Lenovo} & \emph{dual}\\
\hline
{\scriptsize $o_2$} & 9.5 & \emph{Sony} & \emph{single}\\
\hline
{\scriptsize $o_3$} & 12 & \emph{Apple} & \emph{dual}\\
\hline
{\scriptsize $o_4$} & 16 & \emph{Lenovo} & \emph{quad}\\
\hline
{\scriptsize $o_5$} & 19 & \emph{Toshiba} & \emph{single}\\
\hline
{\scriptsize $o_6$} & 12.5 & \emph{Samsung} & \emph{quad}\\
\hline
{\scriptsize $o_7$} & 14 & \emph{Apple} & \emph{dual}\\
\hline
\end{tabular}
\caption{\small Product Table}
\label{tab:product_window}
\end{table}

\begin{table}[h!]
\centering
\scriptsize
\begin{tabular}{|l|*{6}{@{\hskip-0.001mm}c@{\hskip-0.001mm}|}}
\hline
$W$ & $\mathcal{P}_{c_1}$ & $\mathcal{P}_{c_2}$ & $\mathcal{PB}_{c_1}$ & $\mathcal{PB}_{c_2}$\\
\hline
\hline
$[1, 6]$ & $\{o_1, o_3\}$ & $\{o_3, o_4\}$ & $\{o_1, o_3, o_4, o_6\}$ & $\{o_3, o_4, o_5, o_6\}$\\
\hline
$(1, 6]$ & $\{o_3\}$ & $\{o_3, o_4\}$ & $\{o_3, o_4, o_6\}$ & $\{o_3, o_4, o_5, o_6\}$\\
\hline
$(1, 7]$ & $\{o_7\}$ & $\{o_4, o_7\}$ & $\{o_4, o_7\}$ & $\{o_4, o_7\}$\\
\hline
\end{tabular}
\caption{\small Content of Pareto Frontiers and Pareto Buffers During 3 Different Phases of Window of \algname{BaselineSW}}\label{tab:window_baseline}
\end{table}

\begin{table}[h!]
\centering
\scriptsize
\begin{tabular}{|l|*{6}{@{\hskip-0.001mm}c@{\hskip-0.001mm}|}}
\hline
$ W$ & $\mathcal{P}_{U}$ & $\mathcal{P}_{c_1}$ & $\mathcal{P}_{c_2}$ & $\mathcal{PB}_{U}$\\
\hline
\hline
$[1, 6]$ & $\{o_1, o_3, o_4\}$ & $\{o_1, o_3\}$ & $\{o_3, o_4\}$ & $\{o_1, o_3, o_4, o_5, o_6\}$\\
\hline
$(1, 6]$ & $\{o_3, o_4\}$ & $\{o_3\}$ & $\{o_3, o_4\}$ & $\{o_3, o_4, o_5, o_6\}$\\
\hline
$(1, 7]$ & $\{o_4, o_7\}$ & $\{o_7\}$ & $\{o_4, o_7\}$ & $\{o_4, o_7\}$\\
\hline
\end{tabular}
\caption{\small Content of Pareto Frontiers and Pareto Buffers During 3 Different Phases of Window of \algname{FilterThenVerifySW}}\label{tab:window_ca}
\end{table}

In our sliding window framework, upon the expiration of an outgoing object $o_{out}$, for all $c \in \mathcal{C}$, at first \algname{BaselineSW} calls Procedure~\algname{mendParetoFrontierSW} to mend $\mathcal{P}_c$. Because at this point, the alive objects those are exclusively dominated by $o_{out}$, acquire Pareto-optimality. While $o_{in}$ arrives, if $o_{in}$ belongs to $\mathcal{P}_c$, then Procedure~\algname{updateParetoFrontierSW} in \algname{BaselineSW} discards objects that are dominated by $o_{in}$, thereby updates $\mathcal{P}_c$ (Line~\ref{line:BaselineSW-updateParetoFrontierSW}). After that, Procedure~\algname{refreshParetoBufferSW} in \algname{BaselineSW} repairs $\mathcal{PB}_{c}$. Specifically,  $o_{in}$ replaces the alive objects from $\mathcal{PB}_{c}$ that it dominates (Line~\ref{line:BaselineSW-refreshParetoBufferSW}). Thus $\mathcal{PB}_{c}$ remains concurrent with Def.~\ref{def:pareto_frontier_buffer}.

On the contrary, in case of \algname{FilterThenVerifySW}, upon the expiration of an outgoing object $o_{out}$, Procedures~\algname{mendParetoFrontierUSW} and~\algname{mendParetoFrontierSW} together mend $\mathcal{P}_U$ and $\mathcal{P}_c$ for all $U \subseteq \mathcal{C}$, for all $c \in U$. While $o_{in}$ arrives, if $o_{in}$ belongs to $\mathcal{P}_U$, then Procedure~\algname{updateParetoFrontierUSW} discards objects from $\mathcal{P}_U$ that are dominated by $o_{in}$ (Line~\ref{line:FilterThenVerifySW-updateParetoFrontierUSW}). Now for all $c \in U$, if $c$ approves $o_{in}$ as a Pareto-optimal object, then Procedure~\algname{updateParetoFrontierSW} finds out the objects in $\mathcal{P}_c$ dominated by $o_{in}$ and thereby removes them (Line~\ref{line:FilterThenVerifySW-updateParetoFrontierSW}). Lastly, Procedure~\algname{refreshParetoBufferUSW} repairs $\mathcal{PB}_{U}$  so that it includes only the objects which have the potentiality to acquire Pareto-optimality over time with regard to $U$ (Def.~\ref{def:pareto_frontier_buffer} and Theorem~\ref{thm:window1}) (Line~\ref{line:FilterThenVerifySW-refreshParetoBufferSW}).

\begin{figure}[t]
\centering
\pagestyle{empty}
\def\firstcircle{(0,0) circle (0.75cm)}
\def\secondcircle{(0:0.5cm) circle (0.75cm)}
\def\thirdcircle{(0:0.25cm) circle (1.25cm)}
\def\forthcircle{(0:0.25cm) circle (1.75cm)}
\centering
\begin{tikzpicture}
    \begin{scope}
        \firstcircle;
        \secondcircle;
        \thirdcircle;
        \draw \firstcircle;
        \draw \secondcircle;
        \draw \thirdcircle;
        \draw \forthcircle;
        \draw (-2,-2) rectangle (2.5,2);
    \end{scope}
      \node at ( 0:-0.45cm)    {\scriptsize $\mathcal{P}_{c_1}$};
      \node at ( 0:1cm)    {\scriptsize $\mathcal{P}_{c_2}$};
      \node at ( 45:1.25cm)    {\scriptsize $\mathcal{P}_U$};
      \node at ( 45:1.7cm)    {\scriptsize $\mathcal{PB}_U$};
      \node at ( 40:2.5cm)    {\scriptsize $\mathcal{O}$};
\end{tikzpicture}
   \vspace{-1mm}\caption{\small Venn diagram depicting  $\mathcal{P}_{c_1}$, $\mathcal{P}_{c_2}$, $\mathcal{P}_U$, $\mathcal{PB}_U$ and $\mathcal{O}$}
   \label{fig:venn_sw}
\end{figure}
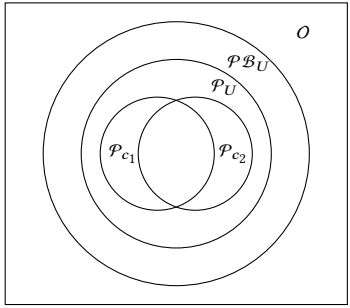

\begin{example}
The executions of \algname{BaselineSW} and \algname{FilterThenVerifySW} on Table~\ref{tab:product_window} and Table~\ref{tab:customer} are briefly explained here. Consider $W$, $in$ and $out$ as 6, 7 and 1, respectively.

While the sliding window is at $[1$, $6]$, $\mathcal{P}_{c_1} = \{o_1, o_3\}$, $\mathcal{P}_{c_2} = \{o_3, o_4\}$, $\mathcal{PB}_{c_1} = \{o_1, o_3, o_4, o_6\}$ and $\mathcal{PB}_{c_2} = \{o_3$, $o_4$, $o_5$, $o_6\}$. Upon the expiration of $o_1 = \langle 17, \emph{Lenovo}, \emph{dual} \rangle$, the window is at $(1$, $6]$. Now \algname{BaselineSW} checks whether $o_1$ belongs to $\mathcal{P}_{c_1}$ and $\mathcal{P}_{c_2}$. Since $o_1$ belongs to $\mathcal{P}_{c_1}$, $\mathcal{P}_{c_1}$ is mended to $\{o_3\}$. Upon the arrival of $o_7 = \langle 14, \emph{Apple}, \emph{dual} \rangle$, the window includes objects correspond to $(1$, $7]$. At this point \algname{BaselineSW} starts checking whether $o_7$ is qualified as an element of $\mathcal{P}_{c_1}$ and $\mathcal{P}_{c_2}$, sequentially. In both $\mathcal{P}_{c_1}$ and $\mathcal{P}_{c_2}$, $o_7$ takes the place of $o_3$ (Line~\ref{line:BaselineSW-updateParetoFrontierSW}). After that, $o_7$ is stored to $\mathcal{PB}_{c_1}$ and $\mathcal{PB}_{c_2}$. Furthermore, for both $\mathcal{PB}_{c_1}$ and $\mathcal{PB}_{c_2}$, \algname{BaselineSW} finds out the objects dominated by $o_7$ and discards them, i.e., $\{ o_3, o_6\}$ and $\{ o_3, o_5, o_6\}$, respectively. As these dominated objects arrives before $o_7$, they could never acquire Pareto-optimality. Now $\mathcal{PB}_{c_1} = \{o_4, o_7\}$ and $\mathcal{PB}_{c_2} = \{o_4$, $o_7\}$ (Line~\ref{line:BaselineSW-refreshParetoBufferSW}) (Def.~\ref{def:pareto_frontier_buffer}). We get that $\mathcal{C}_{o_7} = \{c_1, c_2\}$. The content of Pareto frontiers and Pareto buffers at 3 phases of window of \algname{BaselineSW} is shown in Table~\ref{tab:window_baseline}.

In case of \algname{FilterThenVerifySW}, while the sliding window is at $[1$, $6]$, $\mathcal{P}_{c_1} = \{o_1, o_3\}$, $\mathcal{P}_{c_2} = \{o_3, o_4\}$, $\mathcal{P}_{U} = \{o_1, o_3$, $o_4\}$ and $\mathcal{PB}_{U} = \{o_1, o_3, o_4, o_5, o_6\}$. Upon the expiration of $o_1$, the algorithm checks whether $o_1$ belongs to $\mathcal{P}_{U}$. Therefore, $\mathcal{P}_{U}$ becomes $\{o_3, o_4\}$ while the window is at $(1$, $6]$. Upon the arrival of $o_7$, the window includes objects correspond to $(1$, $7]$. Now \algname{FilterThenVerifySW} starts checking whether $o_7$ can occupy $\mathcal{P}_{U}$. With respect to $U$, $o_7$ dominates $o_3$, i.e., $o_7$ replaces of $o_3$ in $\mathcal{P}_{U}$ (Line~\ref{line:FilterThenVerifySW-updateParetoFrontierUSW}) as well as in both $\mathcal{P}_{c_1}$ and $\mathcal{P}_{c_2}$ (Line~\ref{line:FilterThenVerifySW-updateParetoFrontierSW}). After that, $o_7$ is stored in $\mathcal{PB}_{U}$. Moreover, $o_7$ dominates $o_3$, $o_5$ and $o_6$ in $\mathcal{PB}_{U}$. Since each of these dominated objects in $\mathcal{PB}_{U}$ arrives before $o_7$, they could never been qualified in Pareto-optimality. Therefore, \algname{FilterThenVerifySW} discards them from $\mathcal{PB}_{U}$, i.e., $\mathcal{PB}_{U} = \{o_4, o_7\}$ (Line~\ref{line:FilterThenVerifySW-refreshParetoBufferSW}) (Def.~\ref{def:pareto_frontier_buffer}). Finally we get $\mathcal{C}_{o_7} = \{c_1, c_2\}$. The content of Pareto frontiers and Pareto buffers at 3 phases of window of \algname{FilterThenVerifySW} are shown in Table~\ref{tab:window_ca}. The Venn diagram in Fig.\ref{fig:venn_sw} depicts $\mathcal{P}_{c_1}$, $\mathcal{P}_{c_2}$, $\mathcal{P}_U$ and $\mathcal{PB}_U$. Note that, while \algname{BaselineSW} needs to maintain individual Pareto frontier buffer per user ($\mathcal{PB}_{c_1}$ and $\mathcal{PB}_{c_2}$), a shared Pareto frontier buffer per cluster ($\mathcal{PB}_{U}$) suffices for \algname{FilterThenVerifySW}. In conclusion, along with Pareto frontier maintenance, \algname{FilterThenVerifySW} prunes  comparisons in terms of Pareto buffer maintenance.\closedef
\end{example} 
\section{Experiments}\label{sec:prefquery-exp}
\begin{figure*}[t]
\noindent
\begin{minipage}{0.48\textwidth}
\centering
\begin{subfigure}[b]{0.49\linewidth}
\centering
   \epsfig{file=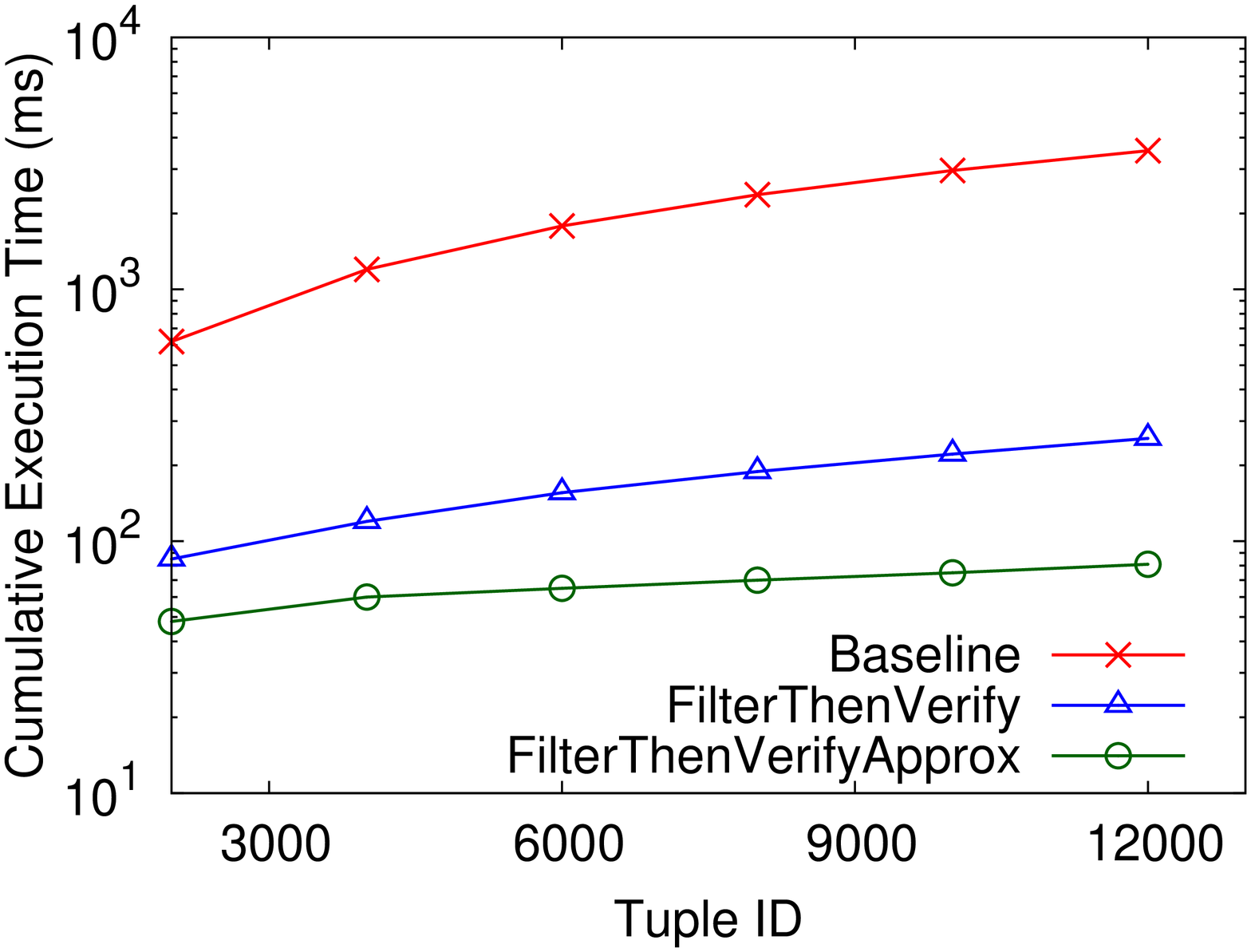, angle=-90, width=42mm,clip=}
   \caption{Execution time}
   \label{fig:time_id}
\end{subfigure}
\begin{subfigure}[b]{0.49\linewidth}
\centering
   \epsfig{file=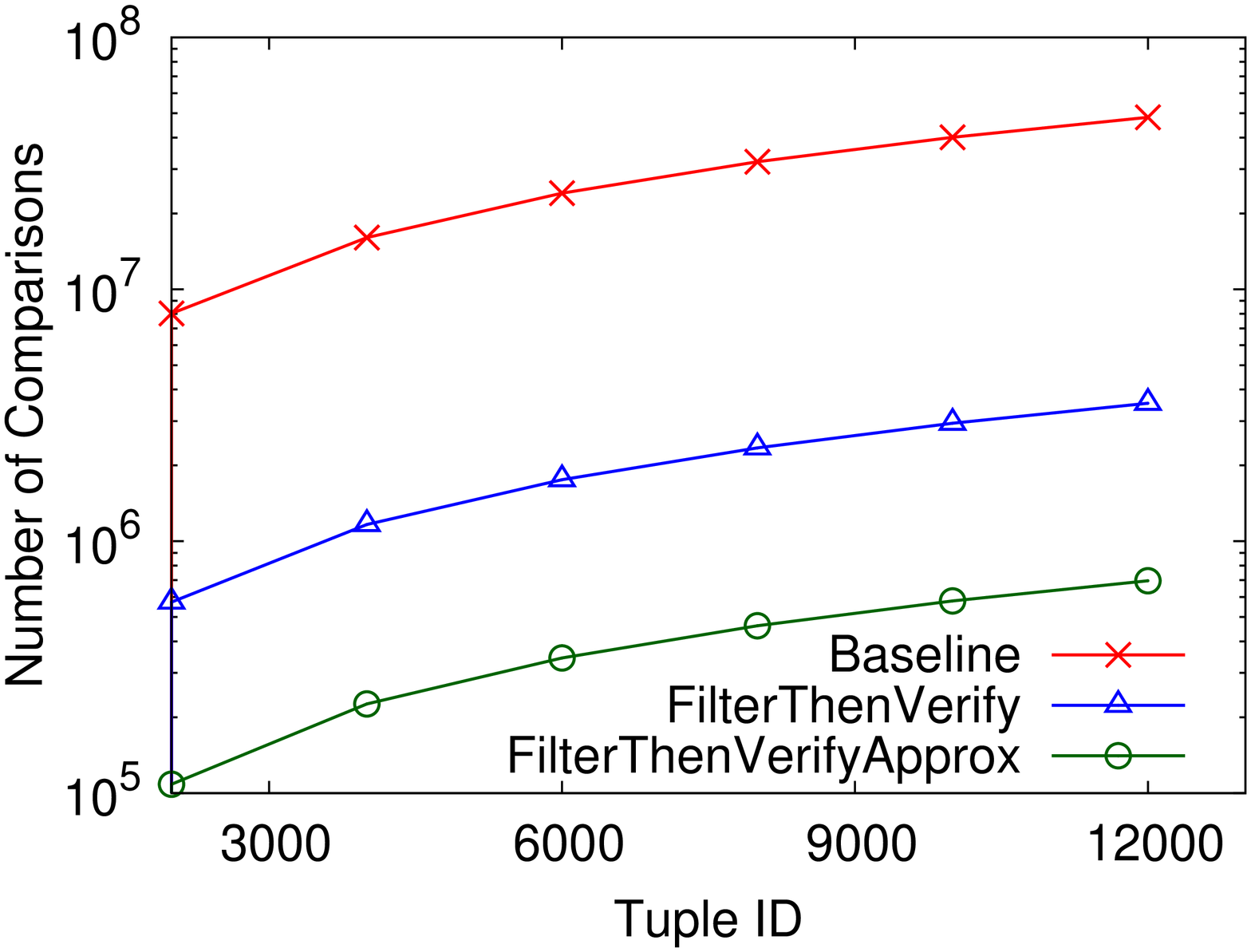, angle=-90, width=42mm,clip=}
   \caption{Object comparisons}
   \label{fig:comparison_id}
\end{subfigure}
\caption{\small Comparison of \algname{Baseline}, \solname{FilterThenVerify} and \solname{FilterThenVerifyApprox} on the movie dataset. Varying $|\mathcal{O}|$, $h$ = $0.55$, $d$ = $4$.}
\end{minipage}
\hspace{1mm}
\noindent
\begin{minipage}{0.48\textwidth}
\centering
\begin{subfigure}[b]{0.49\linewidth}
\centering
   \epsfig{file=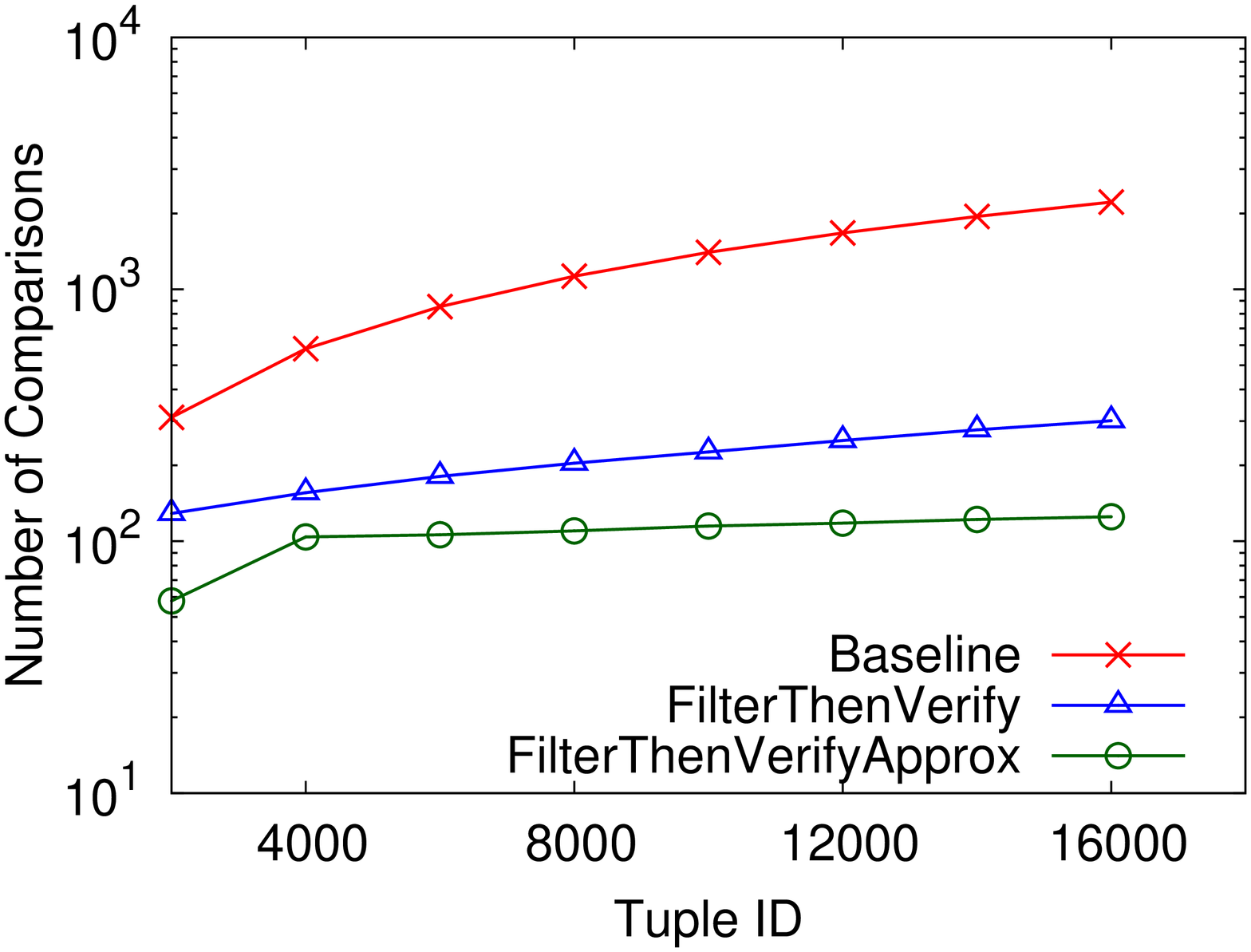, angle=-90, width=42mm,clip=}
   \caption{Execution time}
   \label{fig:time_id_p}
\end{subfigure}
\begin{subfigure}[b]{0.49\linewidth}
\centering
   \epsfig{file=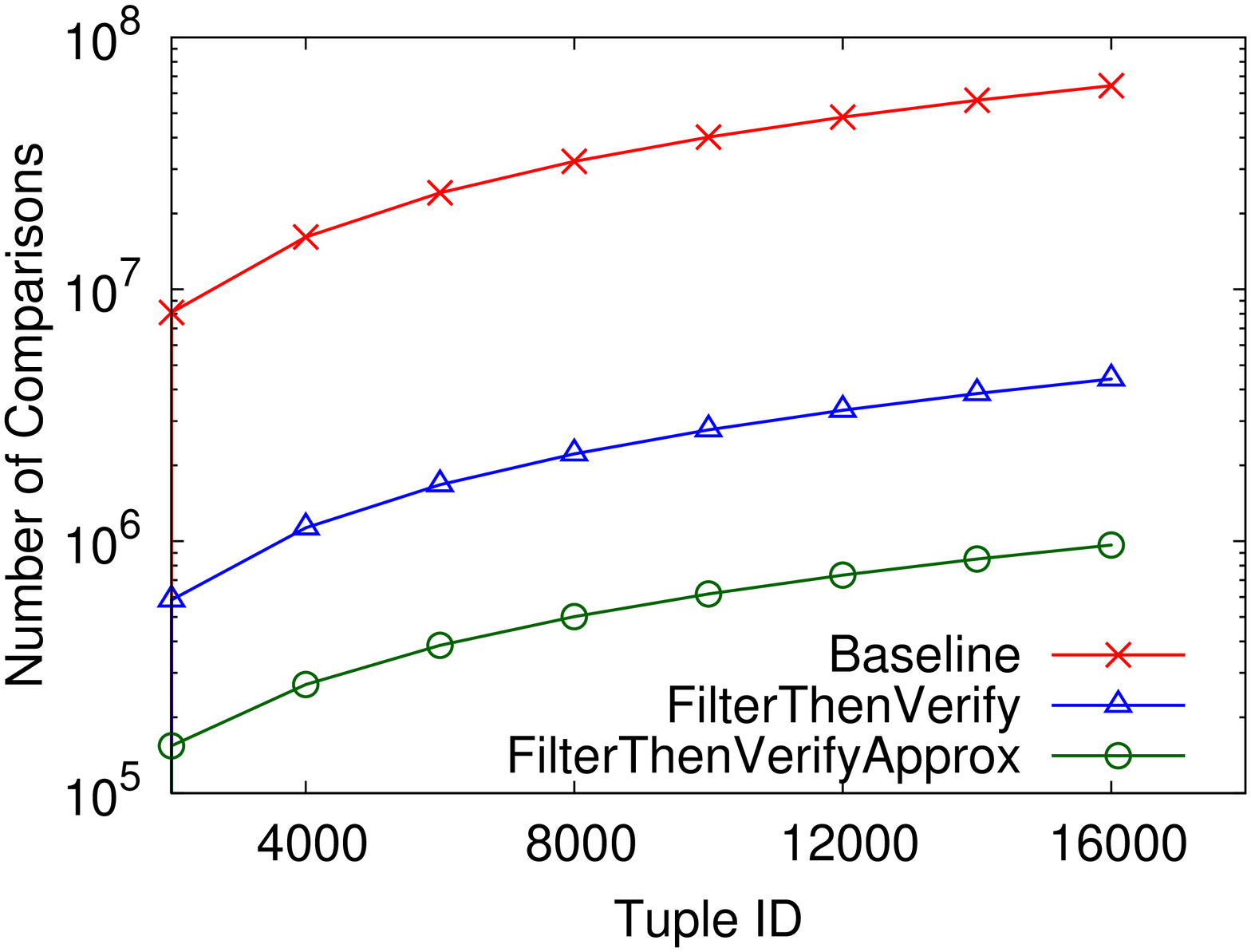, angle=-90, width=42mm,clip=}
   \caption{Object comparisons}
   \label{fig:comparison_id_p}
\end{subfigure}
\caption{\small Comparison of \algname{Baseline}, \solname{FilterThenVerify} and \solname{FilterThenVerifyApprox} on the publication dataset. Varying $|\mathcal{O}|$, $h$ = $0.55$, $d$ = $4$.}
\end{minipage}
\end{figure*}

\begin{figure*}[t]
\noindent
\begin{minipage}{0.48\textwidth}
\centering
\begin{subfigure}[b]{0.49\linewidth}
\centering
   \epsfig{file=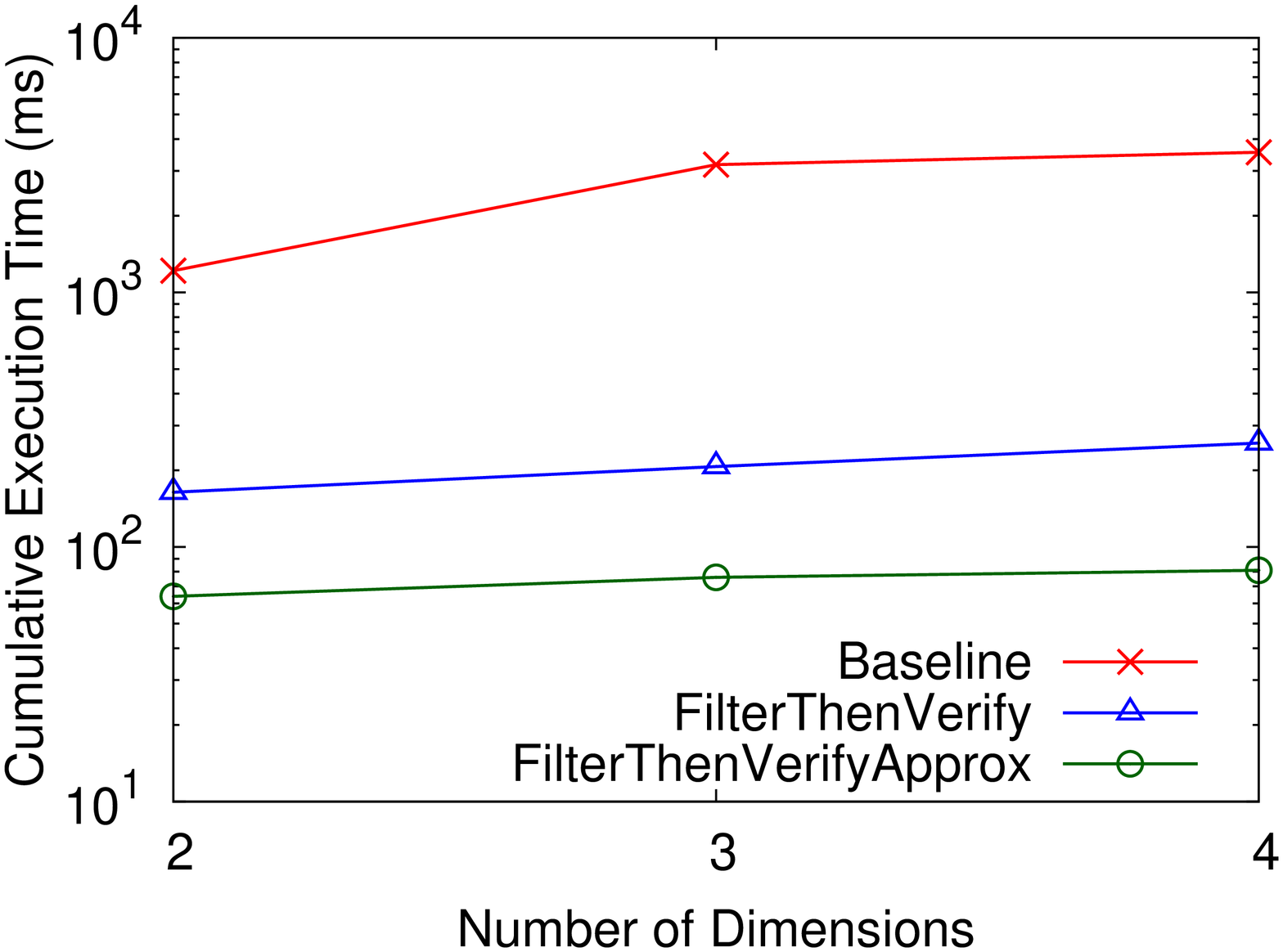, angle=-90, width=42mm,clip=}
   \caption{Execution time}
   \label{fig:time_d}
\end{subfigure}
\begin{subfigure}[b]{0.49\linewidth}
\centering
   \epsfig{file=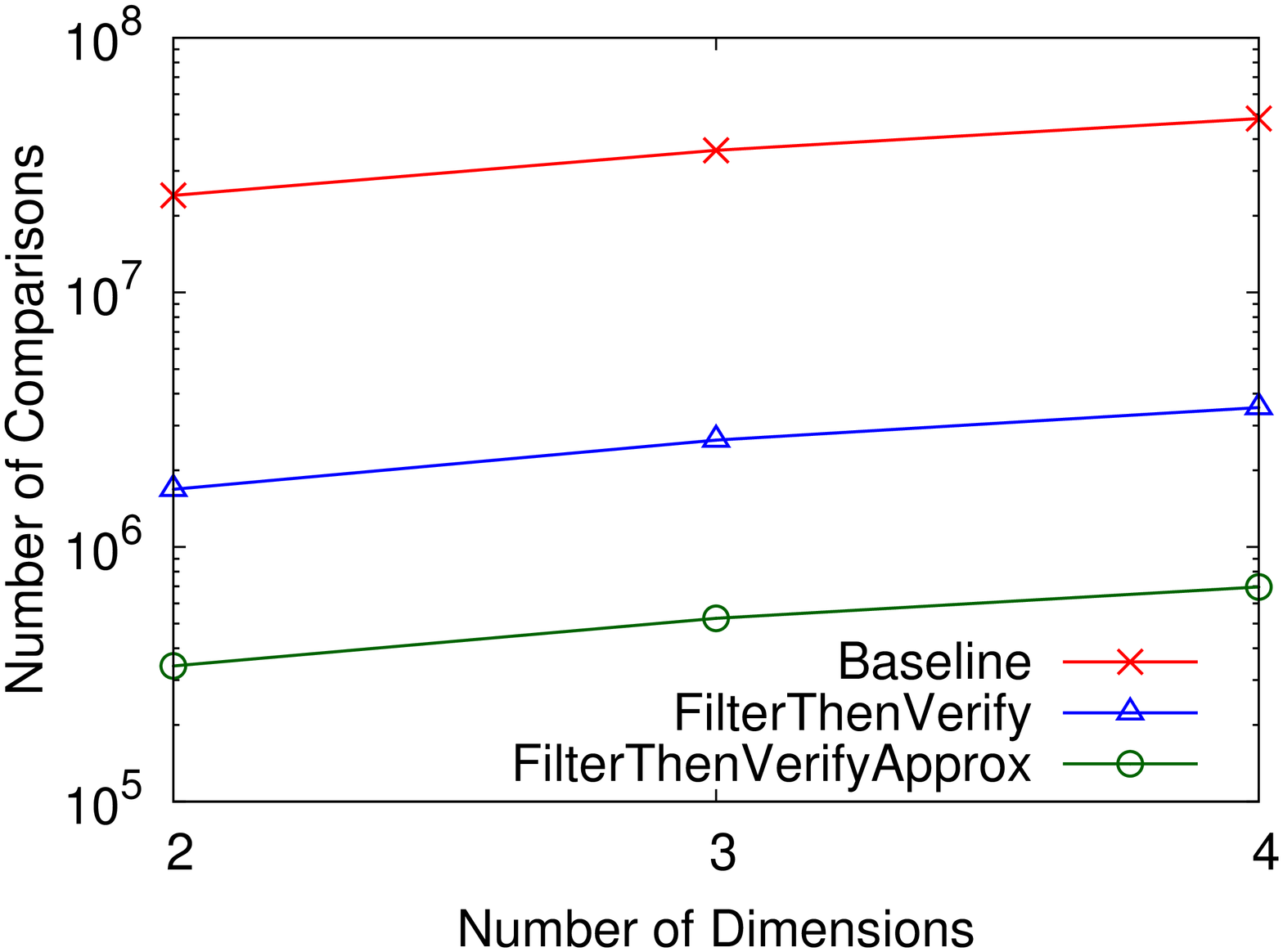, angle=-90, width=42mm,clip=}
   \caption{Object comparisons}
   \label{fig:comparison_d}
\end{subfigure}
\caption{\small Comparison of \algname{Baseline}, \solname{FilterThenVerify} and \solname{FilterThenVerifyApprox} on the movie dataset. Varying $d$, $|\mathcal{O}|$ = $12,749$, $h$ = $0.55$.}
\label{fig:d}
\end{minipage}
\hspace{1mm}
\noindent
\begin{minipage}{0.48\textwidth}
\centering
\begin{subfigure}[b]{0.49\linewidth}
\centering
   \epsfig{file=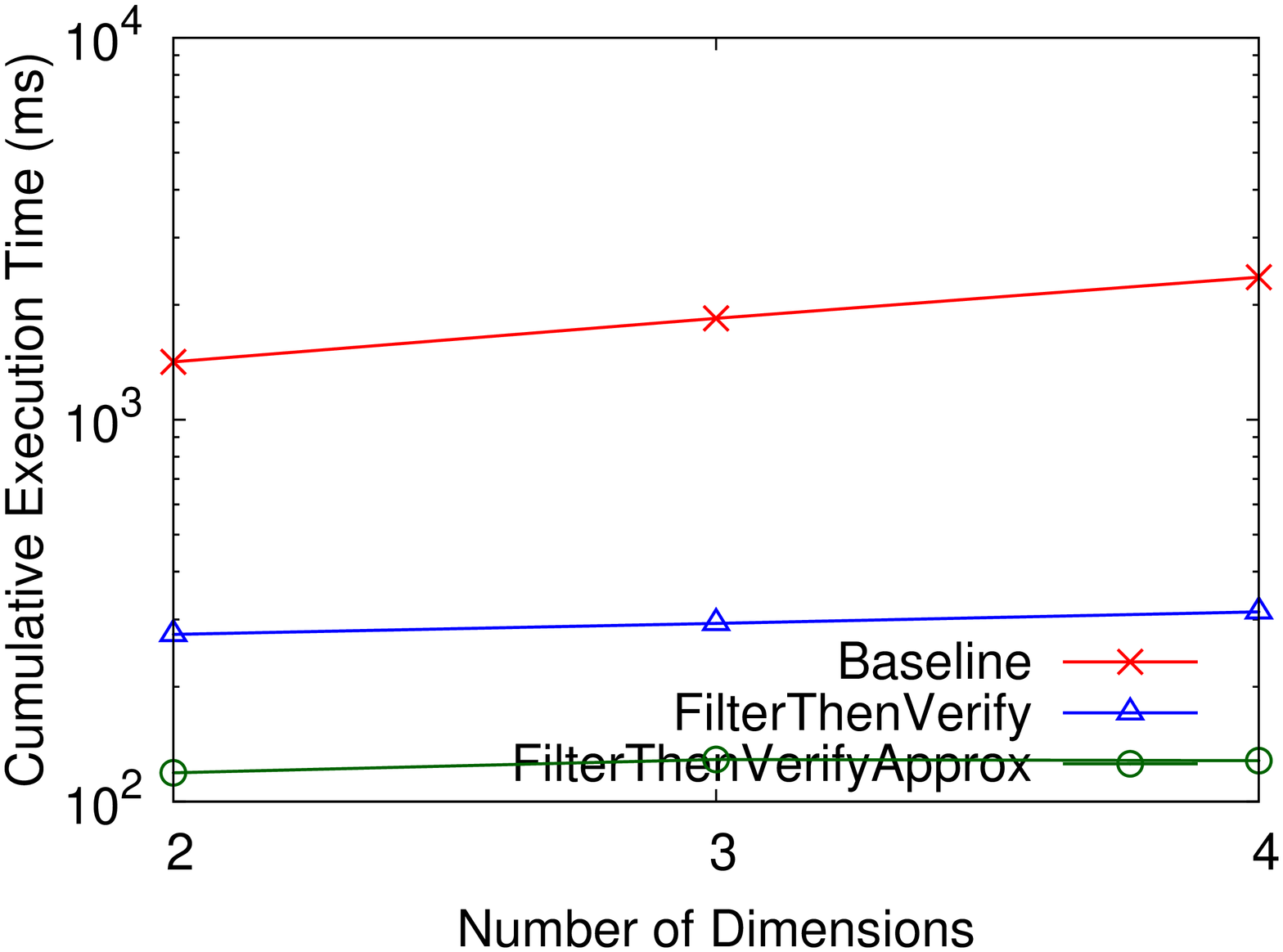, angle=-90, width=42mm,clip=}
   \caption{Execution time}
   \label{fig:time_d_p}
\end{subfigure}
\begin{subfigure}[b]{0.49\linewidth}
\centering
   \epsfig{file=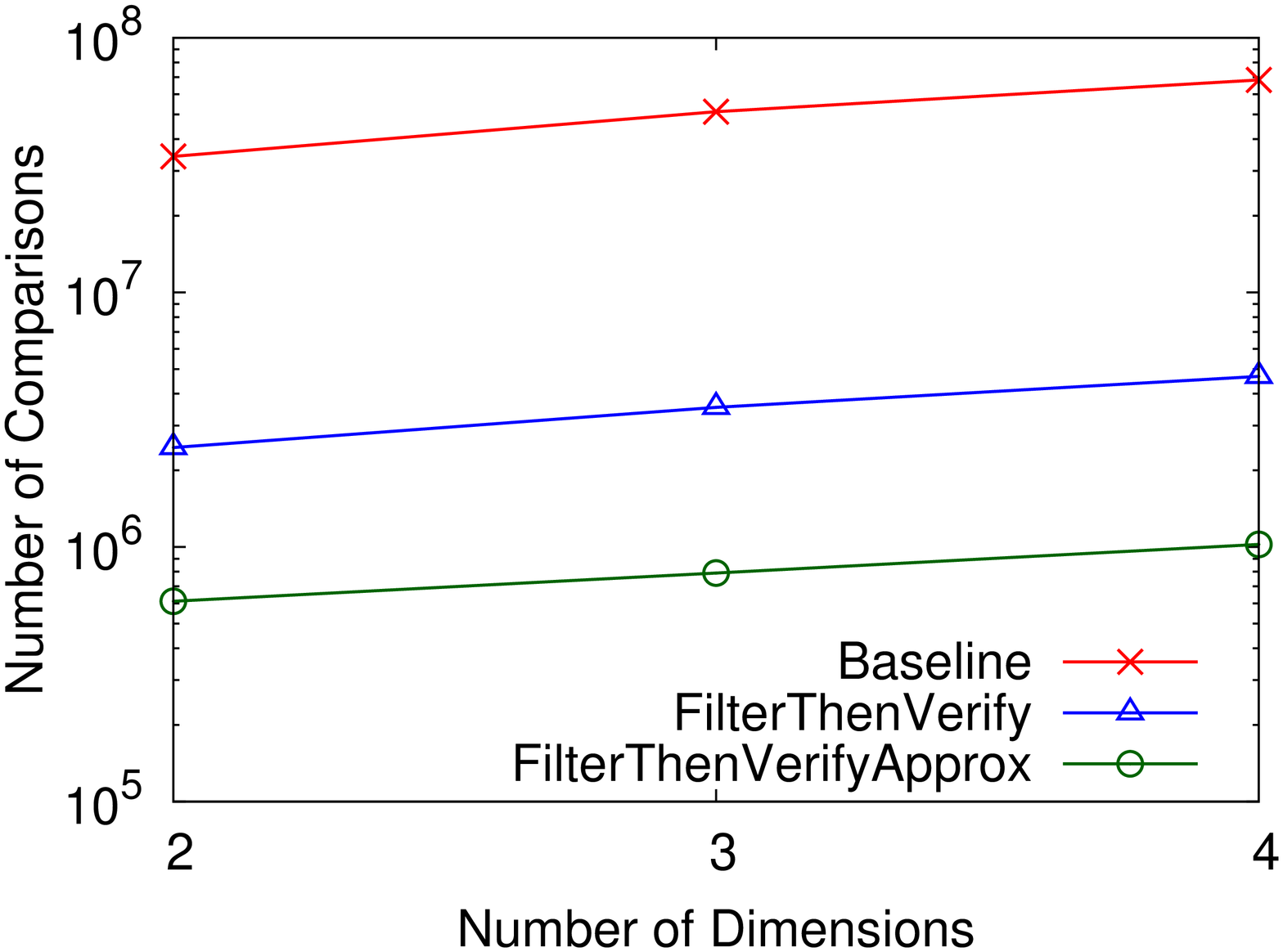, angle=-90, width=42mm,clip=}
   \caption{Object comparisons}
   \label{fig:comparison_d_p}
\end{subfigure}
\caption{\small Comparison of \algname{Baseline}, \solname{FilterThenVerify} and \solname{FilterThenVerifyApprox} on the publication dataset. Varying $d$, $|\mathcal{O}|$ = $17,598$, $h$ = $0.55$.}
\label{fig:d_p}
\end{minipage}
\end{figure*}

\begin{table*}[t]
\centering
\scriptsize
\begin{tabular}{|l|l|l|l|l|l|l|l|l|l|l|l|l|l|}
\hline
\multirow{2}{*}{Dataset} & \multirow{2}{*}{$|\mathcal{O}|$} & \multicolumn{3}{c|}{$h = 0.70$} & \multicolumn{3}{c|}{$h = 0.65$} & \multicolumn{3}{c|}{$h = 0.60$} & \multicolumn{3}{c|}{$h = 0.55$} \\ \cline{3-14}
                &  &  Precision & Recall & F-measure & Precision & Recall & F-measure & Precision & Recall & F-measure & Precision & Recall & F-measure \\ \hline\hline
\multirow{1}{*}{Movie} & \multirow{1}{*}{$12,749$} & $100$	& $95.43$	& $97.67$	& $100$ & $93.93$ & $96.87$	& $99.99$ & $93.28$	& $96.52$	& $99.99$ & $90.46$ &	$94.99$\\ \hline
\multirow{1}{*}{Publication} & \multirow{1}{*}{$17,598$} & $100$ & $96.59$ & $98.27$ & $100$ & $95.85$	& $97.88$ &	$100$ & $95.54$ &	$97.72$ & $100$ & $95.13$ & $97.51$\\ \hline
\end{tabular}
\caption{The precision, recall and F-measure (in percentage) of \solname{FilterThenVerifyApprox}. Varying $h$, $d$=$4$.}\vspace{-4mm}
\label{tab:efficacy}
\end{table*}

\subsection{Experiment Setup}
The algorithms were implemented in Java. The maximal heap size of Java Virtual Machine (JVM) was set to 16 GB. The experiments were conducted on a computer with $2.0$ GHz Quad Core 2 Duo Xeon CPU running Ubontu 8.10.

\textbf{Datasets}\hspace{2mm} Currently there exists no publicly available dataset that captures real users' preferences in partial orders. We thus simulated such partial orders using two real datasets of users' preferences.

\emph{Movie Dataset}\hspace{2mm} We joined the Netflix dataset ({\small \url{netflixprize.com}}) with data from IMDB ({\small \url{imdb.com}}).  The Netflix dataset contains the ratings (ranging from $0$ to $5$) given by users to movies. From IMDB we fetched the movies' attribute values, including actors, directors, genres, and writers. In this way, we found the attributes of $12,749$ Netflix movies. The goal is to, for each particular movie, identify users who may like it according to their preferences on those attributes. The mapping from our problem formulation to this dataset is the following: (i) $\mathcal{O}$ is the set of $12,749$ movies. (ii) $\mathcal{C}$ is the set of users. It includes the $1,000$ most active users based on how many movies they have rated. The excluded less active users is the subject of the well-known \emph{cold-start} problem in recommendation systems and is outside the of scope of this work. (iii) $\mathcal{D} = \{
\smallattrname{actor}$, $\smallattrname{director}$, $\smallattrname{genre}$, $\smallattrname{writer}\}$. (iv) Given the lack of user preference data, for each attribute, the partial order corresponding to a user's preferences is simulated as follows. For two attribute values, the user's preference is based on the \emph{average rating} and the \emph{count} of movies satisfying these attribute values.
More specifically, consider a user $c$ who has rated $m$ movies featuring actor $a$. Suppose the ratings of these movies are $r_1, r_2$, $\ldots$, $r_m$. Given $c$ and $a$, the average rating is $R_a = {\sum_i {r_i} \over m}$ and the count is $M_a = m$. Consider another actor $b$. If $(R_a > R_b$ $\wedge$ $M_a \ge M_b)$ $\vee$ $(R_a \ge R_b$ $\wedge$ $M_a > M_b)$, then $(a, b) \in \succ_c^{\attrname{actor}}$. Intuitively, if user $c$ watches more movies featuring $a$ than $b$ and gives them higher ratings, our simulation assumes the user prefers $a$ to $b$.

\emph{Publication Dataset}\hspace{2mm} We collected from the ACM Digital Library ({\small \url{dl.acm.org}}) $17,598$ publications and their attributes, including affiliations, authors, conferences and topic keywords. The users are the authors themselves. The goal is to notify them about newly published articles. The recommendations are based on the users' preference relations on the attributes. The mapping from our problem formulation to this dataset is the following: (i) $\mathcal{O}$ is the set of papers. (ii) $\mathcal{C}$ is the set of authors. It includes the $1,000$ most prolific authors based on how many publications they have, similar to the $1,000$ most active users in the movie dataset. (iii)  $\mathcal{D} = \{$\smallattrname{affiliation}$, \smallattrname{author}$, $\smallattrname{conference}$, $\smallattrname{keyword}\}$. The domain of attribute \smallattrname{author} is the same $1,000$ authors in $\mathcal{C}$. (iv) Given a user, the partial order on each attribute is simulated based on their preferences on the attribute values. The preference between two values on $\smallattrname{affiliation}$ (and similarly \smallattrname{author}) is based on the \emph{number of collaborations} between the user and the affiliation/author and the \emph{number of citations}. For \smallattrname{conference} and \smallattrname{keyword}, the preference between two values is based on \emph{number of publications} and \emph{number of citations}.
More specifically, consider a user $c$ and an affiliation (or similarly another author) $a$. Suppose $c$ has $p_a$ collaborations with $a$ and has cited articles from $a$ $q_a$ times. If $(p_a > p_b$ $\wedge$ $q_a \ge q_b)$ $\vee$ $(p_a \ge p_b$ $\wedge$ $q_a > q_b)$, then $(a, b) \in \succ_c^{\attrname{affiliation}}$ (or $(a, b) \in \succ_c^{\attrname{author}}$).
With regard to a conference (keyword) $x$, suppose $c$ has $r_x$ publications associated with $x$ and has cited publications associated with $x$ $s_x$ times. If $(r_x > r_y$ $\wedge$ $s_x \ge s_y)$ $\vee$ $(r_x \ge r_y$ $\wedge$ $s_x > s_y)$, then $(x, y) \in \succ_c^{\attrname{conference}}$ (or $(x, y) \in \succ_c^{\attrname{keyword}}$).

\subsection{\textsf{Baseline}, \textsf{FilterThenVerify}, and \textsf{FilterThenVerifyApprox}}\label{sec:prefquery-results}
We conducted experiments to compare the performance of \algname{Baseline}, \solname{FilterThenVerify} and \solname{FilterThenVerifyApprox}. 
For \solname{FilterThenVerify} (resp. \solname{FilterThenVerifyApprox}), users are clustered by the conventional hierarchical agglomerative clustering algorithm~\cite{dmbook} using the similarity functions in Sec.~\ref{sec:cluster} (resp. Sec.~\ref{sec:approx-similarity}) and, for each cluster, it extracts the common preference relation (resp. approximate common preference relation). %%%\algname{FilterThenVerify} clusters users based on their preference relations and reuses the computations for each cluster. 
The experiments use three parameters which are number of objects ($|\mathcal{O}|$), number of attributes ($d$), and branch cut ($h$). 
In hierarchical clustering, the \emph{branch cut} $h$ is a threshold that controls the number of clusters by governing the minimum pairwise similarity that two clusters must satisfy in order to be merged into one cluster. The sequential order of merging clusters is depicted as a tree called \emph{dendrogram}. 
The branch cut thus controls where to cut the dendrogram. In Example~\ref{example:sim_wj}, the set of clusters are $\{\{c_1$, $c_2$, $c_5$, $c_6\}$, $\{c_3$, $c_4\}\}$ for $h \in (0, {3 \over 11}]$. This is because $sim(U_4$,$U_2)$=$0$ where $U_2$=$\{c_3$,$c_4\}$ and $U_4$ is the cluster composed of $c_1$, $c_2$, $c_5$, and $c_6$. 
%%%In other words, given a cluster $U$ and a branch cut $h$, the similarity between two users $c_1$, $c_2 \in U$ is at least $h$, i.e., $sim(U_1$,$U_2) \ge h$. 
%%%For instance, at $h = 1$, from Table~\ref{tab:customer4} we get each user as an individual cluster---$\{\{c_1\}, \{c_2\}, \{c_3\}, \{c_4\}, \{c_5\}, \{c_6\}\}$. 

Fig.\ref{fig:time_id} shows, for each of the three methods on the movie dataset, how its cumulative execution time (by milliseconds, in logarithmic scale) increases while the objects (i.e., movies) are sequentially processed. Fig.\ref{fig:time_id_p} depicts similar behaviours of these methods on the publication dataset.
Fig.\ref{fig:comparison_id} and Fig.\ref{fig:comparison_id_p}, for the two datasets separately, further present the amount of work done by these methods, in terms of number of pairwise object comparisons (in logarithmic scale) for maintaining Pareto frontiers.
The figures show that \solname{FilterThenVerify} and \solname{FilterThenVerifyApprox} beat \algname{Baseline} by $1$ to $2$ orders of magnitude. 
The reason is as follows. With regard to a user $c$, \algname{Baseline} considers all objects as candidate Pareto-optimal objects and compares all pairs. 
On the contrary, \algname{FilterThenVerify} eliminates an object $o$ if the corresponding common preference tuples disqualify $o$.  
\solname{FilterThenVerifyApprox} incurs even less comparisons by benefiting from shared computations for clusters of users.

Fig.\ref{fig:time_d} (Fig.\ref{fig:time_d_p}) shows that the execution time of all these methods increased super-linearly by number of attributes ($d$). Fig.\ref{fig:comparison_d} (Fig.\ref{fig:comparison_d_p}) further reveals that the number of object comparisons also increases similarly. 
This is not surprising because more attributes result in larger Pareto frontiers, which makes it necessary for objects to be compared with more existing Pareto-optimal objects.

Table~\ref{tab:efficacy} reports the precision, recall and F-measure of \solname{FilterThenVerifyApprox} on varying $h$. 
We can observe that, when $h$ got smaller, the recall slowly decreased. This is expected because smaller $h$ results in larger clusters and potentially more approximate common preference tuples for each cluster. Those approximate common preference tuples cause false negatives---the domination and elimination of objects that are instead in the Pareto frontier under the true common preference tuples, which are a subset of the approximate common preference tuples. 
What can be more surprising is the almost perfect precision under the various $h$ values in Table~\ref{tab:efficacy}, i.e., almost no false positives were introduced into the results. 
For a user $c$, an object $o$ becomes a false positive if every single Pareto optimal object that dominates $o$ becomes a false negative. 
As long as one of its dominating objects is not mistakenly filtered out, $o$ will not be mistakenly introduced into the Pareto frontier. 
Therefore, an object is much less likely to become a false positive than a false negative. 
Overall, under the $h$ values in Table~\ref{tab:efficacy}, both precision and recall remain high. 
This may suggest that the thresholds $\theta_1$ and $\theta_2$ (Sec.~\ref{def:approx_common_preference_tuple}) effectively ensure that the approximate common preference relation only includes frequent preference tuples and does not overgrow in size. 

\subsection{\textsf{BaselineSW}, \textsf{FilterThenVerifySW}, and \textsf{FilterThenVerifyApproxSW}}\label{sec:sw_results}
\begin{figure*}[t]
\noindent
\begin{minipage}{0.48\textwidth}
\centering
\begin{subfigure}[b]{0.49\linewidth}
\centering
   \epsfig{file=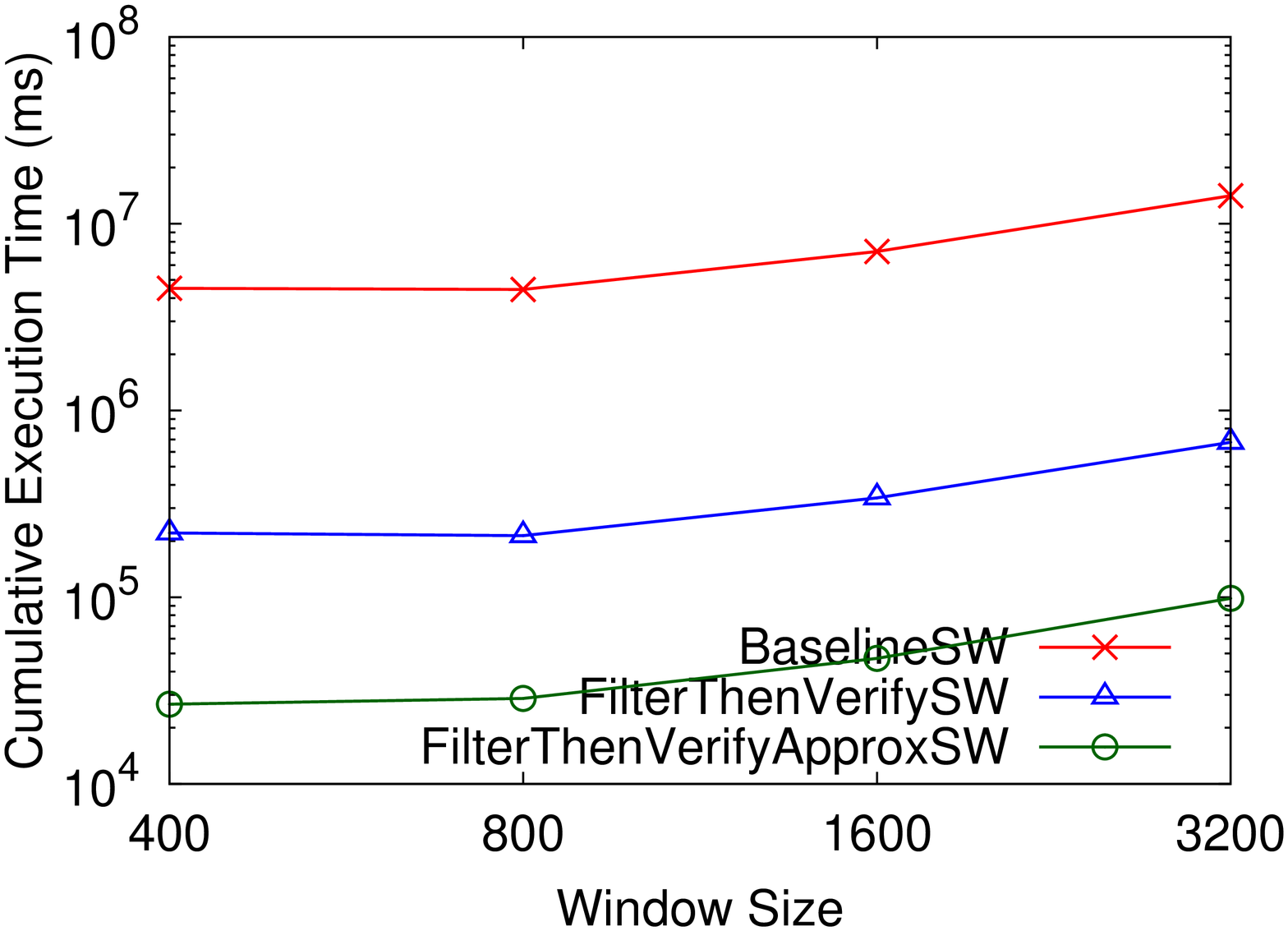, angle=-90, width=42mm,clip=}
   \caption{Execution time}
   \label{fig:time_window_m}
\end{subfigure}
\begin{subfigure}[b]{0.49\linewidth}
\centering
   \epsfig{file=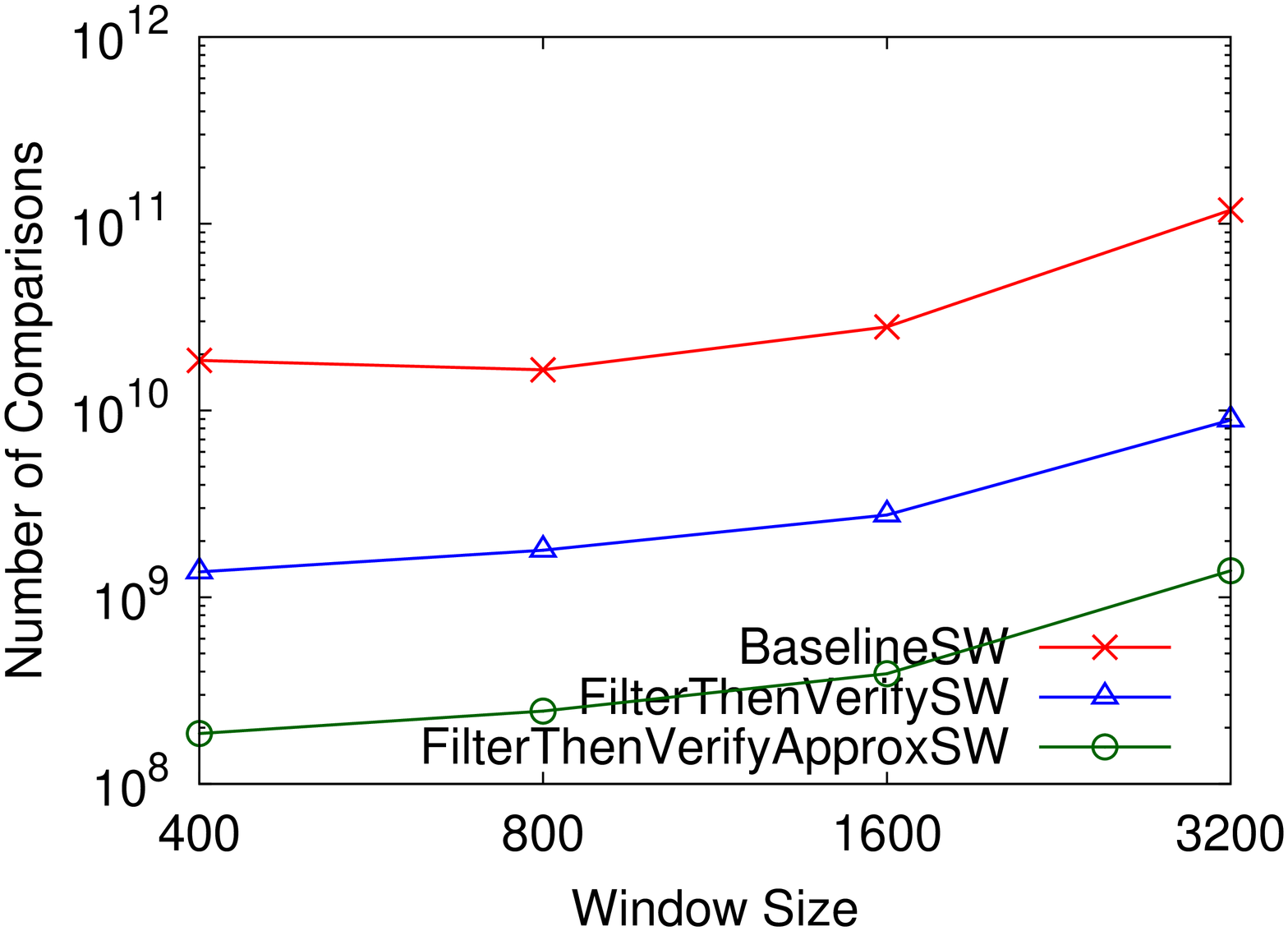, angle=-90, width=42mm,clip=}
   \caption{Object comparisons}
   \label{fig:comparison_window_m}
\end{subfigure}
\caption{\small Effect of window size on the movie dataset. Varying $W$, $|\mathcal{O}|$ = $1$M, $h$ = $0.55$, $d$ = $4$.}
\label{fig:sw_m}
\end{minipage}
\hspace{1mm}
\noindent
\begin{minipage}{0.48\textwidth}
\centering
\begin{subfigure}[b]{0.49\linewidth}
\centering
   \epsfig{file=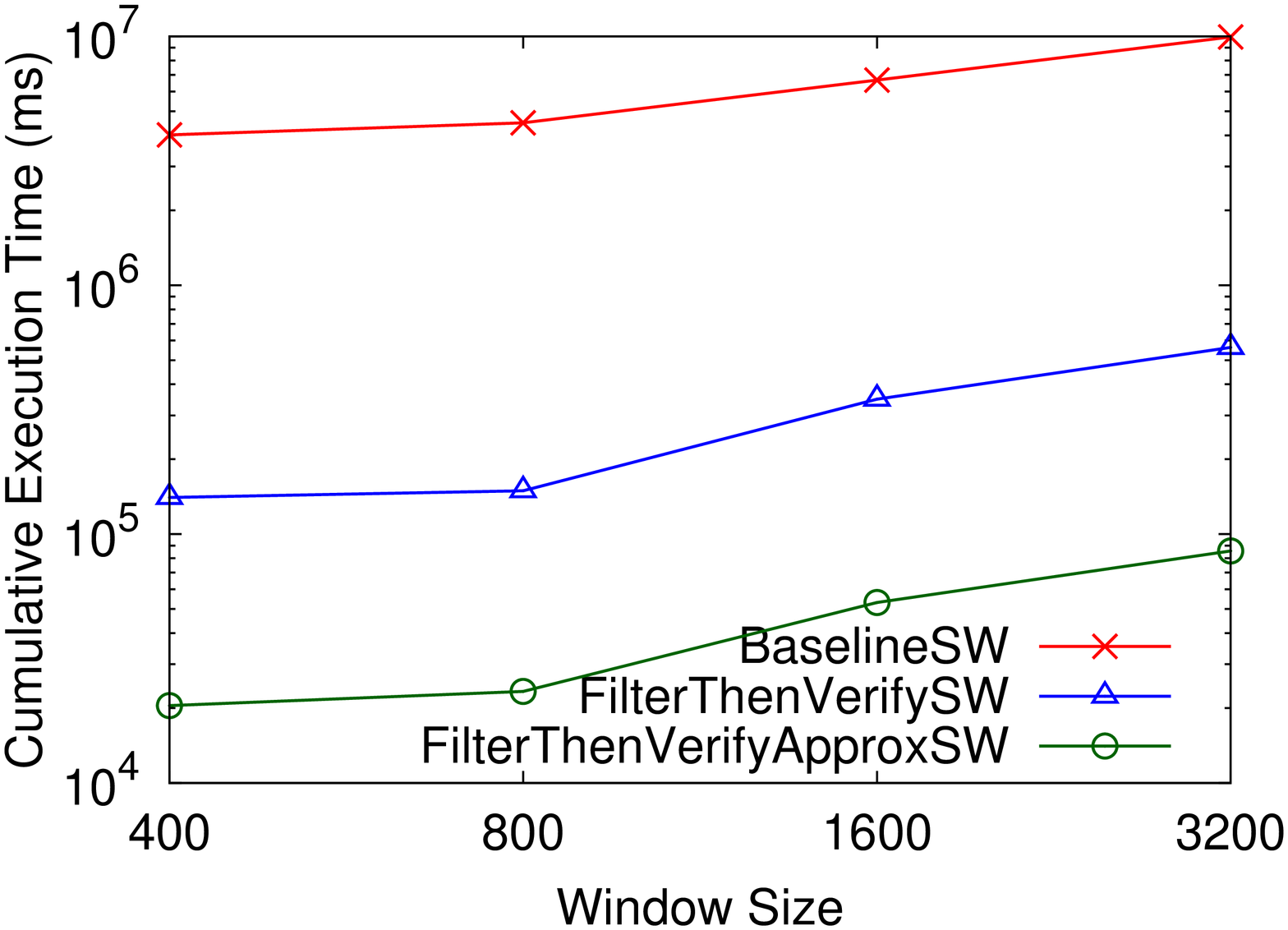, angle=-90, width=42mm,clip=}
   \caption{Execution time}
   \label{fig:time_window_p}
\end{subfigure}
\begin{subfigure}[b]{0.49\linewidth}
\centering
   \epsfig{file=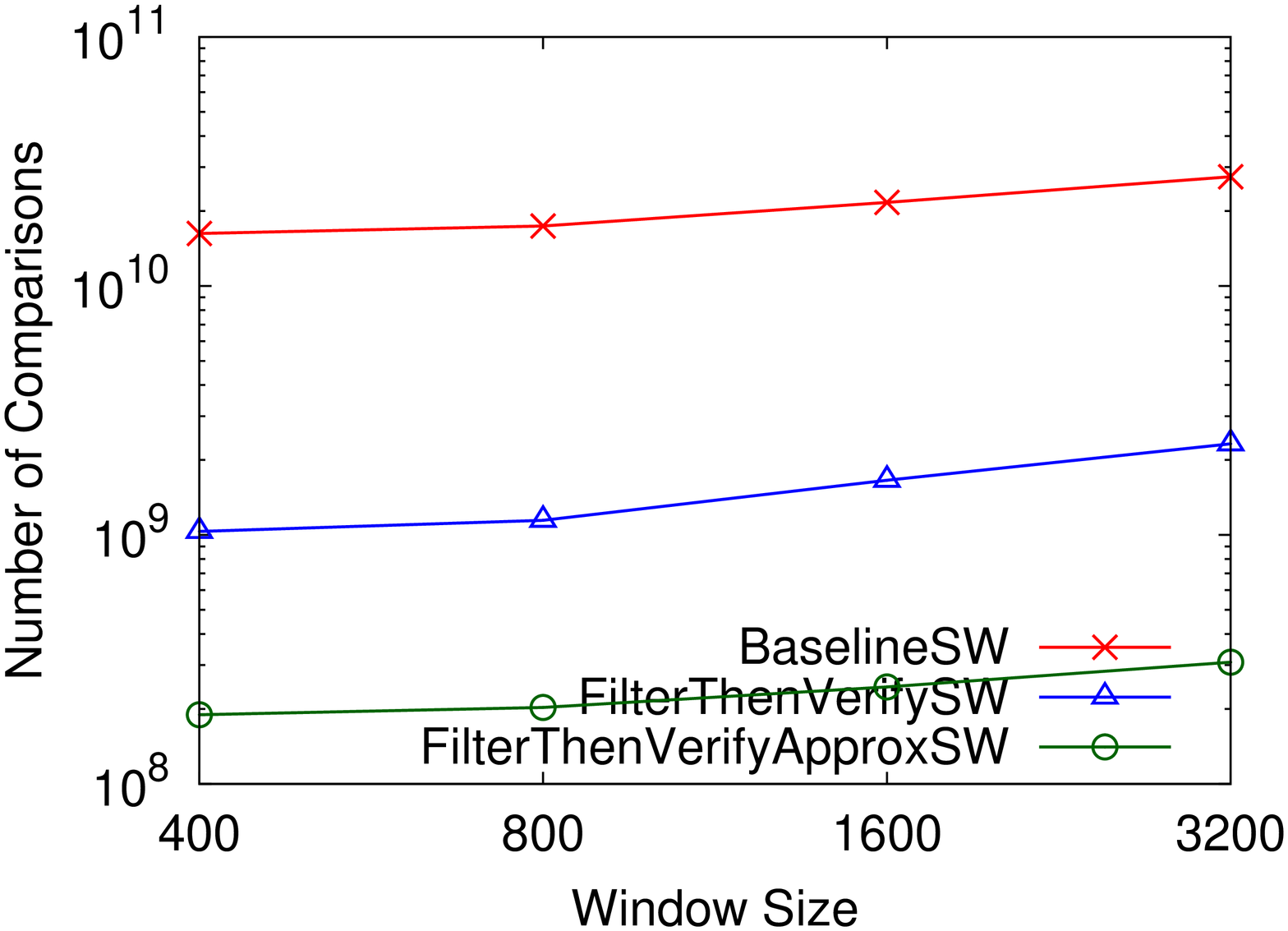, angle=-90, width=42mm,clip=}
   \caption{Object comparisons}
   \label{fig:comparison_window_p}
\end{subfigure}
\caption{\small Effect of window size on the publication dataset. Varying $W$, $|\mathcal{O}|$ = $1$M, $h$ = $0.55$, $d$ = $4$.}
\label{fig:sw_p}
\end{minipage}
\end{figure*}

\begin{figure*}[t]
\noindent
\begin{minipage}{0.48\textwidth}
\centering
\begin{subfigure}[b]{0.49\linewidth}
\centering
   \epsfig{file=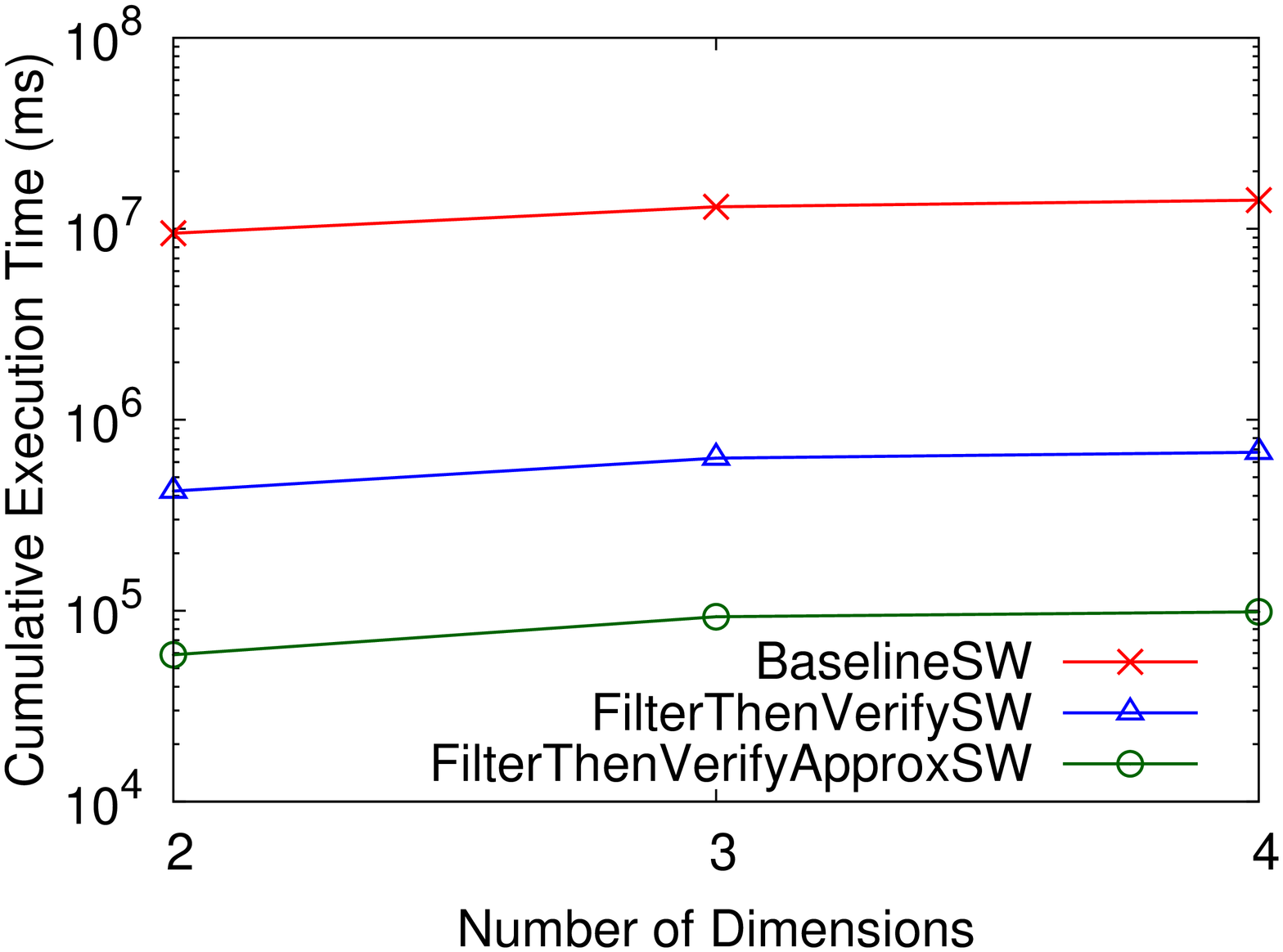, angle=-90, width=42mm,clip=}
   \caption{Execution time}
   \label{fig:time_d_window}
\end{subfigure}
\begin{subfigure}[b]{0.49\linewidth}
\centering
   \epsfig{file=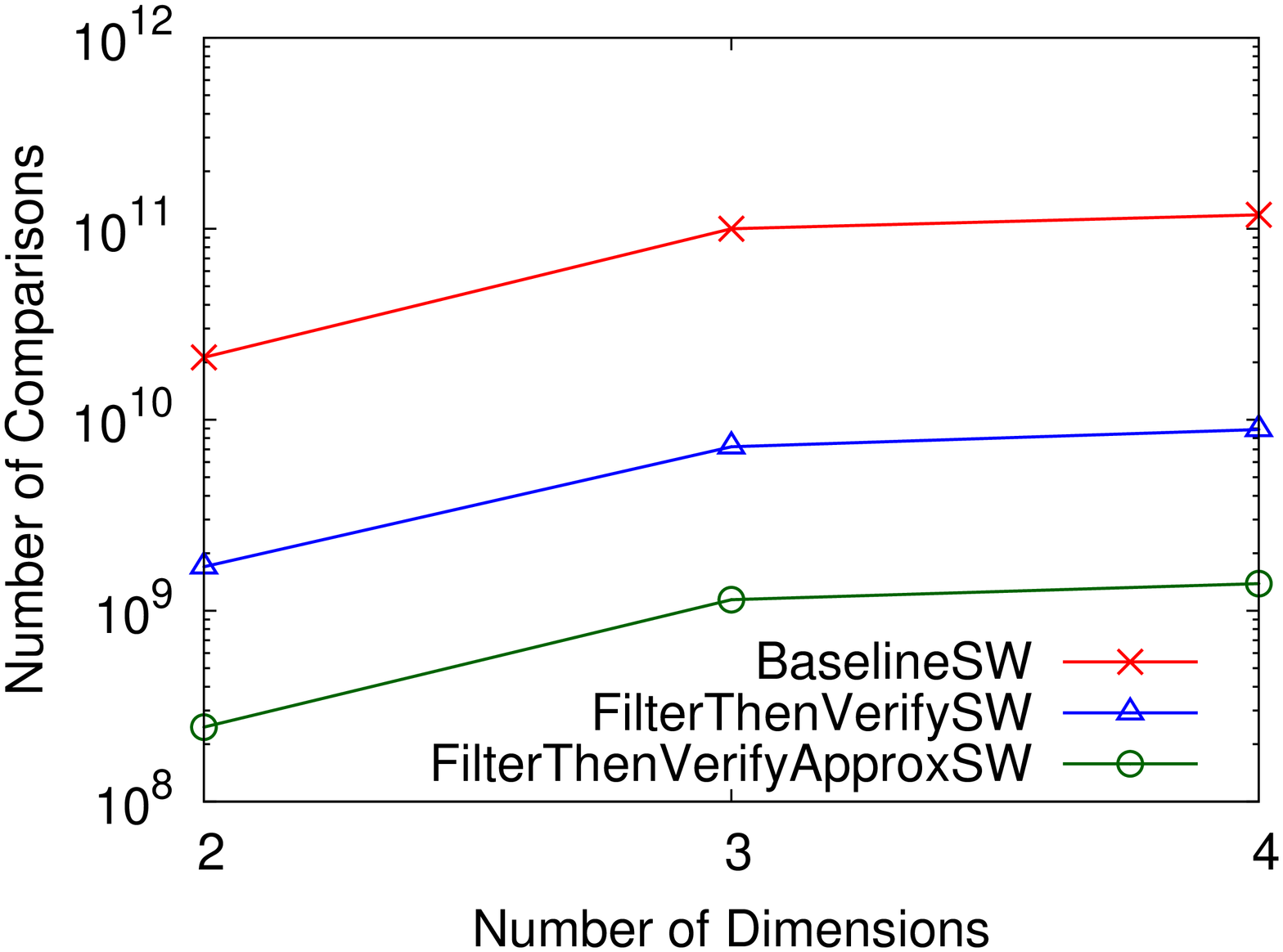, angle=-90, width=42mm,clip=}
   \caption{Object comparisons}
   \label{fig:comparison_d_window}
\end{subfigure}
\caption{\small Comparison of \algname{BaselineSW}, \solname{FilterThenVerifySW} and \solname{FilterThenVerifyApproxSW} on the movie dataset. Varying $d$, $W$ = $3{,}200$, $|\mathcal{O}|$ = $1$M, $h$ = $0.55$.}
\label{fig:d_sw}
\end{minipage}
\hspace{1mm}
\noindent
\begin{minipage}{0.48\textwidth}
\centering
\begin{subfigure}[b]{0.49\linewidth}
\centering
   \epsfig{file=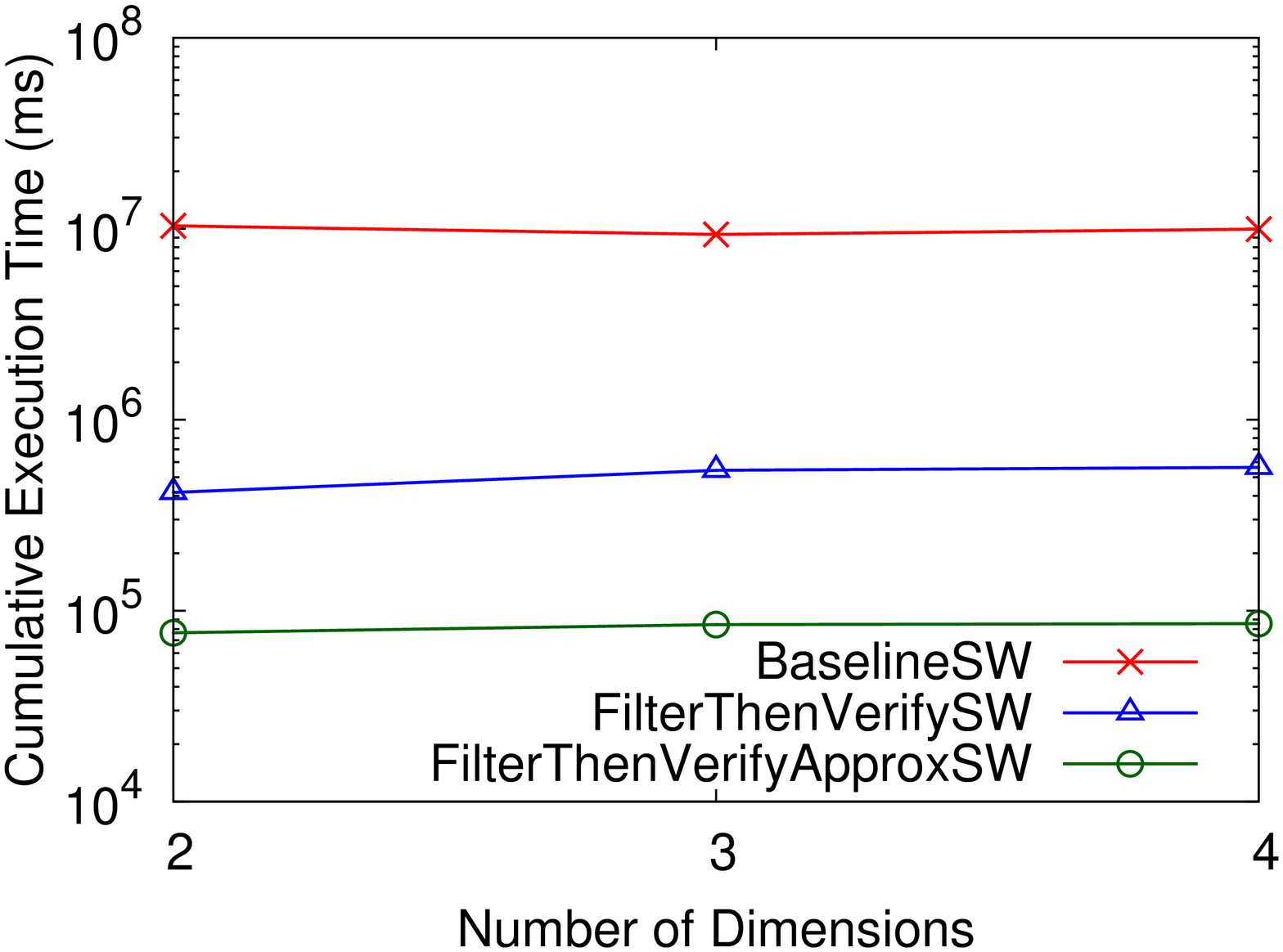, angle=-90, width=42mm,clip=}
   \caption{Execution time}
   \label{fig:time_d_window_p}
\end{subfigure}
\begin{subfigure}[b]{0.49\linewidth}
\centering
   \epsfig{file=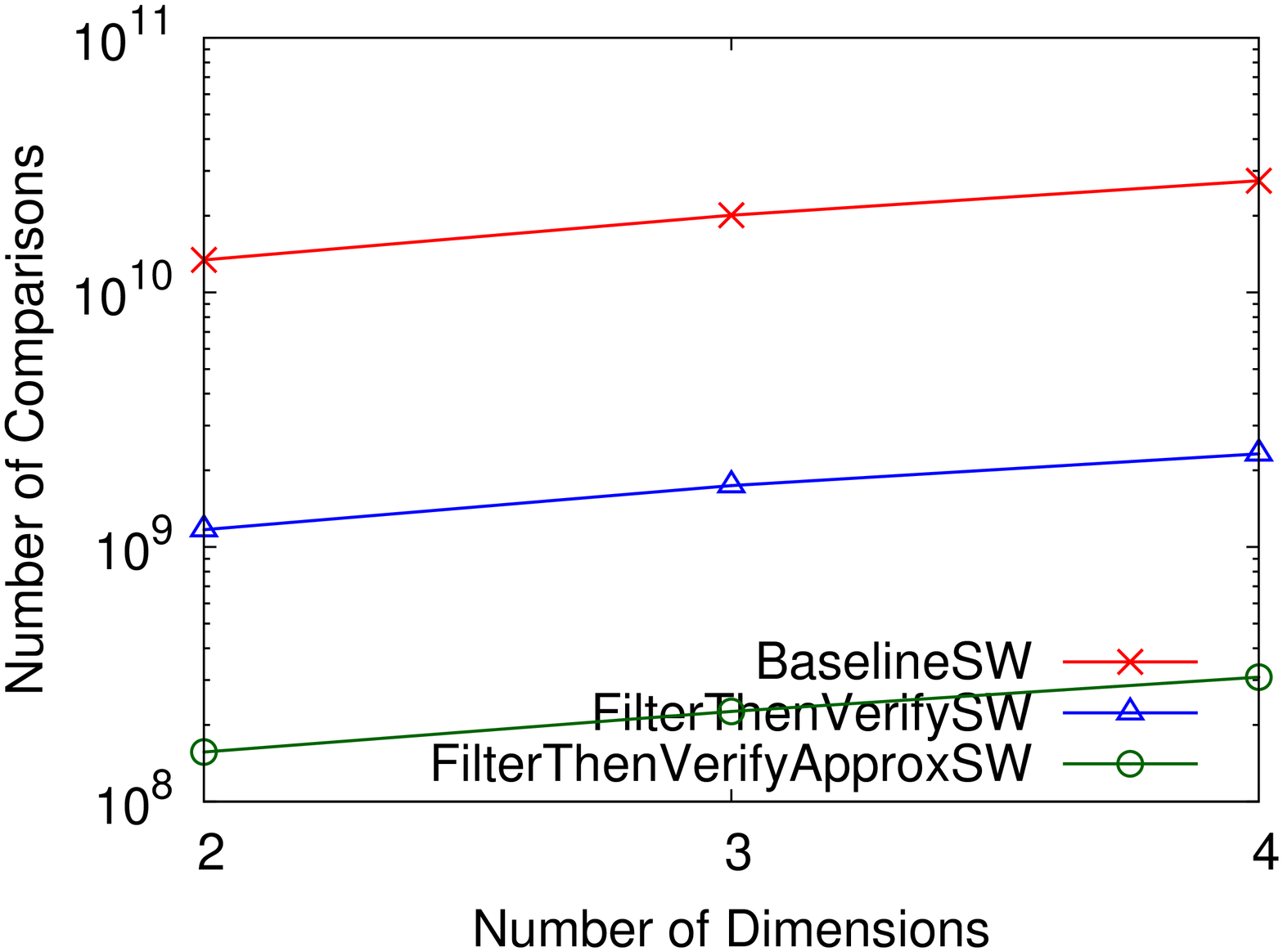, angle=-90, width=42mm,clip=}
   \caption{Object comparisons}
   \label{fig:comparison_d_window_p}
\end{subfigure}
\caption{\small Comparison of \algname{BaselineSW}, \solname{FilterThenVerifySW} and \solname{FilterThenVerifyApproxSW} on the publication dataset. Varying $d$, $W$ = $3{,}200$, $|\mathcal{O}|$ = $1$M, $h$ = $0.55$.}
   \label{fig:d_p_sw}
\end{minipage}
\end{figure*}

\begin{table*}[t]
\centering
\scriptsize
\begin{tabular}{|l|l|l|l|l|l|l|l|l|l|l|l|l|l|}
\hline
\multirow{2}{*}{Data stream} & \multirow{2}{*}{$W$} & \multicolumn{3}{c|}{$h = 0.70$} & \multicolumn{3}{c|}{$h = 0.65$} & \multicolumn{3}{c|}{$h = 0.60$} & \multicolumn{3}{c|}{$h = 0.55$} \\ \cline{3-14}
                  &  & Precision & Recall & F-measure & Precision & Recall & F-measure & Precision & Recall & F-measure & Precision & Recall & F-measure \\ \hline\hline
\multirow{4}{*}{Movie} & $400$ & $100$	& $89.36$	& $94.38$	& $100$	& $87.33$	& $93.24$	& $100$ &	$85.94$	& $92.44$ & $100$ & $81.95$ & $90.08$\\ \cline{2-14}
                  & $800$ & $100$	& $87.87$ & $93.54$	& $100$ &	$85.78$ & $92.34$ &	$100$ & $84.04$ & $91.33$ & $100$ & $80.10$ & $88.95$
\\ \cline{2-14}
                  & $1600$ & $100$ & $88.65$ & $93.98$ & $100$ & $86.58$ & $92.81$ & $100$ &	$85.01$ &	$91.90$ &	$100$ & $81.10$ & $89.56$
\\ \cline{2-14}
                  & $3200$ &  $99.99$ & $94.80$ & $97.33$ & $100$ & $93.08$ & $96.41$ & $100$ & $92.29$ & $95.99$ & $100$ & $88.99$ & $94.17$
 \\ \hline
\multirow{4}{*}{Publication} & $400$ &  $100$ &	$94.58$ & $97.21$	& $100$ & $93.57$ & $96.68$ & $100$	& $92.98$ & $96.36$ & $100$ & $92.06$ & $95.87$
\\ \cline{2-14}
                  & $800$ &  $100$ & $94.79$ &	$97.32$ & $100$ & $93.60$ & $96.70$ & $100$ & $93.01$ &	$96.38$ &	$100$ & $91.98$ & $95.82$
\\ \cline{2-14}
                  & $1600$ & $100$ & $94.62$ & $97.24$ & $100$ & $93.44$ & $96.61$ & $100$ & $92.85$ &	$96.29$ & $100$ & $91.81$ & $95.73$
\\ \cline{2-14}
                  & $3200$ & $100$ & $96.71$ & $98.33$ & $100$ & $95.98$ & $97.95$ & $100$ & $95.67$ &	$97.79$ &	$100$ & $95.27$ & $97.58$
\\ \hline
\end{tabular}
\caption{\small The precision, recall and F-measure (in percentage) of \solname{FilterThenVerifyApproxSW}. Varying $W$ and $h$, $|\mathcal{O}|$=$1$M, $d$=$4$.}
\label{tab:efficacy_m_sw}\vspace{-8mm}
\end{table*}

We further compare the performance of \solname{FilterThenVerifySW} and \solname{FilterThenVerifyApproxSW} with \solname{BaselineSW}. In this regard, we simulated two data streams---movie and publication where $\mathcal{O}$ is composed of duplicated sequence of the corresponding dataset such that $|\mathcal{O}|$=$1$ million. Following~\cite{tao2006maintaining}, we experimented with windows of size $400$, $800$, $1{,}600$, and $3{,}200$, as well as report the cumulative execution times in milliseconds.  In this direction, Fig.\ref{fig:time_window_m} demonstrates the cumulative execution times (by milliseconds, in logarithmic scale) of the aforementioned methods on the movie stream. Fig.\ref{fig:time_window_m} shows that the cumulative execution times increase super-linearly by $W$ as wider window broadens the size of Pareo frontiers. These figures illustrate that both \solname{FilterThenVerifySW} and \solname{FilterThenVerifyApproxSW} outperformed \algname{BaselineSW} by $1$ to $2$ orders of magnitude, which concurs with the comparative behaviours of \solname{FilterThenVerify}, \solname{FilterThenVerifyApprox} and \algname{Baseline}. This concurrence is also applicable for the publication stream (Fig.\ref{fig:time_window_p}).

Fig.\ref{fig:comparison_window_m} (Fig.\ref{fig:comparison_window_p}) further reveals the amount of work done by these solutions, in aspect of compared objects (in logarithmic scale) to maintain Pareto frontiers over sliding window. Moreover, Fig.\ref{fig:time_d_window} (Fig.\ref{fig:time_d_window_p}) depicts the effectiveness of \solname{FilterThenVerifySW} (\solname{FilterThenVerifyApproxSW}) on varying $d$.  Fig.\ref{fig:comparison_d_window} (Fig.\ref{fig:comparison_d_window_p}) clarifies Fig.\ref{fig:time_d_window} (Fig.\ref{fig:time_d_window_p}) through illustrating the number of compared objects. The reason behind the comparative behaviour of \algname{Baseline}, \solname{FilterThenVerify} and \solname{FilterThenVerifyApprox} is also applicable in this case. In addition, \algname{BaselineSW} maintains exclusive Pareto buffer for each user ($\mathcal{PB}_{c}$) while \algname{FilterThenVerifySW} shares a Pareto buffer across users in a cluster ($\mathcal{PB}_{U}$). Therefore, in sliding window protocol, the filter-then-verify approach attains the benefit of clustering in greater extent.

Table~\ref{tab:efficacy_m_sw} demonstrates the precision, recall and F-measure of \solname{FilterThenVerifyApproxSW} on varying $W$ and $h$. We can observe that the recall declines slowly by $h$. Nevertheless, $h$ does not have significant impact on the efficacy of \solname{FilterThenVerifyApproxSW}. Besides, the loss of accuracy is due to false negatives rather than false positives. These behaviors concur with \solname{FilterThenVerifyApprox} and the reasons behind are same as before. In addition, Table~\ref{tab:efficacy_m_sw} reveals that $W$ does not have noticeable impact on efficacy and \solname{FilterThenVerifyApprox} remains effective on varying $W$. 
\section{Conclusion}\label{sec:conclusion}
We studied the problem of continuous object dissemination, which is formalized as finding the users who approve a new object in Pareto-optimality. We designed algorithm for efficient finding of target users based on sharing computation across similar preferences. To recognize users of similar preferences, we studied the novel problem of clustering users where each user's preferences are described as strict partial orders. We also presented an approximate solution of the problem of finding target users, further improving efficiency with tolerable loss of accuracy. Lastly, we performed a thorough experimental study to evaluate the efficiency and effectiveness of the proposed solutions.

{\flushleft \textbf{Acknowledgement}} \hspace{2mm} We would like to thank Fatma Arslan for her contribution in data collection.\vspace{-2mm}
\bibliographystyle{abbrv}
\small
\bibliography{prefquery}
\normalsize
\begin{comment}
\pagebreak
\textbf{Supplemental Material to ``Continuous Monitoring of Pareto Frontiers on Partially Ordered Attributes for Many Users''}
\begin{appendices}
\input{sec-appendix}
\end{appendices}
\end{comment}

\begin{comment}

Notes: 
What is a false +ve?
Comparison w.r.t CPS.

\begin{definition}[Indifferent Tuple]\label{def:indifferent_tuple}
if neither $(x,y)$ nor $(y,x)$ belongs to the relation $\succ_c^d$ then this binary relation is denoted as $\sim_c^d$ (i.e., $(x,y) \in \sim_c^d$, also denoted $x \sim_c^d y$) and $(x,y)$ is called an \emph{indifferent tuple}, i.e., ``user $c$ finds $x$ and $y$ indifferent to each other on attribute $d$''. The relation is irreflexive $((x,x) \notin \sim_c^d$) and symmetric ($(x,y) \in \sim_c^d \Rightarrow (y,x) \in \sim_c^d$).\closedef
\end{definition}

Given two objects $o, o' \in \mathcal{O}$, $c$ neither prefers $o'$ to $o$ nor $o$ to $o'$ if $o$ and $o'$ are not preferred to each other on at least one attribute.
Formally, $o' \sim_c o$, if and only if $(\exists d \in \mathcal{D} : o'.d \sim_c^d o.d)$.
\end{comment}
\end{document}